\documentclass[a4paper,11pt]{article}
\usepackage[utf8]{inputenc}
\usepackage[english]{babel}
\usepackage{amsmath}
\usepackage{graphicx}
\usepackage{hyperref}
\usepackage{todonotes}
\usepackage{natbib}
\usepackage{subfig}
\usepackage{multirow}
\usepackage{rotating}
\usepackage{dcolumn}
\usepackage{CJKutf8}
\usepackage{authblk}

\usepackage{newtxtext}
\usepackage{newtxmath}
\usepackage{eqnarray}

\usepackage[a4paper,% other options: a3paper, a5paper, etc
        left=2.cm,
        right=2.cm,
        top=2.cm,
        bottom=2.cm,
]
{geometry}

\title{Levels of structural change: An analysis of China's development push 1998-2014}
\author[1,2,3,$\dagger$]{Torsten Heinrich}
\author[4,1,2,$\ddagger$]{Jangho Yang}
\author[5,6,7,$\P$]{Shuanping Dai}

\affil[1]{Oxford Martin Programme on Technological and Economic Change, Oxford Martin School, University of Oxford, Oxford OX1 3BD, UK}
\affil[2]{Institute for New Economic Thinking at the Oxford Martin School, University of Oxford, Oxford OX1 3UQ, UK}
\affil[3]{Department for Business Studies and Economics, University of Bremen, 28359 Bremen, Germany}
\affil[4]{Department of Management Sciences, Faculty of Engineering, University of Waterloo, Waterloo, ON, N2L 3G1, Canada}
\affil[5]{School of Economics, Jilin University, 130012, Changchun, China}
\affil[6]{Centre for China Public Sector Economy Research, Jilin University, 130012, Changchun, China}
\affil[7]{IN-EAST Institute of East Asian Studies, Universit\"{a}t Duisburg-Essen, 47057 Duisburg, Germany\authorcr \textcolor{white}{--}}
\affil[$\dagger$]{\tt torsten.heinrich@uni-bremen.de}
\affil[$\ddagger$]{\tt j643yang@uwaterloo.ca}
\affil[$\P$]{\tt shuanping.dai@uni-due.de\authorcr \textcolor{white}{--}}

\begin{document}

\maketitle

\begin{abstract}
We investigate structural change in the PR China during a period of particularly rapid growth 1998-2014. For this, we utilize sectoral data from the World Input-Output Database and firm-level data from the Chinese Industrial Enterprise Database. Starting with correlation laws known from the literature (Fabricant's laws), we investigate which empirical regularities hold at the sectoral level and show that many of these correlations cannot be recovered at the firm level. For a more detailed analysis, we propose a multi-level framework, which is validated empirically. For this, we perform a robust regression, since various input variables at the firm-level as well as the residuals of exploratory OLS regressions are found to be heavy-tailed. We conclude that Fabricant's laws and other regularities are primarily characteristics of the sectoral level which rely on aspects like infrastructure, technology level, innovation capabilities, and the knowledge base of the relevant labor force. We illustrate our analysis by showing the development of some of the larger sectors in detail and offer some policy implications in the context of development economics, evolutionary economics, and industrial organization.


\end{abstract}

\tableofcontents

\section{Introduction}
% TODO: introduction

\begin{figure}[tb!]
\centering
\includegraphics[width=0.75\textwidth]{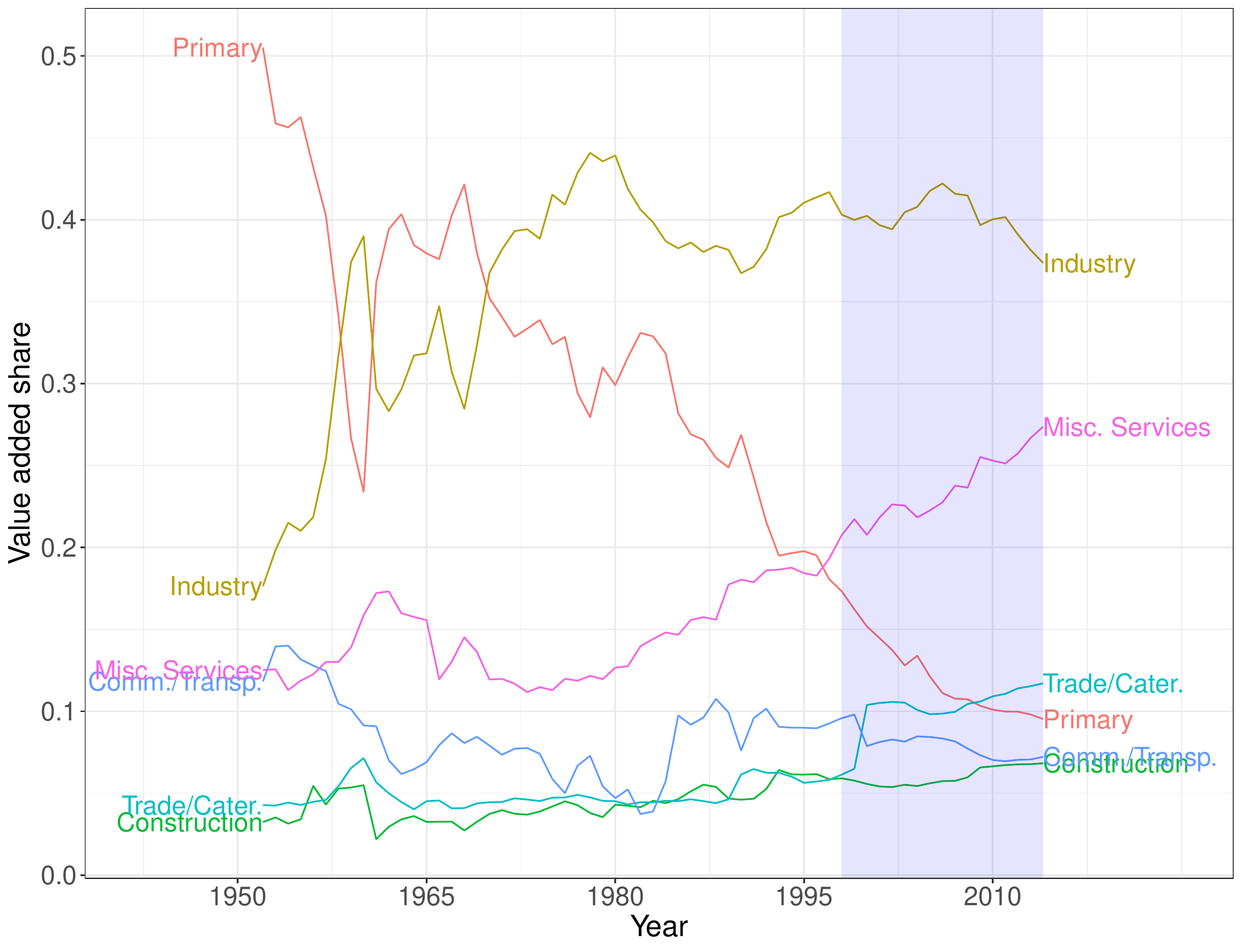}
\caption{Changes in value added shares by sectors (data from \citep{Holz05} until 1999, after that from WIOD). Period covered in our micro-data and sectoral data shaded.}
\label{fig:Dev:longterm}
\end{figure}

Just as the PR China took its first steps towards economic reforms and modernization in the 1970s, a study on the Chinese labor market conjectured that ''the growth rate of non-agricultural employment is a crucial determinant of China's future labour scene'' as unemployment was a real danger for developing economies \citep{Rawski79}. The hypothesis certainly proved correct, but for reasons different from those the author imagined. The role of agriculture in the Chinese economy declined rapidly, its employment share falling below 25\% of the labor force by 2014 while its contribution to value added fell from a third in the 1970s to below 10\% in 2014 (see Fig. \ref{fig:Dev:longterm}). This was accompanied by a period of unprecedented growth, profound economic reorganization, and China's rise into the group of technological advanced societies. For details on the reforms and the resulting changes in economics and policy, see, e.g. \citet{brandt2008china,fan2003structural}. Unemployment was not a major concern, although large parts of the population moved to urban regions; the share of the industrial sector in the economy remained roughly constant in terms of output (value added\footnote{At the sectoral or micro-level, value added is the appropriate equivalent to output variables in macroeconomic models. Gross output would lead to double-counting of intermediate inputs and reflects both activity in the sector and in supplier sectors.}) while the service sector grew at the expense of the agricultural sector (Fig. \ref{fig:Dev:longterm}). 

The details are more complicated, however. Growth of output and growth of employment differ significantly between sectors, labor productivities diverge, and unique dynamics emerge, setting the developing Chinese economy apart from the other extensively studied economies of developed countries. What is more, sectors are also not homogeneous entities but rich ecosystems of a variety of firms the characteristics of and growth dynamics of which differ widely, although firm-level variables can be recovered by regular distributional models with great accuracy. The present paper aims to shed light on the details of this structural change.

We propose a multi-level framework of structural change that connects sectoral level and firm-level, the two levels on which we have empirical data. The framework is stated in equation form, it is testable and it will be used for the econometric analysis in Section \ref{sect:regressions}. Further, we investigate whether a variety of correlation laws known from the literature as Fabricant's laws \citep{Fabricant42,Scott91,Metcalfeetal06} hold at the sectoral and at the micro-level. Assessing whether empirical laws hold across multiple levels of aggregation allows us to understand where the regularities originate; whether they are universal properties of a self-organizing system, that might hold at every level, or emergent patterns, that arise at a specific level.

If the economy is characterized by fractal, self-similar structures, we might assume that empirical observations at aggregated levels are mirrored in more detailed data, at the micro-level. Scaling relationships, for which this is the case, have indeed been found in many economic phenomena, from financial market time series \citep{Marsili02} to price developments \citep{Farmer/Lafond16} to economic geography \citep{Bustosetal12,Haldane19} to industrial organization and firm-level data \citep{Schwarzkopfetal10}. While this view is typical for a complex systems perspective in economics, \citet{Metcalfeetal06}, writing from an evolutionary economic perspective, appear to consider a similar view when forging their empirical findings into a general model of structural change and economic growth. \citet{Martin/Sunley07} offer a comprehensive overview over the complex systems perspective and potential synergies to evolutionary economics. Of course, scaling laws do not have to extend to the micro-level. Sometimes empirical laws emerge from phenomena at intermediate levels, in which case the characterization of these intermediate levels and the associated phenomena are of particular interest.

While we can mostly confirm these laws at the sectoral level, we will show that they do, in fact, fall apart at the micro-level. We will characterize the properties and the dynamic processes at work in this multi-level system of structural change. Further, we address some related questions such as the nature of distributions of variables at the firm level and the autocorrelation and dispersion of the sectoral composition of the Chinese economy. We also show the development of some of the larger sectors in detail and offer some policy implications in the context of development economics, evolutionary economics, and industrial organization.

China's structural reforms were clearly a driving factor in China's structural change. There were extensive privatization efforts,  private enterprises were legalized and encouraged, rural industry was privatized, and state-owned enterprises (SOEs) were reformed. 
The PR China also increasingly encouraged foreign investments starting in the late 1970s, first promoting joint venture type ownership structures before allowing more direct foreign investments in Special Economic Zones, several of which were established starting in the 1980s. In 1988, the constitution was revised to legalize the private sector; private enterprises were now considered to be complementary to the socialist economy. Amendments in 1999 and 2004 further protected and encouraged private entrepreneurship and strengthened the private sector. Significant reforms of SOEs and TVEs (Township and Village Enterprises) were undertaken starting in the 1990s.
Efforts were also made to stimulate innovation and to bolster R\&D \citep{Hu/Jefferson08}; this included tax incentives and grants, improved formal procedures and legal framework in economic policy, enterprise law and patenting, consolidation and privatization of research institutes as well as some focused public research programs in basic and high-tech research (such as \textit{Program 863} and \textit{Program 973}). There was also a focus on modernization of manufacturing sectors that aided productivity growth in this field \citep{brondino2019productivity}.

However, China's development and transformation to a market economy was in many ways not spontaneous or unorganized, different perhaps from other transformation economies in Eastern Europe and Central Asia. There was structural change towards a market oriented system, but the transformation was government-organized and the economy remains partly under the control of the planned economic system \citep{shen2019toward}. For example, early reforms in the 1970s and 1980s encouraged local management and stakeholders to assume a more active role in defining the firms' objectives and strategy, first in agricultural TVEs, later in other SOEs \citep{Naughton08}.\footnote{Very early in the reform, TVEs were for instance permitted to expand from agriculture into food processing \citep{Naughton08}.} Before 1998, the ownership type of the firms was rarely changed and state ownership was typically retained. The more radical and substantial SOEs reform in 1998 allowed privatizing SOEs and layoffs of substantial numbers of employees. %who had an "iron rice bowl". 
TVEs were subject to a similar reform; the TVEs share of China's GDP dropped from 65.1\% in 1993 to 9.1\% in 2000. 

This had multiple obvious effects for the industry structure: Large SOEs quickly lost ground, a multitude of private start-ups were established and the firm population became much younger across the board (see Section \ref{sect:results} and in particular Figure \ref{fig:Dev:sectoral:Age}). The layoffs likely also fueled the rise of the services sector the employment share of which grew from 30\% in 2000 to 45\% in 2014. 

China's entry into the WTO in 2001 brought increased international integration of the Chinese economy, reduced tariff uncertainty, and fueled productivity growth of especially those regions and sectors with more reliance on exports and exposure to foreign investment \citep{erten2019exporting,brandt2017wto}. Unlike most countries, China also increasingly relies on domestic inputs for exportation, despite its deepening global engagement \citep{kee2016domestic}. 

Finally, a four trillion Yuan fiscal stimulus program in the wake of the financial crisis in 2008, while averting larger fallouts of the crisis, may have extended the life-span of many struggling firms, SOEs in particular \citep{yan2020business}.

It is to be expected that these processes would have had an impact on the dynamics of structural change at various levels. The fact that economic growth was rapid in China in the 1990s and 2000s, reaching growth rates around 10\%, probably accelerated the structural change brought about by modernization and the reforms towards a market economy. At the same time, the Chinese case is unique in that detailed micro-level data are available that permit us to investigate structural change not just at the sectoral but also at the firm level (see Section \ref{sect:results}).

The remainder of the paper is organized as follows: Section \ref{sect:literature} reviews previous literature contributions, Section \ref{sect:model:evol} offers some considerations on modeling of structural change at the sectoral and the micro-level with a focus on empirical applicability. Section \ref{sect:results} discusses empirical regularities, Section \ref{sect:regressions} fits some of the models introduced in Section \ref{sect:model:evol}. Section \ref{sect:conclusion} concludes.

\section{Literature review: Structural change in China and the world} 
\label{sect:literature}

Economists and statisticians first started to categorize economic activity in sectors of different characteristics in the last years of the 19th century. Year-books in Australia and New Zealand began distinguishing primary - agricultural, pastoral, mineral production - and secondary - derived manufacturing - sectors, later adding a tertiary sector for services; from there the terms spread through the commonwealths and around the world \citep{Fisher39}. Other scholars had, at the time already recognized that parts of the economy and of the industry may be subject to different fluctuations \citep{Robertson1915} and different dynamics (involving, e.g., decreasing and increasing returns \citep{Clapham1922}). However, it was \citet{Fabricant42}, who offered the first detailed study covering 50 industry sectors in the US over a period of 40 years. %...

The rise and fall of industrial and sectoral shares indicate fundamental structural change in the long-term economic growth \citep{baumol1967macroeconomics,maddison1983comparison}, with industrialization, urbanization, and technological change being crucial factors in structural change \citep{syrquin1988patterns}. Maintaining high rates of economic growth has been linked to keeping the sectoral composition balanced \citep{chenery1975patterns,rostow1960stages}.

\subsection{Regularities}

\citet{Fabricant42} identified a variety of correlation laws in sectoral accounts:
\begin{itemize}
  \item positive correlation of output growth and labor productivity growth,
  \item negative correlation of output growth and growth of wage bill per output,
  \item positive correlation of output growth and capital growth,
  \item positive correlation of output growth and employment growth, 
  \item positive correlation of value added per output growth and wage growth,
  \item positive correlation of value added per output and output price,
  \item positive correlation of output growth and capital intensity growth,
  \item negative correlation of the wage bill per unit of output and labor productivity,
  \item negative correlation of value added growth and output price growth.
\end{itemize}

Other scholars have attempted to confirm and extend the correlation laws \citep{Scott91}. The first law, correlation of output and productivity growth, corresponds to Verdoorn's law and has sparked discussions on the causality behind this relation (cf. \citet{Scott91}) while the other laws were less influential in the literature. Besides genuine increase in efficiency the correspondence may be due to either labor and capital reallocation or due to characteristics of the sector. Such characteristics may involve different rates of technological change, economies of scale, or other aspects.
% 4 options, differential tech change, labor reallocation to high-tech/high-productivity/CI sectors, increased efficientcy, economies of scale

\subsection{Productivity growth and structural change}

General equilibrium approaches in neoclassical or neo-Keynesian tradition have generally favored the factor re-allocation explanation. TFP growth rates across industrial sectors are conjectured to function as a valid predictors of labor moving among sectors. Employment share drops in different rates among the sectors; specifically, labor tends to move away from technology intensive sectors and towards sectors with low growth \citep{ngai2007structural}. 

Evolutionary scholars, on the other hand, often view sectors as interdependent systems with specific characteristics  \citep{Nelson/Winter75,Montobbio02}. \citet{fagerberg2000technological}, for instance, argues that structural change on average has not been conducive to productivity growth, which instead may be determined by technological change. \citet{Metcalfeetal06} offer an extensive analysis of structural change and economic growth which they then try to relate back to the macroeconomic level via income and demand.\footnote{It should be noted that this analysis may be affected by the presence of heavy-tailed distributions as it uses the variance of growth rates, which may not exist as argued in Section \ref{sect:results:distributions} and \ref{sect:regressions} below.} Crucially, Metcalfe et al. show that there is a wide variety of compositions of growth rates of employment on the one hand and output on the other, which they find to be correlated although the autocorrelations decay with the time lag.

Another strand of literature is devoted to the investigation of the sources of innovation that drives technological change and arguably, on occasion, structural change. \citet{Schumpeter43} argued that only large firms are able to afford innovative research and put the desire to create temporary competitive advantages at the centre of entrepreneurship. Later evolutionary thinkers \citep{Nelson/Winter82} extended this to formal and stochastic models with innovative research and imitation creating a complex dynamic of cutting-edge technological progress and the diffusion of technologies to the rest of the industry. This was found to be able to conveniently recover business cycles \citep{Silverberg/Lehnert93} and highly skewed firm size distributions \citep{Kwasnicki98} in simulation models. The introduction of stochasticity coincidentally made innovation by small firms possible and likely.

\citet{Freeman/Perez88} conjectured technological paradigms to be at the centre of such long-wave processes. This would imply long periods of relative structural stability interspersed by periods of relatively rapid replacement of the technological infrastructure. While structural change is not necessarily bound to technological change, an interaction between these processes may be assumed. \citet{Pyka/Saviotti13} model industry sector life cycles and economic development such that the number of industry sectors increases when earlier sectors are sufficiently developed to support the emergence of new ones.

Contrary to Schumpeter's hypothesis on the innovative capabilities of large firms, it appears that small firms may on average be slightly more innovative \citep{Nooteboom94}. It has been shown that there are economies of scale from local knowledge spillovers (Jacobs externalities) that are different from simple Marshallian agglomeration effects \citep{Beaudry/Schifferauova09}. In light of this, path-dependence would be expected to lead to divergent development in the economies of regions, but because of regional specialization in industries also between sectors.

\subsection{The transformation economy and economic development in China}

A substantial literature has addressed productivity growth in China. While productivity growth has implications for structural change, most analyses that consider this question are limited to the classical three-sector model \citep{fisher1935clash,clark1940conditions}. Few contributions attempt to investigate structural change from a micro-perspective. Recent examples include \citep{hsieh2009misallocation,brandt2012creative,ding2016determinants}, where firm level data is applied to calculate total factor productivity (TFP). %, but they tend not to view the changes within a structural change framework.

Another branch of literature \citep{Duschl/Peng15,yu2015institutional} investigates the distribution of growth rates in China (computed using value added growth or sales growth) in the tradition of Bottazzi and Secchi's analysis \citep{Bottazzi/Secchi06,Bottazzi/Secchi11} where the distribution is conjectured to belong to the family of Asymmetric Exponential Power (AEP) distributions. However, there is also strong evidence that both growth rates and many other quantities at the firm level are heavy-tailed \citep{Axtell01,Heinrich/Dai16,Yangetal19,Heinrichetal19}, such that the tails could be severely underestimated by AEP fits.

What is not in doubt is that China experienced impressive growth accompanied by significant increases in productivity. This mirrors the development in other developing countries.

\citet{fan2003structural} construct a multi-sector model to investigate how production factors reallocate among urban industry, urban services, agriculture, and rural enterprises in China. They find that labor movements from low-productivity sectors to high-productivity sectors have significantly contributed to China’s structural change. 
% labor reallocation low prod -> high prod
%
\citet{dong2009labor} find that China experienced a more synchronized pace of job destruction and creation, compared to the other transition economies while \citet{wang2008productivity} finds no regional divergence of productivity trends. 
% no regional divergence
% \citep{kamal2013labor}

Two main points of contention in the literature on structural change in China are (1) if growth could have been even faster if resources were not misallocated and (2) which factors were driving the growth in output and the increase in labor productivity.

Question (1) is based on the conjecture that productivity dispersion indicates misallocation of resources, as there are evidently many firms in the same country and sector with widely different productivities. Should resources not be reallocated to the most productive firms, if the less productive ones refuse to learn and copy better technologies? And indeed, evidence for dispersion and thus misallocation is then found for China as well \citep{hsieh2009misallocation,brandt2012creative,song2011growing}. What is more, standard measures of productivity dispersion (the variance, the Olley-pakes gap, the inter-quantile range) may make it appear that dispersion is increasing over time. Would this not be bad, even if there was an unavoidable baseline misallocation due to bounded rationality? However, labor productivities (and total factor productivities, TFPs) follow heavy-tailed distributions, which makes these dispersion measures unreliable \citep{Yangetal19}. Other dispersion measures, that are robust with respect to heavy tails, such as the scale parameter of L\'{e}vy alpha stable fits, show no sign of increasing, at least not relative to the median of the distribution \citep{Yangetal19,Heinrichetal19}.

%The mainstream about China’s structural change is to look at the changes of the labor shares of agricultural, industrial and services sectors in its GDP, and most agrees that China’s labor productivity was making a remarkable improvement but does not reach what it should be \citep{brandt2008china,song2011growing}. The driving forces for the improvement and constraints usually are attributed to, such as factor reallocation, technological change, and institutional frictions and so on.
% changing labor allocation
% driving forces: factor reallocation, technological change, and institutional frictions and so on.

No consensus has emerged with respect to the sources of productivity growth in China. 
\citep{yang2010sources} suggest that structural change is the primary reason for productivity growth. 
%\citep{tombe2019trade}
%\citep{li2012end}
% 
\citet{nabar2013sector} claim that the low labor productivity growth in services is a unique compared to other transformation economies. They suggest that credit and labor market frictions are the primary inhibiting reasons. \citet{brandt2013factor} come to similar conclusions with respect to total factor productivity (TFP) growth in China and argue that labor mobility restrictions a distorted incentive structure are at fault. However, \citet{brondino2019productivity}, applying the ''growing subsystems'' developed by \citet{pasinetti1988growing}, argues that aggregate productivity growth in China in 1995-2002 was driven by technological progress rather than sectoral reallocation of labor. 
% low LP growth in services?? compare Fig. \ref{fig:MAC:long-term-absolute_growth}
% low TFP growth in services
%      ...because of labor market rigidities
% LP/TFP growth because of structural change or because of technological progress
% 
%Market-oriented institutional changes stimulated productivity growth. %-> ??
\citet{wang2008productivity} find that shifts in firm ownership can explain a substantial part of the productivity growth after 1985. Reforms not only increased efficiency by encouraging private entrepreneurship, also SOEs exhibited significant productivity growth. \citet{fu2008productivity} demonstrate that some reforms in SOEs have improved SOEs' productivity level during the period of 1986-2003 even when taking macroeconomic growth into account. Especially corporate governance reforms have led SOEs to make improvements in innovation, technological change and adoption of new technologies. 
% Changes in the composition of firm types increase LP
%              ... but there are other factors too
%
\citet{dekle2012quantitative} argue that tax reduction also helped labor mobility for productivity growth. 
% taxes
%
Institutional frictions may also have jeopardized productivity growth. \citet{chen2011structural} find that structural change does not significantly contribute the continuing TFP growth since 2001, because the privileges state-owned institutions enjoy discouraged private capital. 
% institutions

\section{Modelling structural change at multiple levels}      
\label{sect:model:evol}

With the period 1998-2014, we are investigating an interesting case involving both economic development (China still being a developing country), technological change, and transformation to a market economy. If we aim to address questions of deep structural change in this context, we must be able to relate multiple dimensions - employment, output, and productivity - of economic structure as well as multiple levels - the sectoral and the firm level. In the present section, we offer some considerations in this regard, starting with a sector level model (Section \ref{sect:model:evol:sector}), before moving to the firm-level (Section \ref{sect:model:evol:firm}) and adding some technical considerations regarding variable densities at the firm level (Section \ref{sect:model:evol:densities}).

\subsection{Sector-level dynamics}
\label{sect:model:evol:sector}

We aim to study structural change at both the sectoral level and the firm-level and will for this start with a framework inspired by \citet{Metcalfeetal06}

Labor productivity $Q_t$ is the quotient of output $Y_t$ and employment $L_t$ at time $t$, 

$$Q_t = Y_t/L_t.$$

Let disaggregated - sectoral and firm-level - quantities be denoted by lower case letters, labor productivity $q_t$, value added $y_t$ and employment $l_t$. We will omit time index $t$ for now. Hence, for sector $k$,

$$\begin{array}{r l}
q_k &= y_k/l_k \\
    &= \frac{s_{Y,k} Y}{s_{L,k}L} 
\end{array}$$
where $s_{Y,k}$ is the value added share of sector $k$ and $s_{L,k}$ is its employment share.  
For growth rates denoted as $\dot{Q}$, $\dot{Y}$, $\dot{L}$, etc., and using the approximation 

\begin{equation}
\label{eq:logapproximation}
\log\left(\frac{x_{t+1}}{x_{t}}\right)\approx\frac{x_{t+1}-x_{t}}{x_{t}}
\end{equation}
that holds in the vicinity of zero ($\frac{x_{t+1}-x_{t}}{x_{t}}\approx 0$), we have approximately

$$\dot{Q} = \dot{Y} - \dot{L}$$
$$\dot{q_k} = \dot{s_{Y,k}} - \dot{s_{L,k}} + \dot{Y} - \dot{L}$$
$$\dot{q_k} - \dot{Q} = \dot{s_{Y,k}} - \dot{s_{L,k}},$$
where the left side has unit money per employee while the right-hand side terms are shares. The three elements can then be compared to one another by sectors. While two of the three terms could be statistically independent or could depend on one another in a relationship that could take any number of functional forms, \citet{Metcalfeetal06} conjecture that there is a linear relationship between $\dot{L}$ and $\dot{Y}$ and similarly on disaggregated levels between $\dot{l}$ and $\dot{y}$, although they find wildly different slopes between $0.1$ and $1.2$.

While \citet{Metcalfeetal06} have reasonably long data series with 39 observations each for the sectoral accounts of the United States, we have to work with much shorter time series of less than 15 years at the sectoral level and considerably less for the firm-level. The wide range of different coefficient values in \citet{Metcalfeetal06} may also indicate that the relationship is, in fact, not linear. Further, and this is also true for Metcalfe et al.s \citep{Metcalfeetal06} study, fitting along time series leads to resampling problems as growth rates can reasonably be assumed to be autocorrelated. We therefore choose to remain agnostic with regard to the connection of the two growth dimensions, employment and value added (as a proxy for output) growth. We will show the relationships that we find between these variables and others and report correlation coefficients.

\subsection{Relating sector- and firm-level dynamics}
\label{sect:model:evol:firm}

We now move on to the micro-level. Let $y_{i,t}$ and $l_{i,t}$ denote value added and employment in firm $i$, which belongs to sector $k$, at time $t$. Let $s_{Y,i,t}$ be the share of the firm in the economy's total value added (hence, output), while $s_{L,i,t}$ is the firm's share in total employment. It is not convenient to take the same approach following \citet{Metcalfeetal06} as for the sector above, not least because output or value added can be negative at the firm level. This brings the singularity of the growth rates around level zero into the domain, which makes it inconvenient and counter-intuitive to work with growth rate. Since the approximation \ref{eq:logapproximation} holds only in the vicinity of zero and the expression will be arbitrarily far away from zero in the vicinity of the singularity at $x_t=0$, the above approximation will also not hold any longer.

We are now interested in the dynamic development of these variables. Since we are agnostic with respect to the relation between the variables, employment and value added, and the respective growth rates, we will in the following use the abstract term $x_{i,t}$.  $x$ stands for either employment or value added, but can also be applied to other accounts. We will use $dx_{i,t}=x_{i,t}-x_{i,t-1}$ for the first difference, $\dot{x_{i,t}}=dx_{i,t}/x_{i,t-1}$ for the growth rate, $X_t=\sum_j x_{j,t}$ for the economy's total, $\dot{X_t}$ for the growth of the economy total, $s_{i,t}=x_{i,t}/X_t$ for the share in the economy's total, $ds_{i,t}$ for the first difference of the share, and $\dot{s_{i,t}}$ for the growth of the share.

Consider the standard replicator equation \citep{Nowak06,Mulderetal01}

\begin{equation}
\label{eq:standard-replicator}
  ds_{i,t}/dt=s_{i,t}\left(f_{i,t}-\phi_t\right)
\end{equation}
where $f_{i,t}$ is the fitness of $i$ at time $t$ and $\phi_t=\sum_j s_{j,t} f_j$ is the economy average of the fitness term. The dynamical system given by the shares $s$ and fitnesses $f$ for all firms has the desirable characteristic that the dynamic it defines leads to shares that always sum to 100\%. Fitness is, however, an abstract quantity that denotes the firm's evolutionary success in terms of realized growth in variable $x$. Since relative fitness $(f_{i,t}-\phi_t)$ is not directly observable, this is simply an identity stating that relative fitness is equal to the growth rate 

\begin{equation}
\label{eq:standard-replicator-identity}
  \frac{ds_{i,t}/dt}{s_{i,t}}=\left(f_{i,t}-\phi_t\right).
\end{equation}

In theoretical models (e.g., those in \citet{Nelson/Winter82}), it can be identified with existing variables that may be exposed in the model, such as capabilities of the firm, market share, or productivity. The model can also be empirically fitted to understand which quantities impact it.

The same dynamical system can equally be applied to the sector shares instead of the firm shares: 

\begin{equation}
%\label{eq:standard-replicator}
  ds_{k,t}/dt=s_{k,t}\left(f_{k,t}-\phi_t\right).
\end{equation}

Further, the replicator system can equivalently be written in terms of absolutes $x$ instead of shares $s$:

\begin{equation}
%\label{eq:standard-replicator}
  dx_{k,t}/dt= X_ts_{k,t}\left(f_{k,t}-\phi_t\right) + s_{k,t}\left(X_{t}-X_{t-1}\right)
\end{equation}
where the first term simply scales the dynamics of the sectoral shares to the macro-level quantity $X_t$ and the second term accounts for the growth of the macro level quantity. This reduces to the approximation\footnote{Note that this again only holds approximately with approximation \ref{eq:logapproximation} for non-infinitesimal differences. To be exact, the the second term would have to be multiplied by $X_{t-1}/X_{t}$. However, this factor vanishes in the infinitesimal limit and is small compared to errors in the empirical data.}

% \begin{equation}
% %\label{eq:standard-replicator}
%   dx_{k,t}/dt=X_ts_{k,t}\left(\frac{X_t-X_{t-1}}{X_{t-1}}\right) + X_ts_{k,t}\left(f_{k,t}-\phi_t\right)
% \end{equation}

\begin{equation}
\label{eq:absolute-replicator}
  dx_{k,t}/dt=X_ts_{k,t}\left(\dot{X_t} + f_{k,t}-\phi_t\right)
\end{equation}
which corresponds to 

\begin{equation}
\label{eq:absolute-replicator-1}
%\label{eq:standard-replicator}
  \frac{dx_{k,t}/dt}{X_ts_{k,t}}=\dot{x_{k,t}}=\dot{X_t} + \dot{s_{k,t}}.
\end{equation}

With equation \ref{eq:absolute-replicator-1}, we obtain the functional form of the replicator in absolute terms (i.e., growth of $x$ rather than shares) as an additive combination of aggregate growth and dynamics of sectoral shares. Applying this equation equivalently to the firm level, we obtain 

\begin{equation}
\label{eq:firm-replicator}
  dx_{i,t}/dt=X_ts_{k,t}s_{i,t}\left(\dot{X_t} + \dot{s_{k,t}} + f_{k,t}-\phi_t\right)
\end{equation}
and 

\begin{equation}
\label{eq:regression-replicator}
  \frac{dx_{i,t}/dt}{X_ts_{k,t}s_{i,t}}=\dot{x_{i,t}}=\dot{X_t} + \dot{s_{k,t}} + \left(f_{i,t}-\phi_t\right).
\end{equation}

While the anchoring of growth at different levels is useful, it should be noted that this is a very simple model and only a first approximation. In effect, the growth rates at higher levels are there as additive versions of capacity boundary terms. A standard equation with capacity boundary term \citep{Nowak06,Heinrich17} could, for instance, take the form

\begin{equation}
  dx_{i,t}/dt=x_{i,t}\left(f_{i,t}-\phi_t\right)\left(1-\frac{\sum_{i\in k}x_{i,t}}{z_k}\right)
\end{equation}
where $z_k$ is the sectoral capacity boundary. We choose not to work with this form for the regressions below, since it is difficult to estimate capacity boundaries and since the multiplicative form introduces additional complications.\footnote{We cannot simply take the logarithm since some values may be negative. Otherwise we would have a compound product regressor.} 

The functional form in Eq. \ref{eq:regression-replicator} can be employed as a regression equation in Section \ref{sect:regressions} below, since we know $\dot{x_{i,t}}$, $\dot{X_t}$, and $\dot{s_{k,t}}$ as well as several terms that can reasonably be assumed to be related to $\left(f_{i,t}-\phi_t\right)$. We avoid relating the two absolute variables employment and value added as well as other quantities that are highly correlated with either (capital, wage bill, gross output, returns, revenue, etc., see figures \ref{fig:heatmap:micro}, \ref{fig:heatmap:macro:level:extensive}). Other variables such as labor productivity, labor productivity change, and firm age remain. Different combinations of these variables can be attempted. In cases in which capacity boundaries and other idiosyncratic effects dominate the growth rates of higher aggregation levels, the regression should find that these growth rates are less significant than fixed effects for sectors and (for the macro level) years.

The regression analysis will serve a twofold purpose:

\begin{enumerate}
  \item It verifies the consistency of the model by showing that $\dot{s_{k,t}}$ is a predictor for $\dot{s_{i,t}}$.
  \item More importantly, it allows us to shed more light on the differences between the dynamics governing employment and those governing the development of value added (and closely correlated variables).
\end{enumerate}

\subsection{Distributions of firm level variables}
\label{sect:model:evol:densities}

As detailed below in Section \ref{sect:results:distributions}, the relevant variables at the firm level are heavy-tailed. This is true for employment (cf. \citet{Heinrich/Dai16}), for labor productivity (cf. \citet{Yangetal19}), and value added (see Section \ref{sect:results:distributions}). While employment is strictly positive and the heavy tail therefore only occurs on one side, value added and labor productivity are two-sided. We find that L\'{e}vy alpha-stable distributions \citep{Nolan98,Nolan18} with tail indices $\alpha<2$ are good empirical models for these variables (cf. \citet{Yangetal19} and Section \ref{sect:results:distributions}). First differences of heavy-tailed variables will also be heavy-tailed, but will always be two-sided. Quotients of two heavy-tailed variables may yield different types of variables \citep{Rathieetal16}, but in our case, for both labor productivity, $l_{i,t}=y_{i,t}/l_{i,t}$, and value added growth, $\dot{y_{i,t}}=\frac{y_{i,t}-y_{i,t-1}}{y_{i,t-1}}$, we obtain another heavy-tailed distribution. In fact, these quotients themselves are fitted very well by L\'{e}vy alpha-stable distributions  with tail indices $\alpha<2$.

L\'{e}vy alpha-stable distributions follow the characteristic function\footnote{This is the density function in frequency space. General L\'{e}vy alpha-stable distributions do not have a closed form in the quantity domain except for some special cases including the Normal, the Cauchy, and the L\'{e}vy/inverse gamma.}
\begin{eqnarray}
\varphi(h)=\mathop{\mathbb{E}}[e^{(ihx)}]  ={\begin{cases}
	e^{(-\gamma^\alpha|h|^\alpha[1+i\beta \text{tan}\left({\tfrac {\pi \alpha }{2}}\right) \operatorname {sgn}(h)\left((\gamma|h|)^{1-\alpha}-1 \right))] + i\delta h)}
	&\alpha \neq 1
	\\e^{(-\gamma|h|[1+i\beta{\tfrac {2}{\pi}} \operatorname {sgn}(h)\log(\gamma|h|)] + i\delta h)}&\alpha =1
	\end{cases}}
\label{eq:solution:levy:fourierdomain}
\end{eqnarray}
where $h$ is the frequency, the equivalent to variable $x$ in frequency space. 
It can be parametrized as $S(0,\alpha, \beta, \gamma, \delta$) where the four variables stand for the tail index ($\alpha$), the skew parameter ($\beta$), the scale ($\gamma$) and the location shift ($\delta$). For details on the fitting, see \citet{Yangetal19,Heinrichetal19}.

For value added, value added change, value added growth, and labor productivity, the distribution densities can be seen in figure \ref{fig:density:VA}.

L\'{e}vy alpha-stable distributions are special in that they are the only continuous distribution class that fulfills the stable criterion: Adding two (or arbitrarily many) L\'{e}vy alpha-stable distributions yields another L\'{e}vy alpha-stable distributions. The distribution class is the attractor of the generalized central limit theorem. While summing over short tailed random variates leads to a normal distribution (a member of the L\'{e}vy alpha-stable distribution class with $\alpha=2$), summing over  heavy tailed variates yields heavy tailed members of the L\'{e}vy alpha-stable class. From an evolutionary perspective, it would appear that natural L\'{e}vy alpha-stable distributions (e.g., when encountered in form of value added or labor productivity, etc.) may be the result of aggregation processes occurring in economic systems.

The L\'{e}vy alpha-stable distributions are furthermore the entropy maximizing distributions for constraint $h^{\alpha}=\overline{\alpha}$, i.e. finiteness of the statistical moment\footnote{E.g, the first central moment is the mean, the second moment the variance, etc.} of order $\alpha$ \citep{Frank09}. All moments of order $>\alpha$ will be infinite.

This has consequences for quantities that follow these distributions and for statistics performed on such quantities: While for samples drawn from L\'{e}vy alpha-stable distributions, sample moments can be computed, they will diverge with the sample size, and are therefore contaminated by information that does not belong to the moment. Operating with sample moments $m$ when the corresponding distribution does not have finite moment $m$ will lead to invalid and misleading conclusions. 

Note that aggregates over samples from a distribution $S$ are essentially linear functions of the mean of $S$. If $S$ is L\'{e}vy alpha-stable distributed with $\alpha\leq 1$ so that the mean does not exist, the aggregate will become volatile. Even if the distribution has $\alpha > 1$ such that the mean is finite,  L\'{e}vy alpha-stable moments may converge slowly and the aggregate may still show a volatile behavior compared to quantities that are subject to the classical central limit theorem without heavy-tailed influences. This is highly relevant in our case, as it applies to sectoral as well as macro-level value added (output) and value added changes as well as other quantities. Because of autocorrelation, the quantities will remain relatively stable over short time periods. However, across countries, sectors, and long time spans, the volatility is high. This is in stark contrast to quantities that are based on aggregations of Normally distributed or other short-tailed random variates, e.g., effective monthly work hours per employee, or prices of homogeneous goods at any one point in time.

Finally, in regression analyses involving heavy-tailed (especially L\'{e}vy alpha-stable distributed) quantities, heavy-tailedness may in many cases be inherited by the residuals of the regression. In such event, the assumptions of the OLS regression methodology (normality of residuals) are violated and the OLS results become invalid. Such cases require a \textit{robust regression} approach that allows for heavy-tailed errors.%; software packages for this exist \citep{R::heavy}.

\iffalse
\subsection{Differences between employment and output dynamics}

Differences between growth patterns in value added (or output) and in employment are to be expected. Some technologies are labor intensive, some are capital intensive. And while substitution is usually difficult, if not impossible in the short-run, sectoral characteristics do depend on this and may change over time. The time frame of the analysis was characterized by widespread modernization, consolidation and layoffs in large SOEs, a decline of the importance of agriculture, and the rise of the service sector, which is notoriously labor intensive. 

While we should therefore see a fair amount of labor moving from agriculture and SOEs towards industry and services, labor mobility is constrained by mismatches of available and required skills as well as geographic distance. In the short run, the there are hard limits to available labor in skilled jobs, which may impair growth of some industries even if both demand and investment capital is available. Expanding the skilled labor force in these industries will require a public or coordinated private effort.

Value-added, on the other hand, reacts quickly to shocks and expands and contracts with both demand, labor supply, and investment capital. It is nevertheless the value added that convinces investors, replenished funds, and provides industries with the infrastructure for further development.
\fi

\begin{figure}[h!]
\centering
\subfloat[Growth rates]{\includegraphics[width=0.5\textwidth]{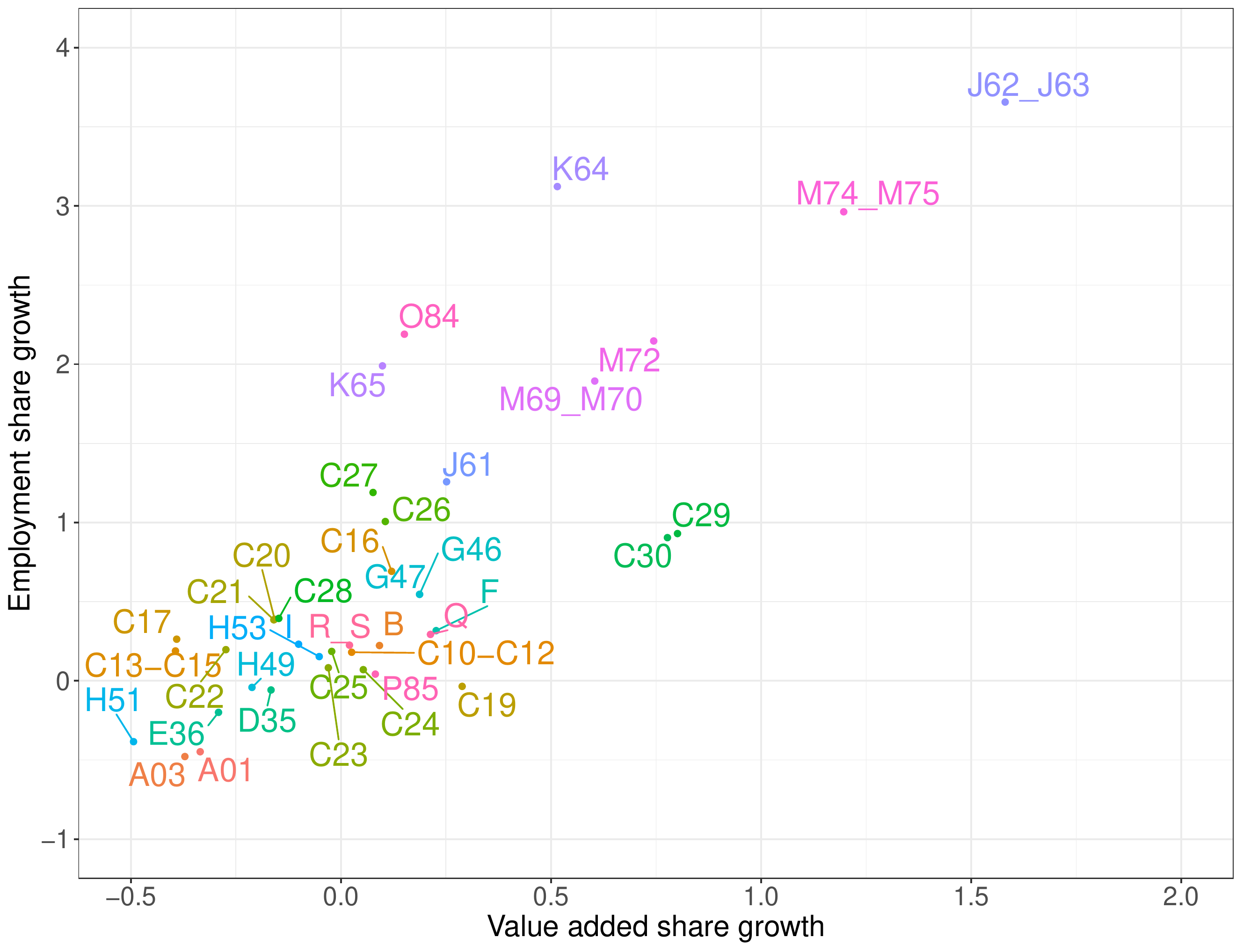}}
\subfloat[First differences]{\includegraphics[width=0.5\textwidth]{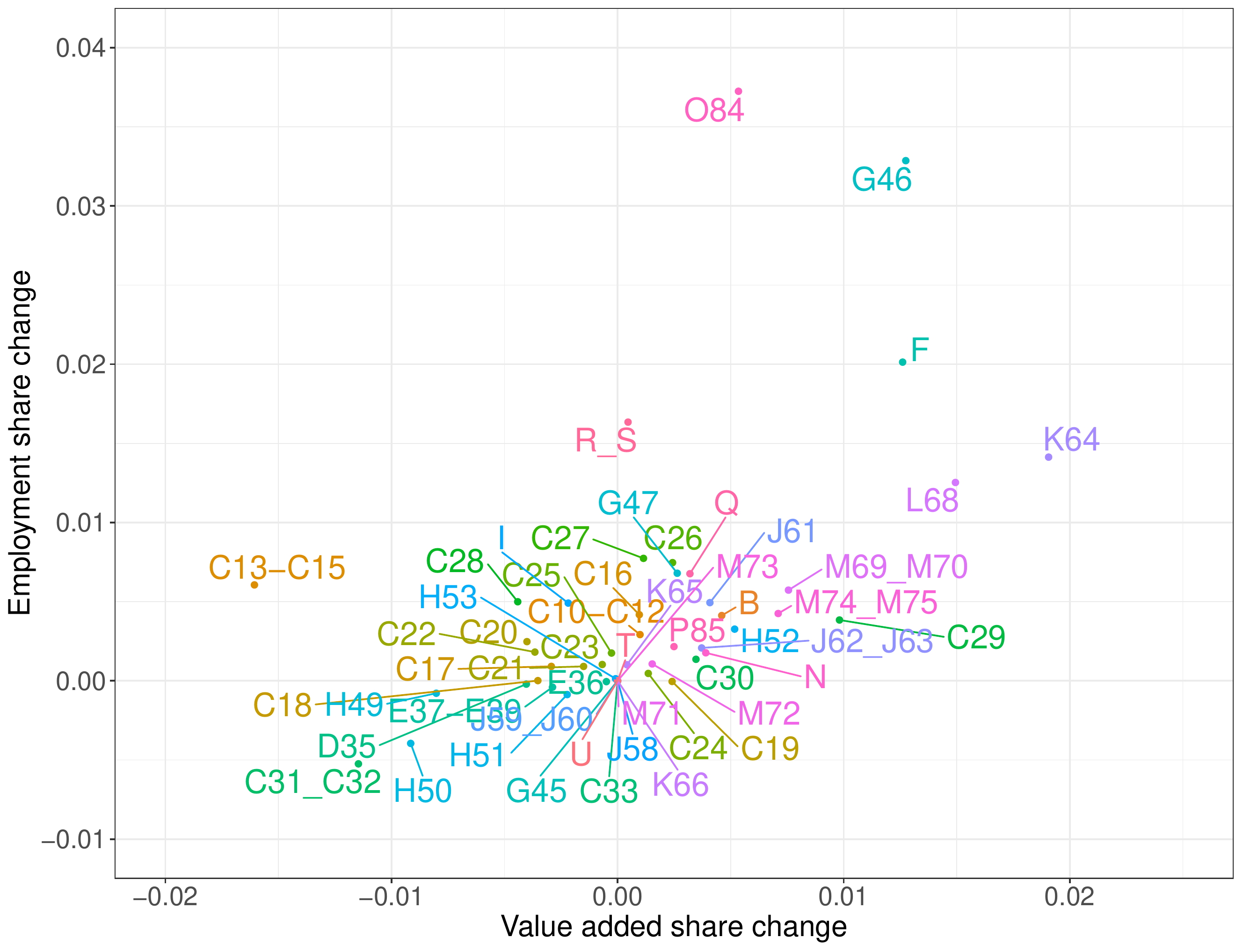}}
\caption{Changes (Growth rates and first differences) in employment shares vs. changes in value added shares by sectors over the period 2000-2014.}
\label{fig:MAC:long-term-growth}
\end{figure}

\begin{figure}[h!]
\centering
\subfloat[All sectors]{\includegraphics[width=0.5\textwidth]{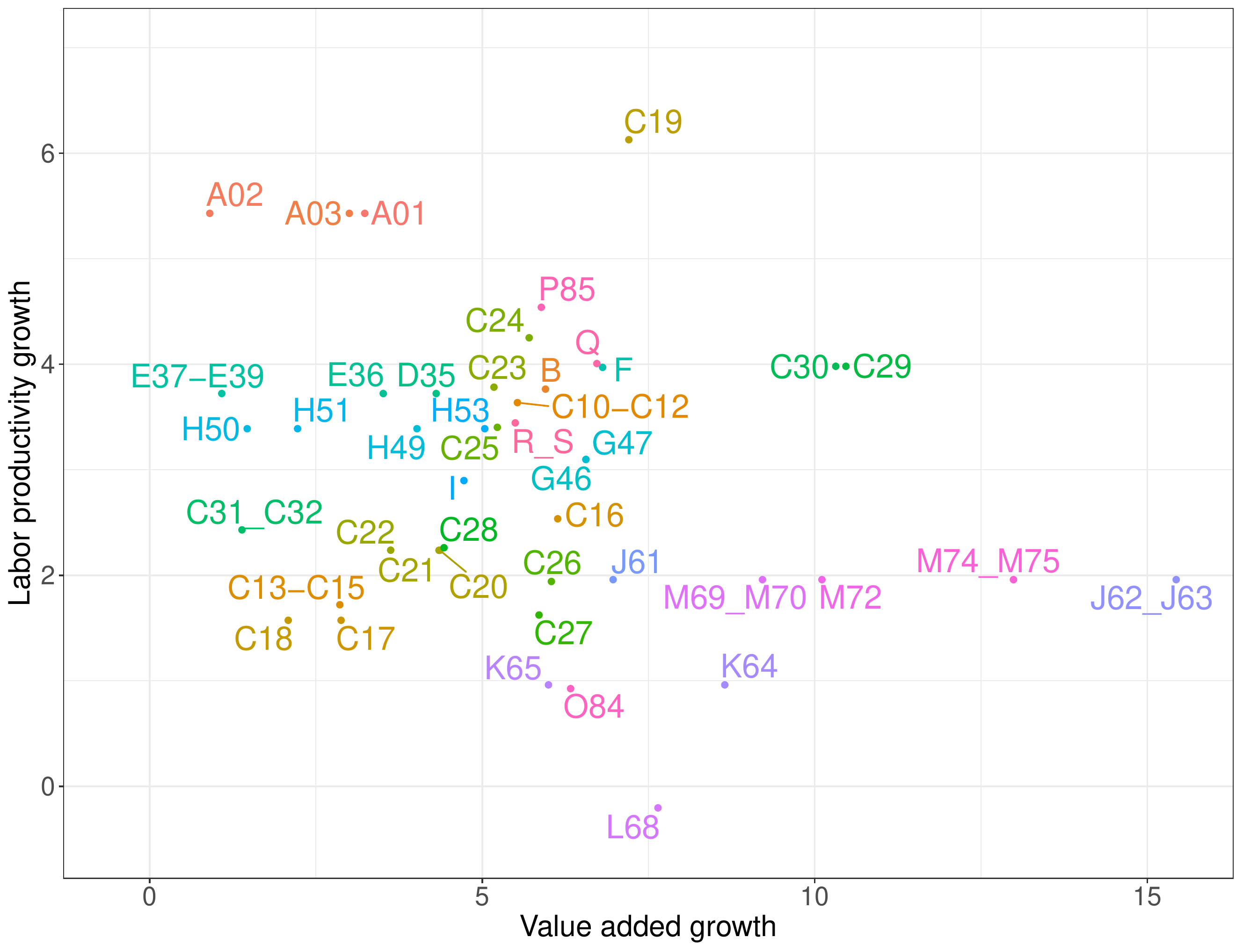}}
\subfloat[Industry sectors only]{\includegraphics[width=0.5\textwidth]{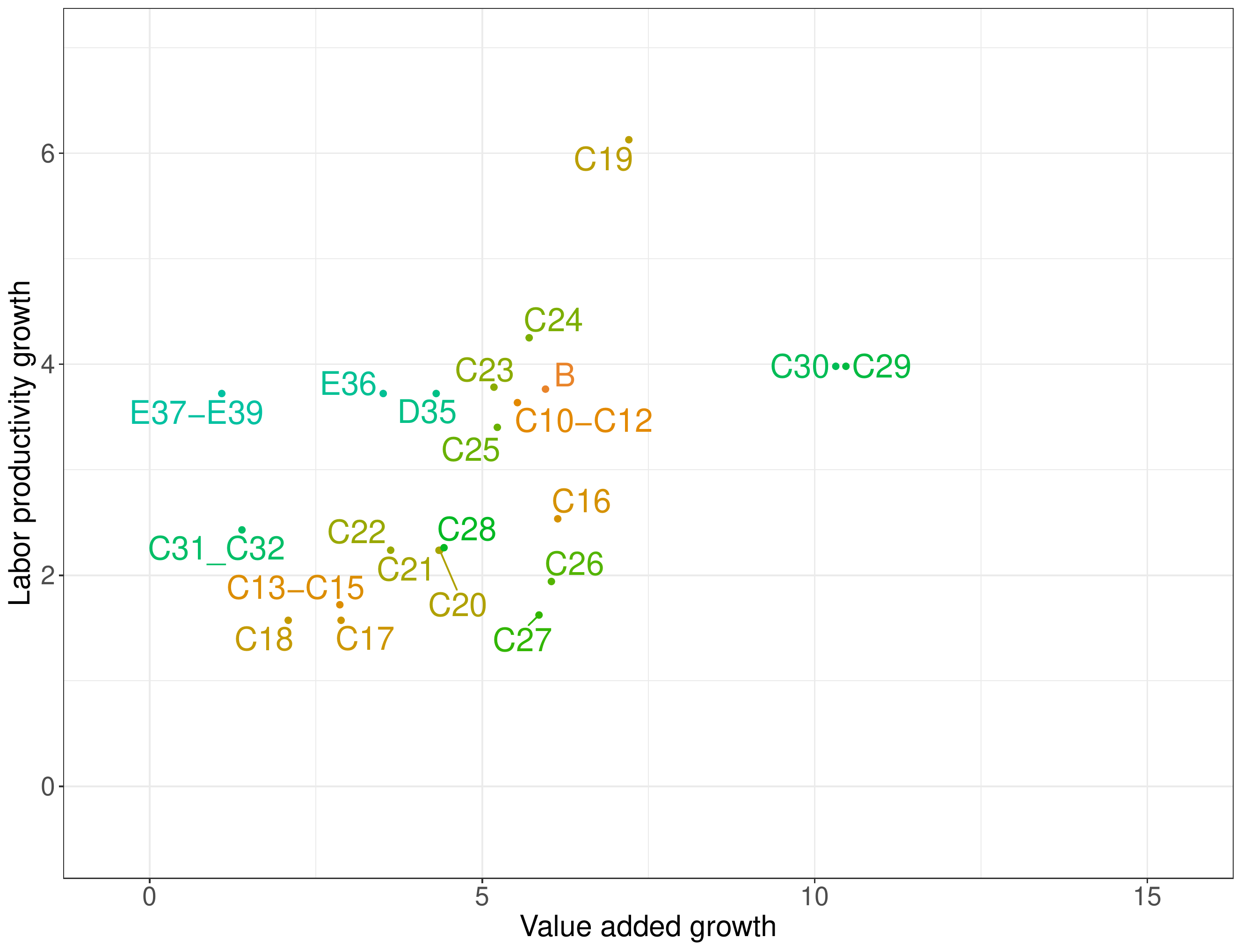}}
\caption{Growth rates of value added (absolute value, not shares) vs. labor productivity growth by sectors over the period 2000-2014. While value added growth and productivity growth seem roughly negatively related for all sectors, in industry sectors only, the variables appear to be roughly positively associated.}
\label{fig:MAC:long-term-absolute_growth}
\end{figure}

\begin{figure}[h!]
\centering
\subfloat[Employment shares]{\includegraphics[width=0.5\textwidth]{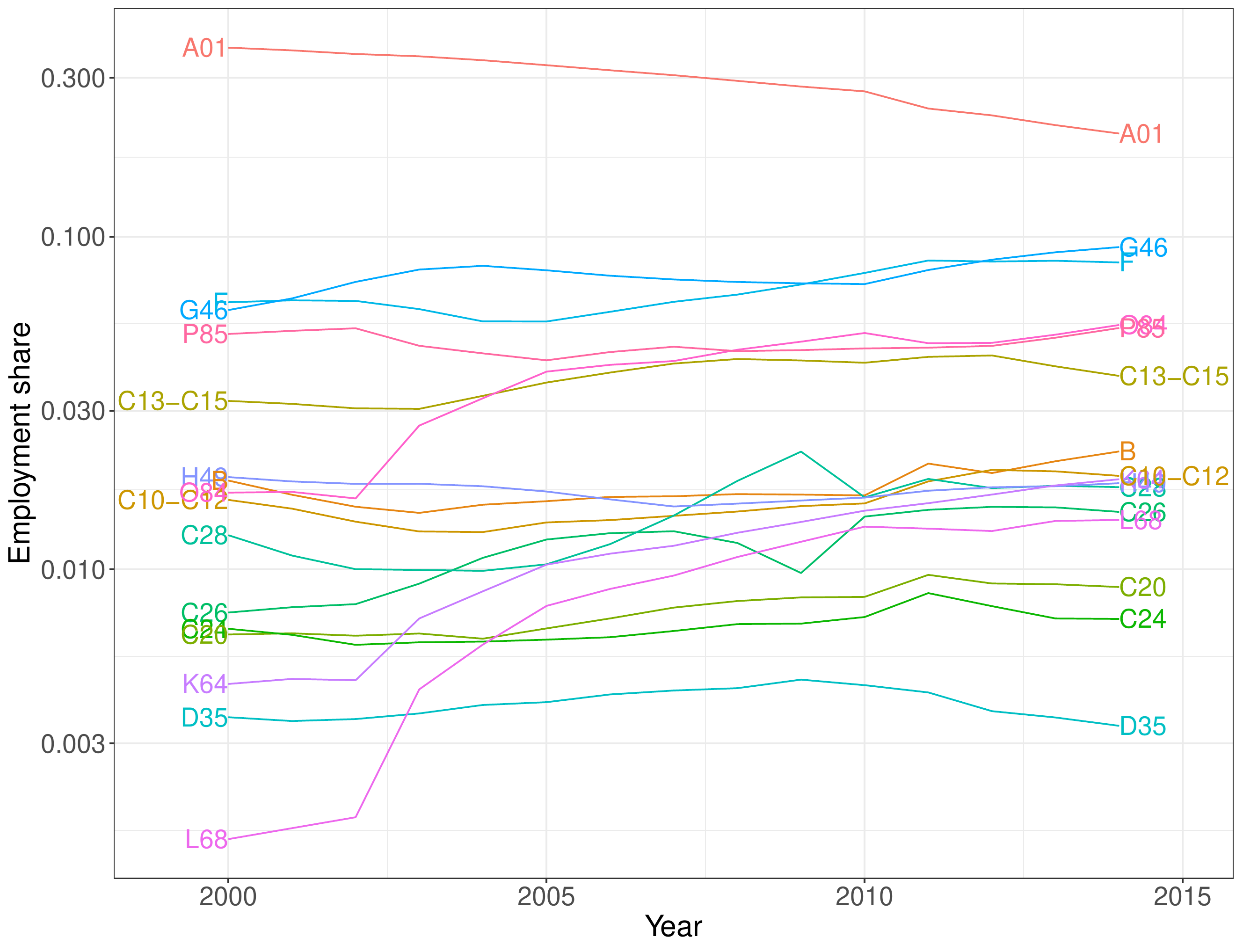}}
\subfloat[Value added shares]{\includegraphics[width=0.5\textwidth]{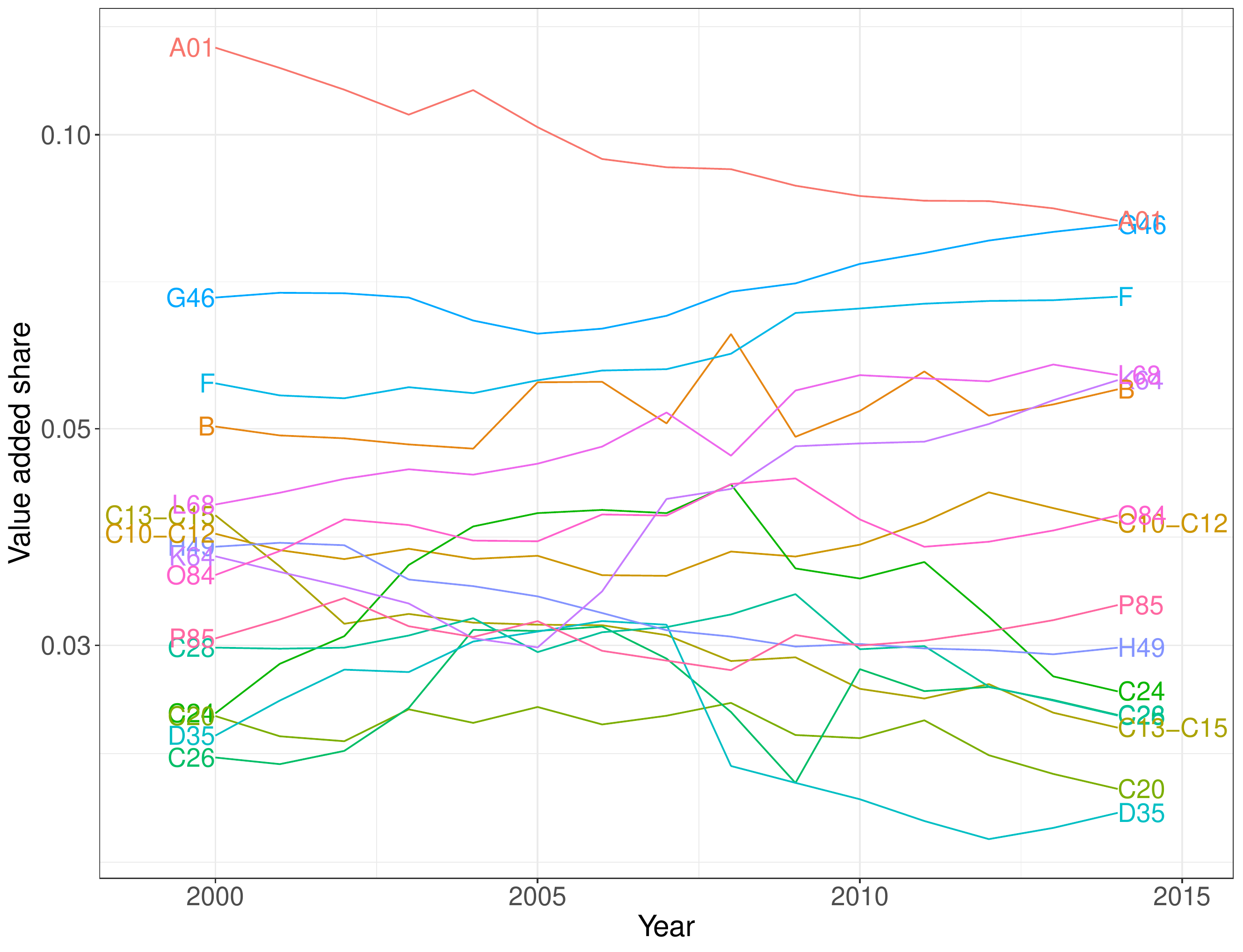}}
\caption{Development of employment and value added shares by sector}
\label{fig:Dev:sectoral:MACRO}
\end{figure}

\begin{figure}[tb!]
\centering
\includegraphics[width=0.85\textwidth]{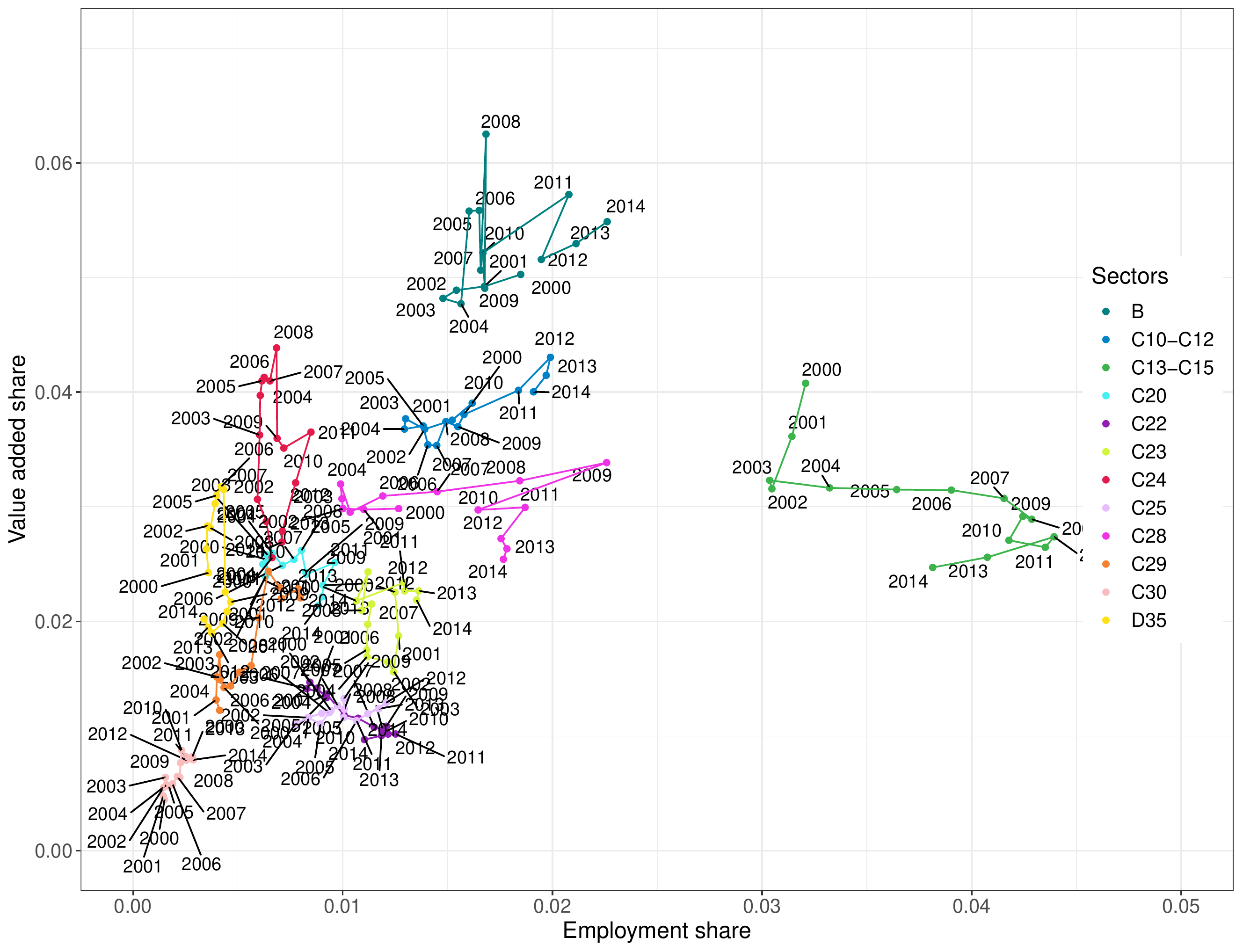}
\caption{Development of employment vs. value added by sector}
\label{fig:Dev:2d:EMPL:VA}
\end{figure}

\begin{figure}[tb!]
\centering
\includegraphics[width=0.85\textwidth]{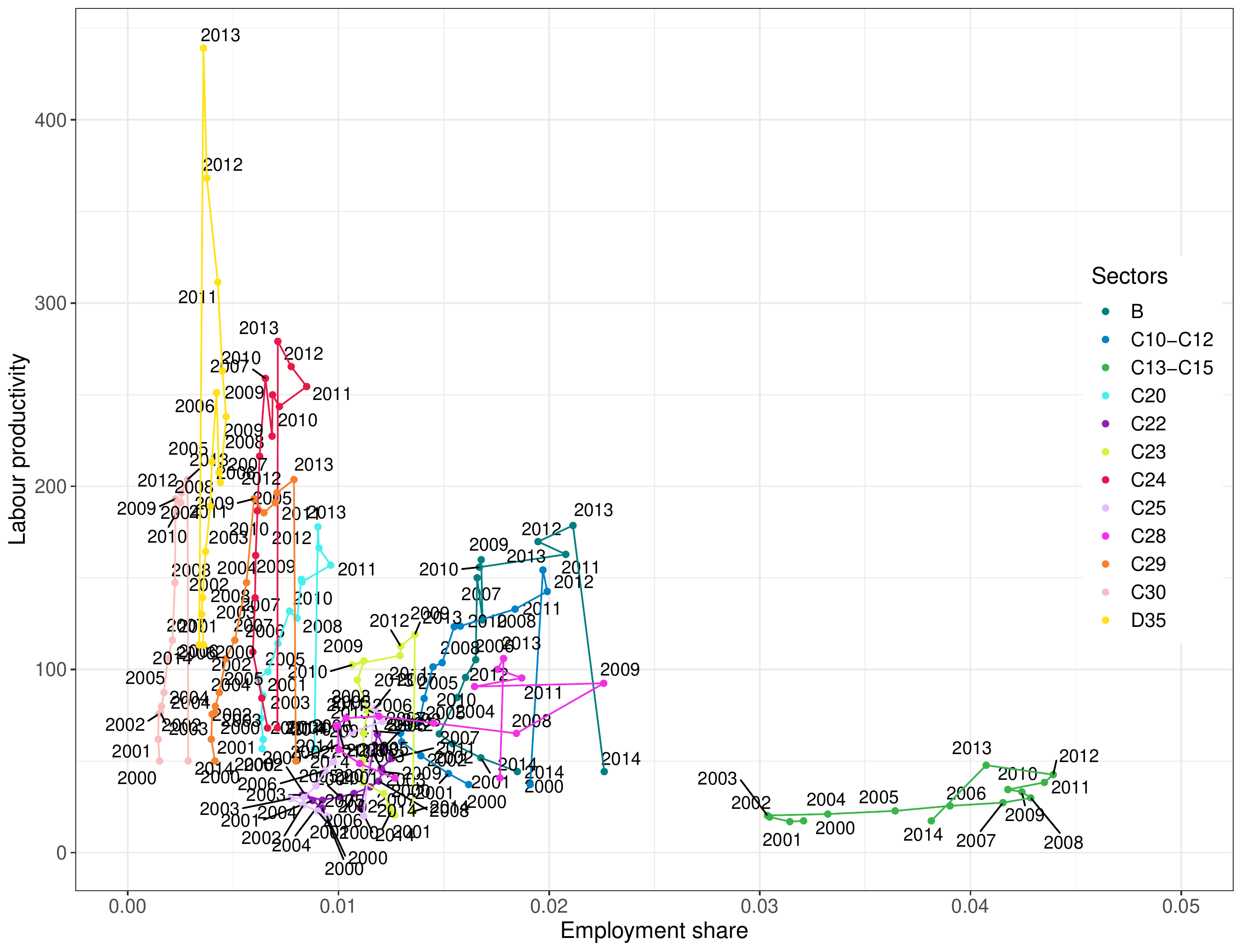}
\caption{Development of employment vs. labor productivity by sector}
\label{fig:Dev:2d:EMPL:LP}
\end{figure}

\begin{figure}[tb!]
\centering
\includegraphics[width=0.85\textwidth]{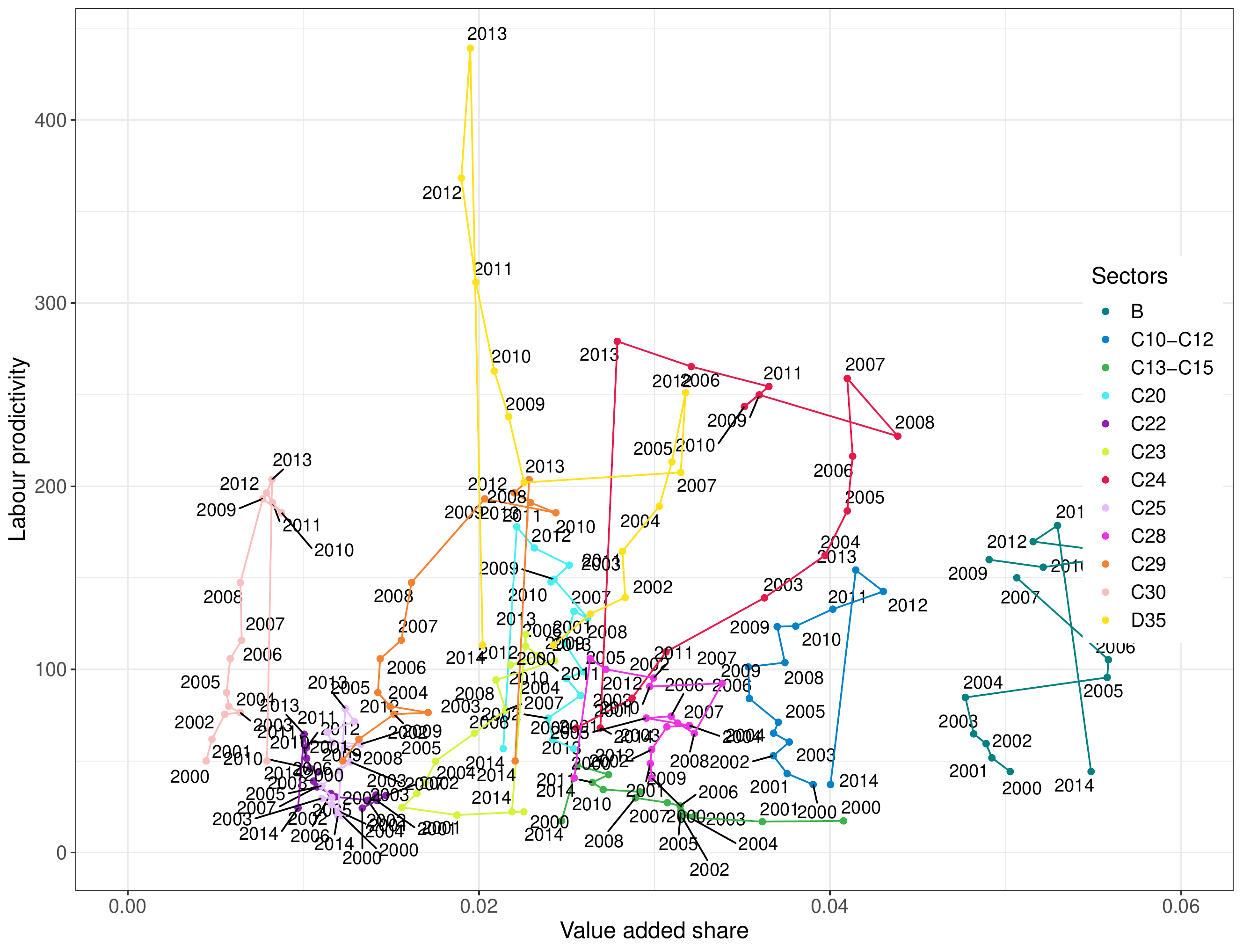}
\caption{Development of value added vs. labor productivity by sector}
\label{fig:Dev:2d:VA:LP}
\end{figure}

\begin{figure}[tb!]
\centering
\includegraphics[width=0.85\textwidth]{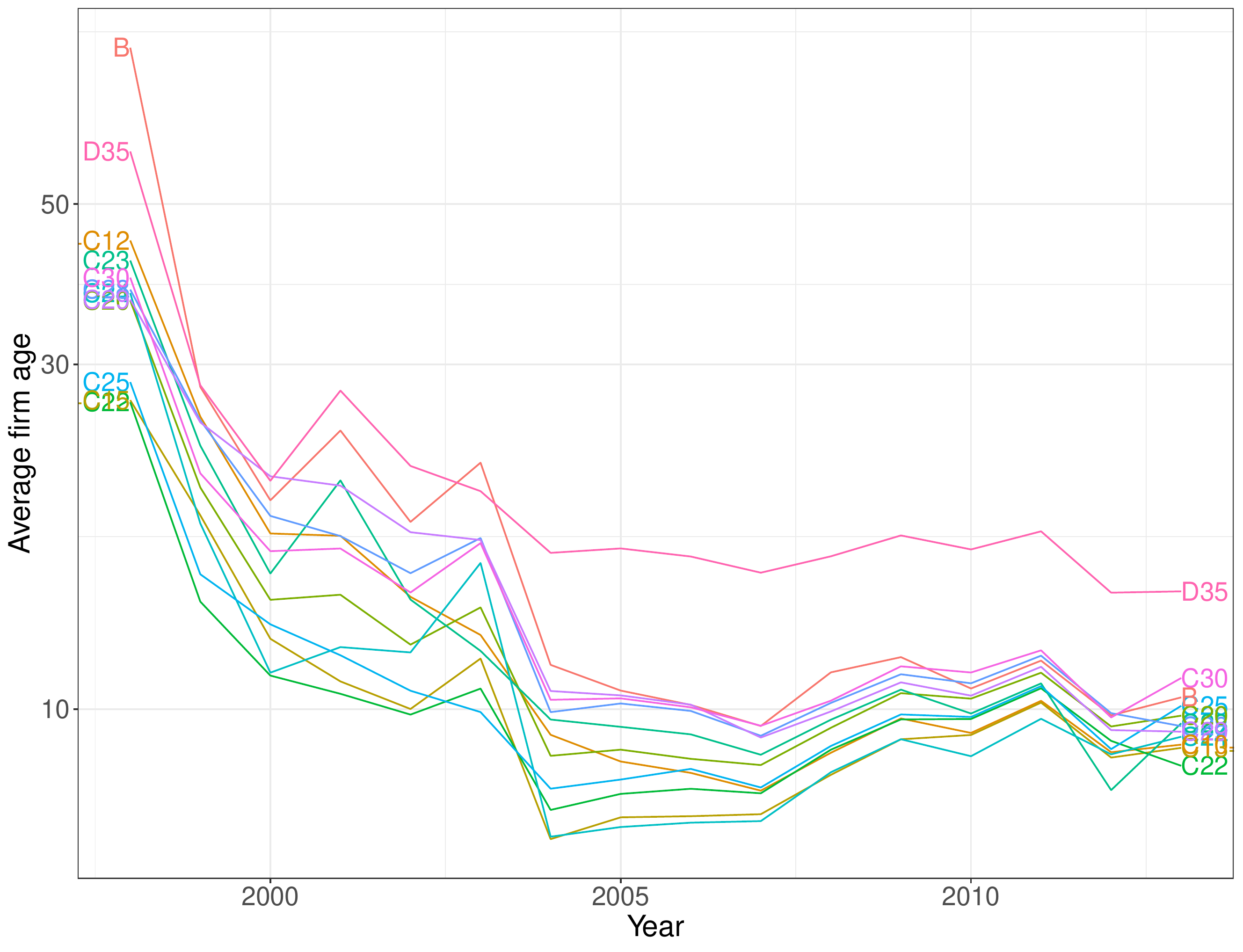}
\caption{Development of average firm ages by sector in firm level data}
\label{fig:Dev:sectoral:Age}
\end{figure}

\section{Empirical evidence} 
\label{sect:results}

In the model section above (Section \ref{sect:model:evol}), we conjectured that economic growth is neither homogeneous across different sectors nor does it align when measured in different variables. Specifically changes and growth rates in employment on the one hand and output or value added on the other differ wildly. In this section, we will show this in detail.

\subsection{Data}

We use firm level data from the \textit{Chinese Industrial Enterprise Database (CIE DB)} for the years 1998-2013 and sector level data from the \textit{World Input-Output Database (WIOD)} for the years 2000-2014. 

The CIE DB includes various micro-data for several hundreds of thousands of firms per year. It is, however, subject to a quickly changing panel (the overwhelming majority of firms are not present for all years), changes in data collection and in the variables recorded. For a more detailed discussion of problems with this data set, see \citet{Brandtetal14}. The database does, however, also offer excellent coverage that is absolutely unique for a country and time period at this development stage. With the years of the opening to foreign capital, the transition to a market economy, the appearance of numerous start-ups, and the closure of a significant number of state-owned enterprises, it also covers a rather crucial period of economic development and economic transition in the PR China. The database includes primary sector classifications following different versions (1994 and 2002) of the Chinese Gu\'{o}Bi\={a}o (\begin{CJK*}{UTF8}{gbsn}国标\end{CJK*}). At the 2-digit level that we work with, most sectors are consistent, but we omit some few sectors the designation of which changes for different years in the database. For details, see Table \ref{tab:sectorcodes} in Appendix \ref{app:sectorcodes}. 

The WIOD includes a range of sector level accounts for the PR China; it employs the ISIC Revision 4 sector classification (International Standard Industrial Classification). For details, see \citet{Timmeretal15wiod}. Throughout the present paper, we will also report sectors according to ISIC Rev.4. The correspondence table between Gu\'{o}Bi\={a}o 1994/2002 and ISIC Rev.4 can be found in in Appendix \ref{app:sectorcodes}.

The variables we employ and the corresponding symbols mostly follow the usual standard in economics. However, for convenience, we include a variable and symbol table in in Appendix \ref{app:variables}.

\subsection{Sectoral dynamics}
\label{sect:results:sectoral}

Fig. \ref{fig:Dev:longterm} shows the long-term development of value added $Y$ in the PR China in six aggregated sectors as reported in the statistical yearbooks and summarized in \citet{Holz05}. The six sectors are 1. primary, i.e., agriculture, 2. industry, 3. construction, 4. trade and catering, 5. communication and transportation, and 6. other services. As the figure shows, the contribution of industry sectors has remained remarkably stable since the 1970s at around 40\% while the tertiary sectors have been growing rapidly at the expense of agriculture. This is consistent with the pattern reported in the literature and seen across developing economies around the globe for both value added and employment \citep{Timmeretal15}. 

Our sectoral and firm-level data sets cover a relatively recent part, 1998-2014. We therefore can not expect to see major shifts to or away from industry, but structural change within and between industry sectors may be present. Figure \ref{fig:Dev:sectoral:MACRO} shows the employment and value added shares of the 16 largest sectors between 2000 and 2014 while the bivariate development and changes (employment share vs. value added share) of the 12 largest industry sectors\footnote{We only show industry sectors since service (real estate L68, defence O84, education P85, ...) and agriculture (A01) are much more volatile (see Fig. \ref{fig:Dev:sectoral:MACRO}) and would make the dynamics in the industry sectors difficult to see.} are depicted in Figure \ref{fig:Dev:2d:EMPL:VA}. Both figures reveal a certain volatility while the general levels remain unsurprisingly stable. 

In particular, Fig. \ref{fig:Dev:2d:EMPL:VA} shows that there is no direct connection between changes in value added and employment, at least in the short term\footnote{In the long term, Fabricant's laws hold at the sectoral level; see Section \ref{sect:results:fabricant}.}, they seem almost independent from one another. What is remarkable, though, is that the movements performed by the sectors in Fig.  \ref{fig:Dev:2d:EMPL:VA} tend to be clockwise, i.e., growth in value added precedes growth in in employment (and similar for decline). However, they are far from regular.

15-year aggregates in Fig. \ref{fig:MAC:long-term-growth} do reveal a trend that may suggest that in the long run and on average, employment and value added growth have a positive relationship (cf. \citet{Fabricant42,Metcalfeetal06}). However, these are again driven by changes in the primary and tertiary sectors, the growth and decline of which far outpaces the industry sectors in the period covered by the data set. Industry sectors, shown in yellow to green color shade are in both panels of Fig. \ref{fig:MAC:long-term-growth} clustered around the origin with no clear dependence between the dynamics of the two variables within industry (i.e., without primary and tertiary sectors) even in the aggregate. The sectors on the growing side tend to be the more sophisticated industries (special purpose machinery C29, transportation equipment C30, IT and optical equipment C26, fuel C19). 

Fig. \ref{fig:Dev:2d:EMPL:LP} and \ref{fig:Dev:2d:VA:LP}, depicting the bivariate development of labor productivity and value added and employment shares respectively, do not reveal any systematic short run\footnote{As the correlations in Section \ref{sect:results:fabricant} suggest, the connection emerges in the average across many years.} connection between these variables. While the three variables are locked in the identity $q_i=y_i/l_i$, the development patterns appear to be highly sector specific. Similarly, as shown in Fig. \ref{fig:MAC:long-term-absolute_growth} the relationship between (absolute) value added growth and labor productivity growth takes a different form in industry sectors (positive slope) and in the whole economy (negative or no slope).

There is a clear signature of the events in the recent economic history of China. The average firm ages dropped significantly in the late 1990s and early 2000s in all industry sectors as seen in Fig. \ref{fig:Dev:sectoral:Age}. This corresponds to the dissolution of some of the SOEs and the simultaneous emergence of startups during this time period. The difference is less pronounced for sector D35 (electricity and gas supply), a sector that should be expected to include larger firms because of natural monopolies and public companies because of strategic importance. Furthermore, aggregated data in Fig. \ref{fig:Dev:sectoral:MACRO} shows turbulence in time periods 2003 and 2008, 2009.

\begin{figure}[tb!]
\centering
\includegraphics[width=0.85\textwidth]{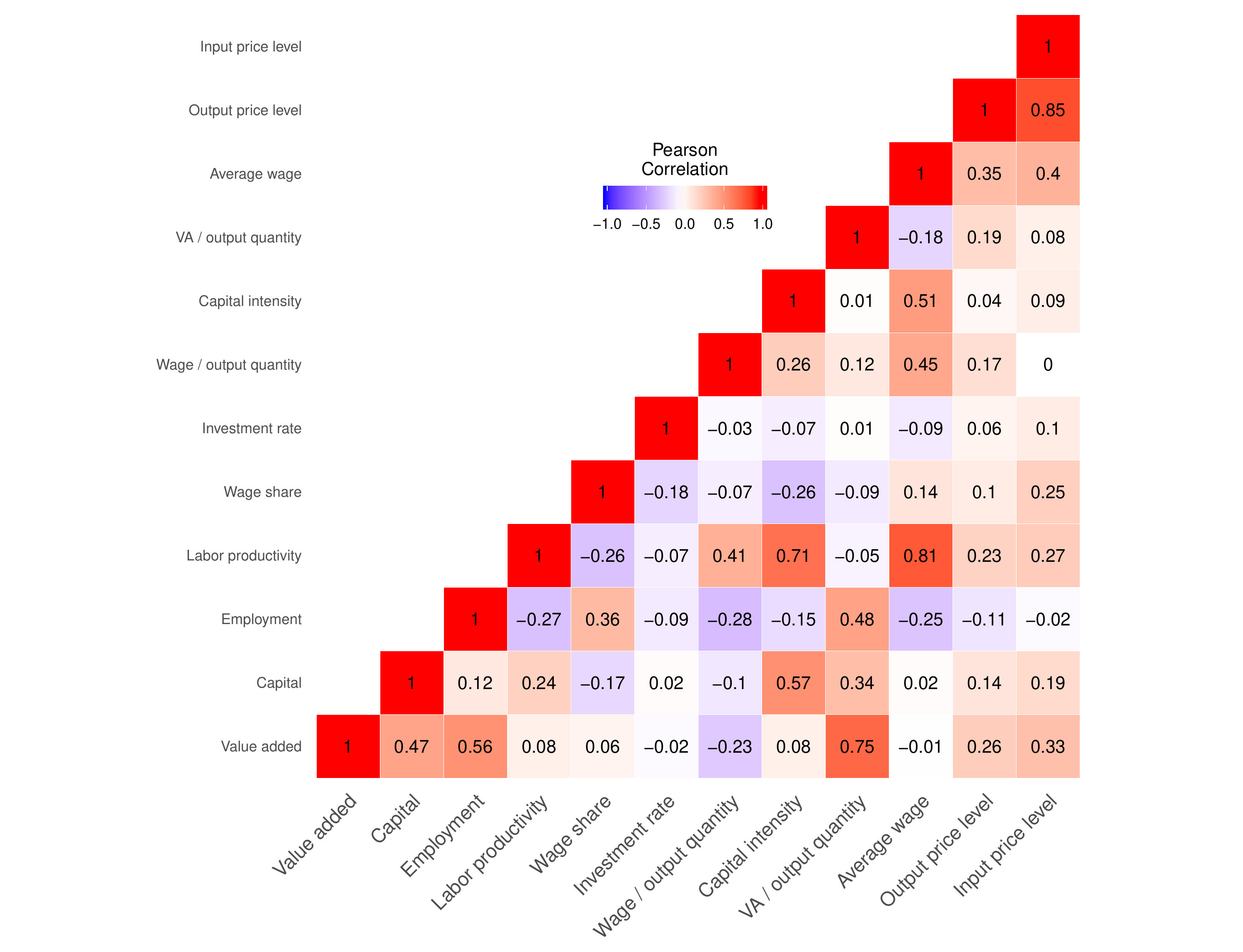}
\caption{Correlations in macro-level data (levels)}
\label{fig:heatmap:macro:level}
\end{figure}

\begin{figure}[tb!]
\centering
\includegraphics[width=0.85\textwidth]{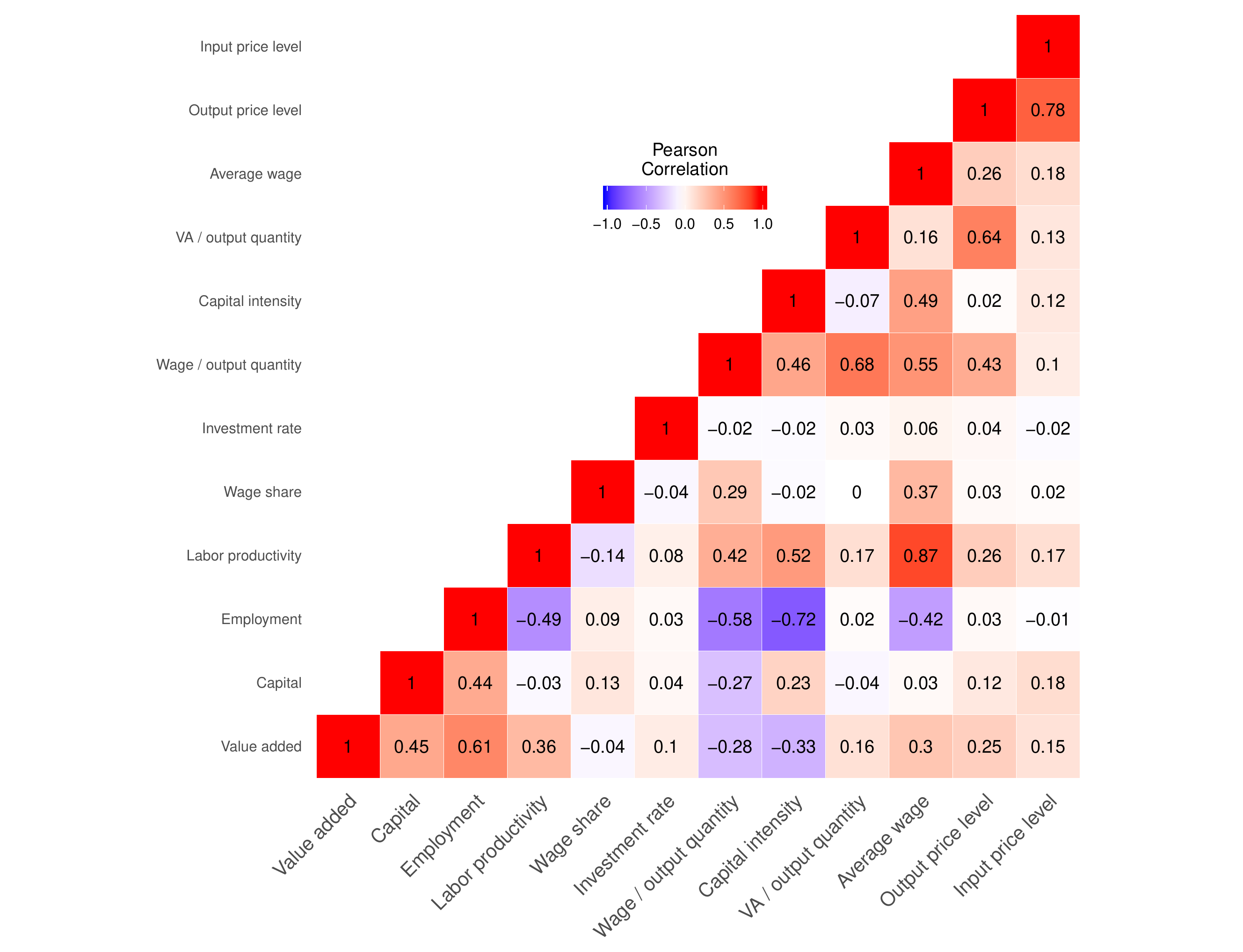}
\caption{Correlations in macro-level growth rate data.}
\label{fig:heatmap:macro:growth}
\end{figure}

\begin{figure}[tb!]
\centering
\includegraphics[width=0.95\textwidth]{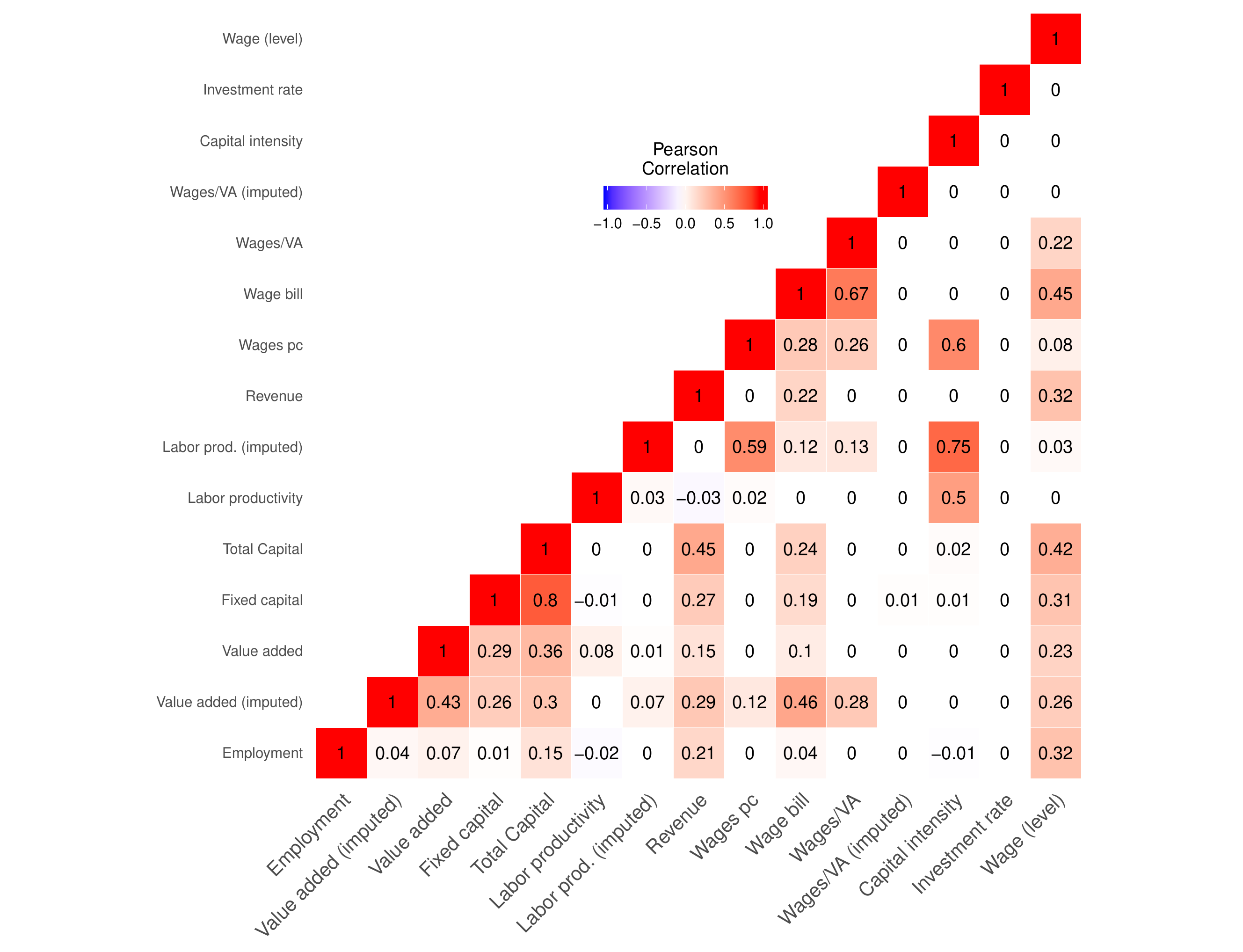}
\caption{Correlations in firm-level data. All rows and columns except for the wage level in the last column represent first differences.}
\label{fig:heatmap:micro}
\end{figure}

\subsection{Correlation laws} %Fabricant's laws
\label{sect:results:fabricant}

Next, we investigate correlation laws between quantities at both the firm-level and the sectoral level, the latter corresponding to Fabricant's laws \citep{Fabricant42,Scott91,Metcalfeetal06}. 

Figures \ref{fig:heatmap:macro:level} and \ref{fig:heatmap:macro:growth} show the correlation matrices between the quantities included in one or the other correlation law proposed in the literature as discussed in Section \ref{sect:literature} above. While Fig. \ref{fig:heatmap:macro:growth} shows the correlations between growth rates, in which the laws are generally expressed in the literature, we include the correlations between the levels in Fig \ref{fig:heatmap:macro:level} to convey an idea of what dependencies exist in the data. Intuitively, in the levels size variables (employment, value added, capital) are correlated. Also, hall marks of sophisticated industry sectors (labor productivity, capital intensity, average wage, and input price) are correlated among each other while being negatively correlated to employment and wage share of output.

Figure \ref{fig:heatmap:macro:growth} confirms most of Fabricant's laws:
\begin{itemize}
    \item Value added growth (i.e., output growth) and labor productivity growth are positively correlated.
    \item Value added growth is negatively correlated to the growth of wage per output quantity (i.e., output growth dominates wage growth).
    \item Value added growth is positively correlated with both employment and capital growth.
    \item Growth of value added per unit of output and wage costs per unit of output are correlated.
    \item Growth of value added per unit of output is correlated with growth of output prices.
\end{itemize}

Indeed, these correlation laws hold with impressive regularity and have been confirmed in many different countries \citep{Fabricant42,Scott91,Metcalfeetal06}, times and contexts at the sectoral level, including in our case. However, we shall show below that they disintegrate at the micro-level, i.e. when observing firms. 

Some other aspects proved to be different already at the sectoral level in the present case: 

\begin{itemize}
    \item Capital intensity $k_i/l_i$ growth and value added growth are strongly negatively correlated, not positively. This is an interesting and counter-intuitive aspect, which arises mainly because the fast growing services sectors have lost capital intensity by expanding employment while at the same time boosting value added growth.
    \item The correlation between the wage bill per unit of output and the labor productivity (output per worker) is positive, not negative. This implies that wages are growing with output, but may be a special effect of the China's rapid total growth in the time frame covered by the study. It can be seen that the correlation between labor productivity growth and the growth of the wage share of income is indeed negative, which implies that profits per output %if not returns on capital 
    were growing even faster.
    \item There is no negative correlation between output price growth and value added growth. The rationale behind this correspondence rests on the law of demand and supply holding in isolation at the output market of the sector in question. This may quite possibly not hold in the context of rapid demand-driven growth if wages and therefore all factor and output prices increase.
    \item Factor price correlation laws attributed by Scott \citep{Scott91} to Salter, negative correlations between factor prices changes (wage and input prices) on the one hand and value added growth because of retirement of old, inefficient techniques, do not hold.
\end{itemize}

At the micro-level, considering data for individual firms from the CIE DB, many of these correlation laws fall apart, however. We consider first differences instead of growth rates because many quantities can take negative values at the micro-level, leading to counter-intuitive and misleading growth rate computations. The correlations are reported in Fig. \ref{fig:heatmap:micro}. Increase in value added is still associated with increase in capital. Higher value added will normally feed into profits and then into the capital stock unless the profits are paid out to capital owners. Both value added and reinvestment will also be autocorrelated, therefore this correspondence is intuitive. However, no correlation exists between increases in value added and in either labor productivity or employment. The correlation between increases in value added and those in wages per value added (the latter used as proxy for wages per output quantity) is either zero (for value added computed as the difference between output and intermediate inputs) or positive (for value added imputed as the sum of wages and profits).\footnote{The large difference between the two ways of computing value added is due to biased data. Intermediate inputs data are only available for a fraction of the data set and this fraction is evidently not unbiased.} Other remarkable correlations include those between labor productivity and capital intensity as well as wages.

What does this mean for our analysis? Two interpretations are possible. Either micro-level data is noisy enough to mask the true effect, be that because of idiosyncratic shocks or because of data collection issues, while everything averages out nicely at the aggregated, sectoral level. This explanation is doubtful, especially in the light of heavy tailed data in the relevant quantities  (see \ref{sect:results:distributions}). Heavy-tailed quantities may exactly not average out in the aggregate as the law of large numbers does not apply. Even for existing, non-infinite moments in heavy-tailed data, the convergence may be quite slow. 

On the other hand, Fabricant's law may not be a property of the firm, but one of the industry sector. Industry sectors share a lot of infrastructure: the same labor force, largely the same capabilities and technologies, as well as procedural knowledge. They also share the same communication and transportation networks and the same input and output markets. Labor mobility of skilled labor within an industry is typically quite high, especially if the industry is concentrated in a geographical region. 
The fortunes of an individual company may grow and fall; the company may choose to hire or to expand at particular times. However, the labor force and infrastructure available to the sector remains the same and only changes slowly. As a consequence, it would not be expected that Fabricant's laws manifest themselves at the firm level, except for the correspondence of output and capital growth. It should, however, be expected that they are evident on average, across a longer time frame, for the sectors of an economy.

This is consistent with the hypothesis that Fabricant's correlation of labor productivity growth and output (or value added) growth is rooted in different rates of technological change or in economies of scale at the sector level \citep{Scott91}, while other explanations such as factor substitution or increased personal efficiency of employees would suggest that there should be an observable effect at the micro-level. We will return to this point in the regression analysis in Section \ref{sect:regressions}.

While this is an interesting result, our analysis is, of course, limited to a very particular case: The economy of the PR China in the decades when its development reached the highest levels of growth, and when the fastest and most intense processes of transformation towards a market economy took place. As pointed out, this may have an effect on some of the usual correlation laws that may be absent in such fast growing developing economies. It is also possible that the micro-level may show more regular characteristics in slow-growing developed economies.

%\clearpage

\begin{figure}[h!]
\centering
\subfloat[Employment]{\includegraphics[width=0.5\textwidth]{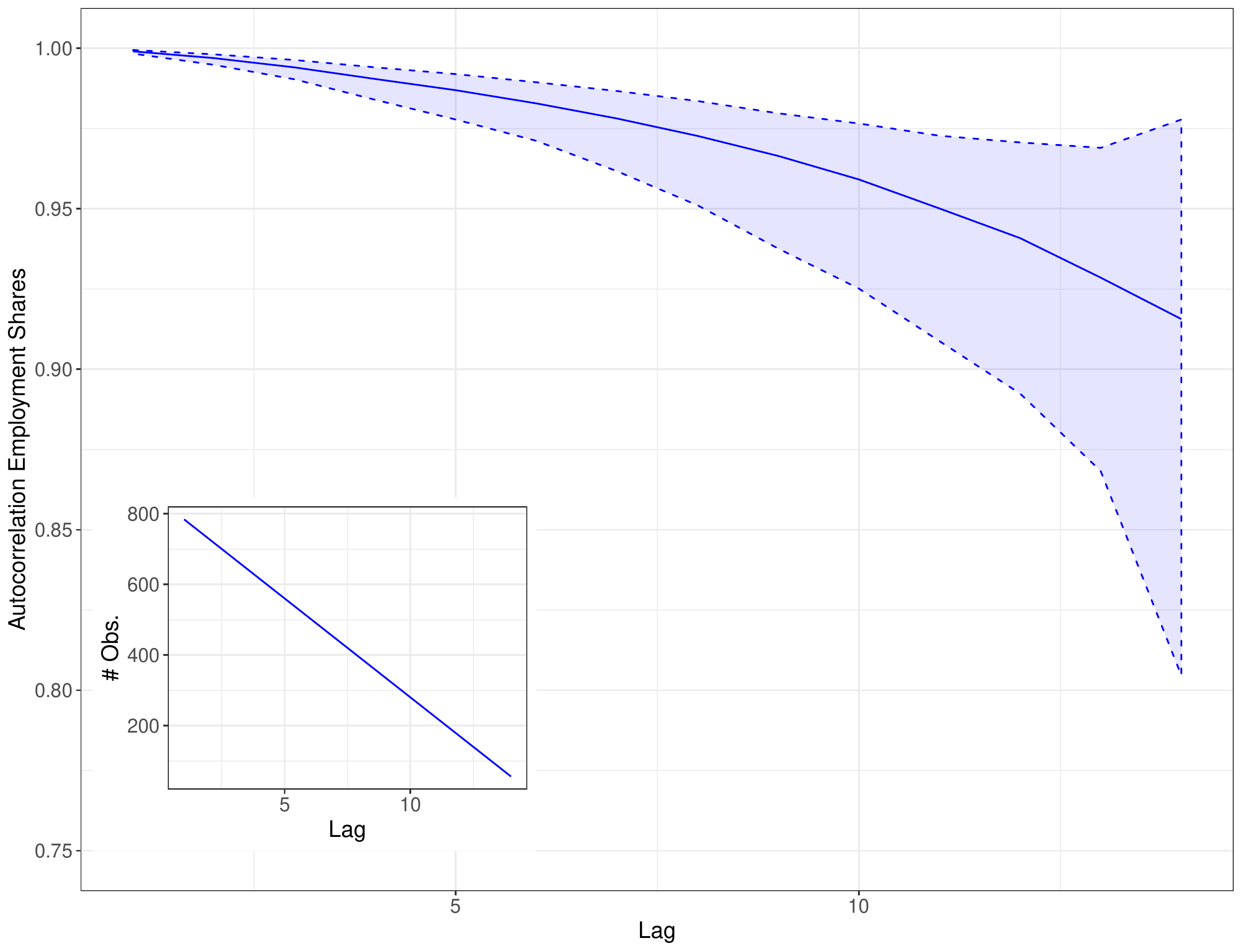}}
\subfloat[Value added]{\includegraphics[width=0.5\textwidth]{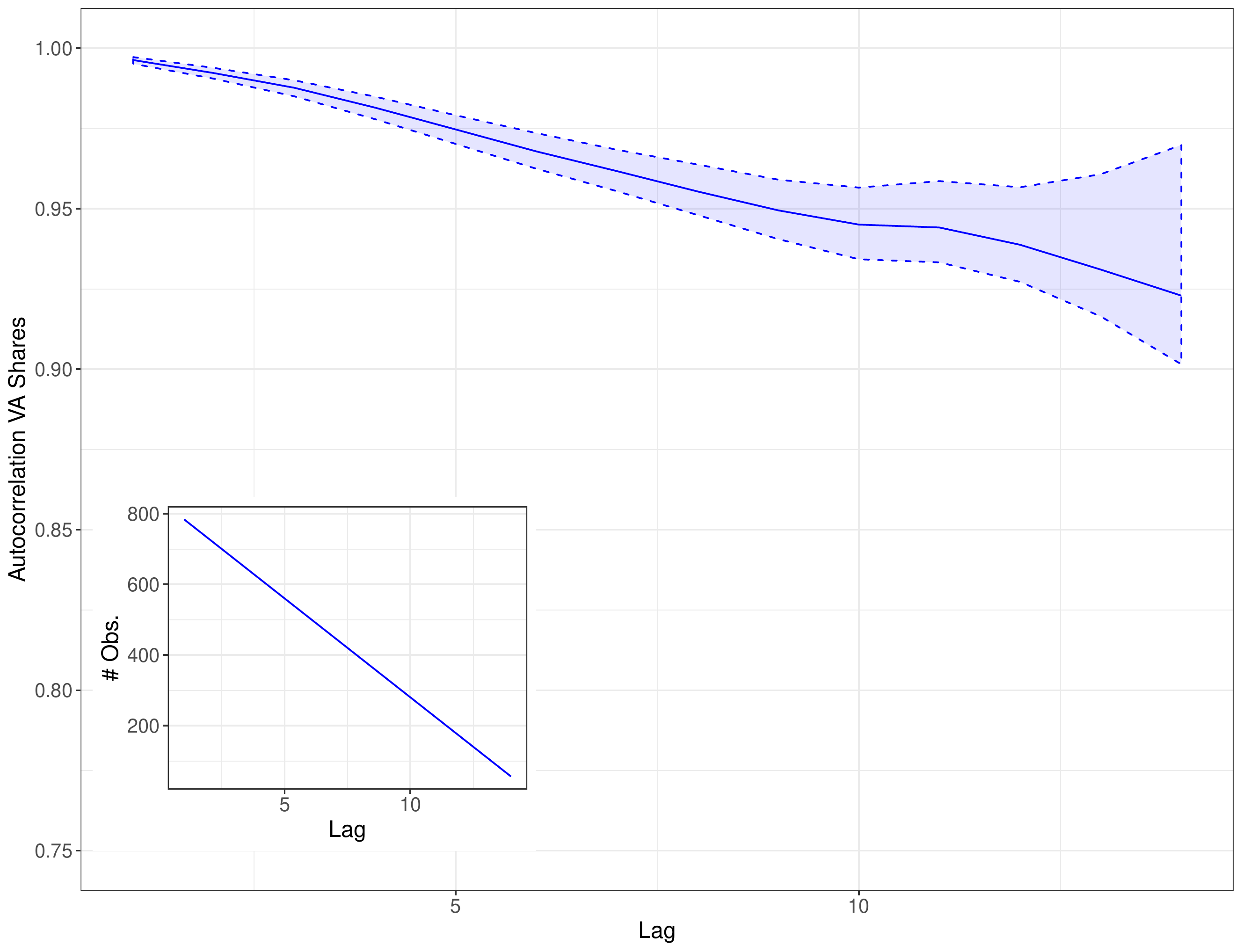}}
\caption{Autocorrelations of sectoral shares}
\label{fig:MAC:shares}
\end{figure}

\begin{figure}[h!]
\centering
\subfloat[Employment]{\includegraphics[width=0.5\textwidth]{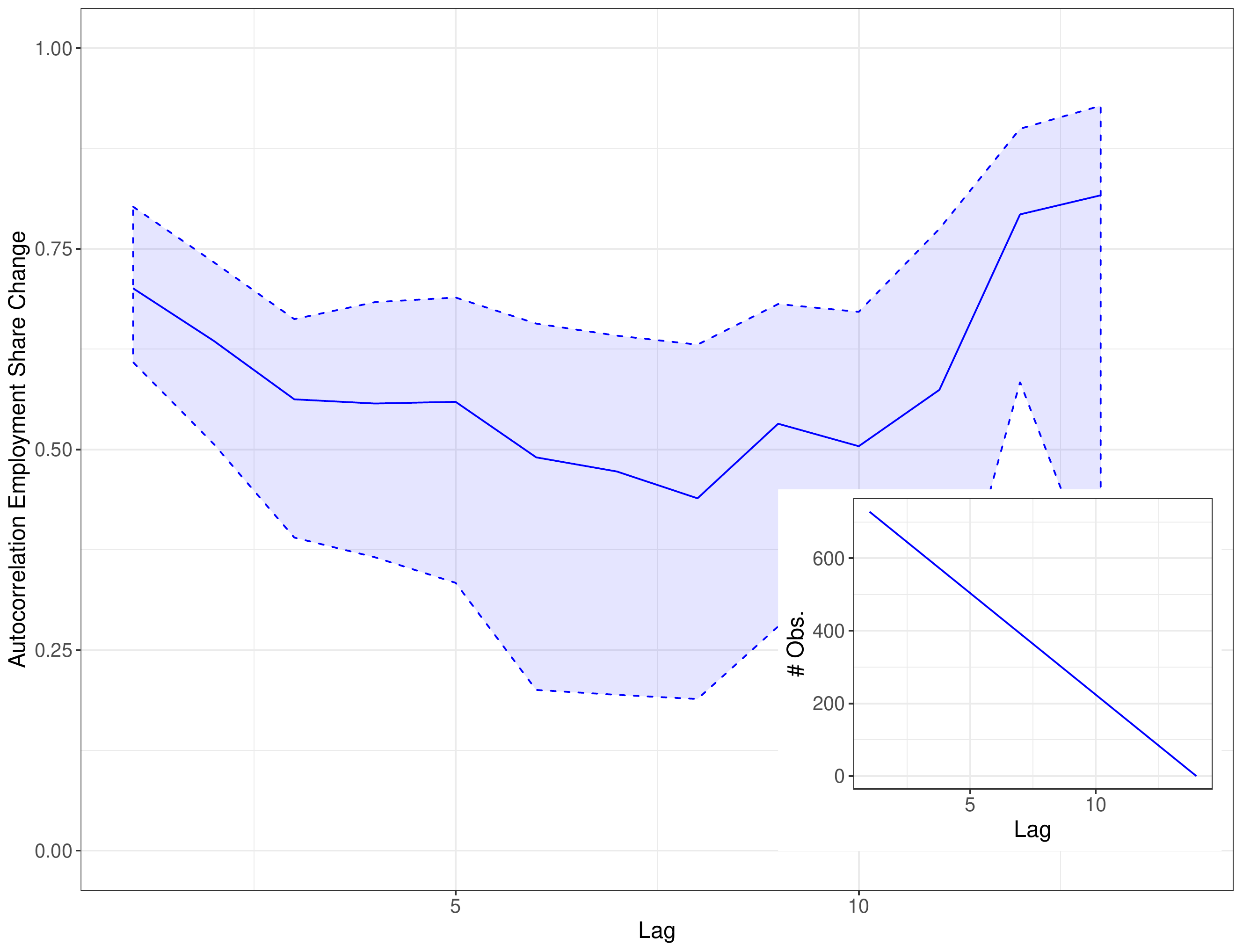}}
\subfloat[Value added]{\includegraphics[width=0.5\textwidth]{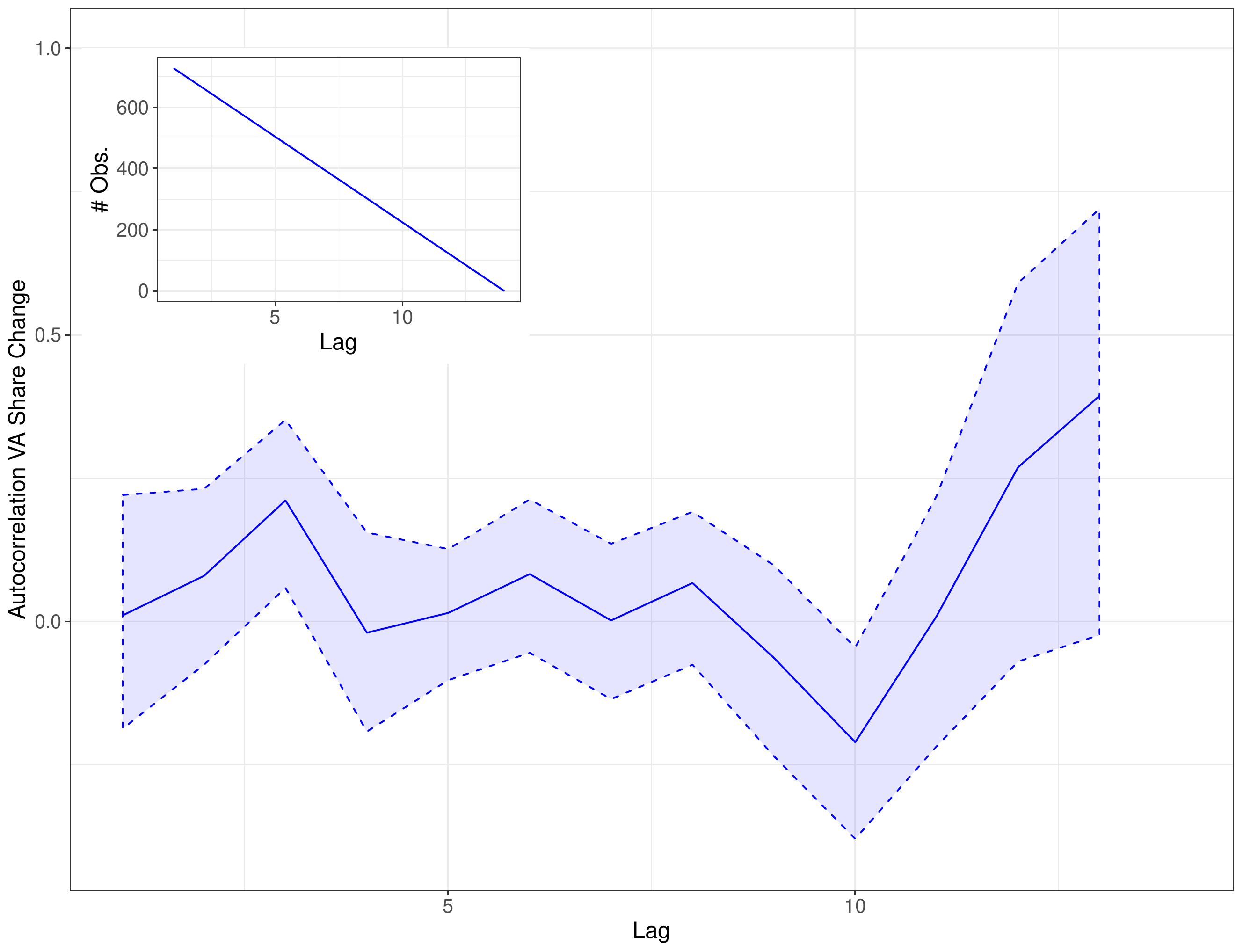}}
\caption{Autocorrelations of the changes in sectoral shares}
\label{fig:MAC:shares:diff}
\end{figure}

\begin{figure}[h!]
\centering
\subfloat[Employment]{\includegraphics[width=0.5\textwidth]{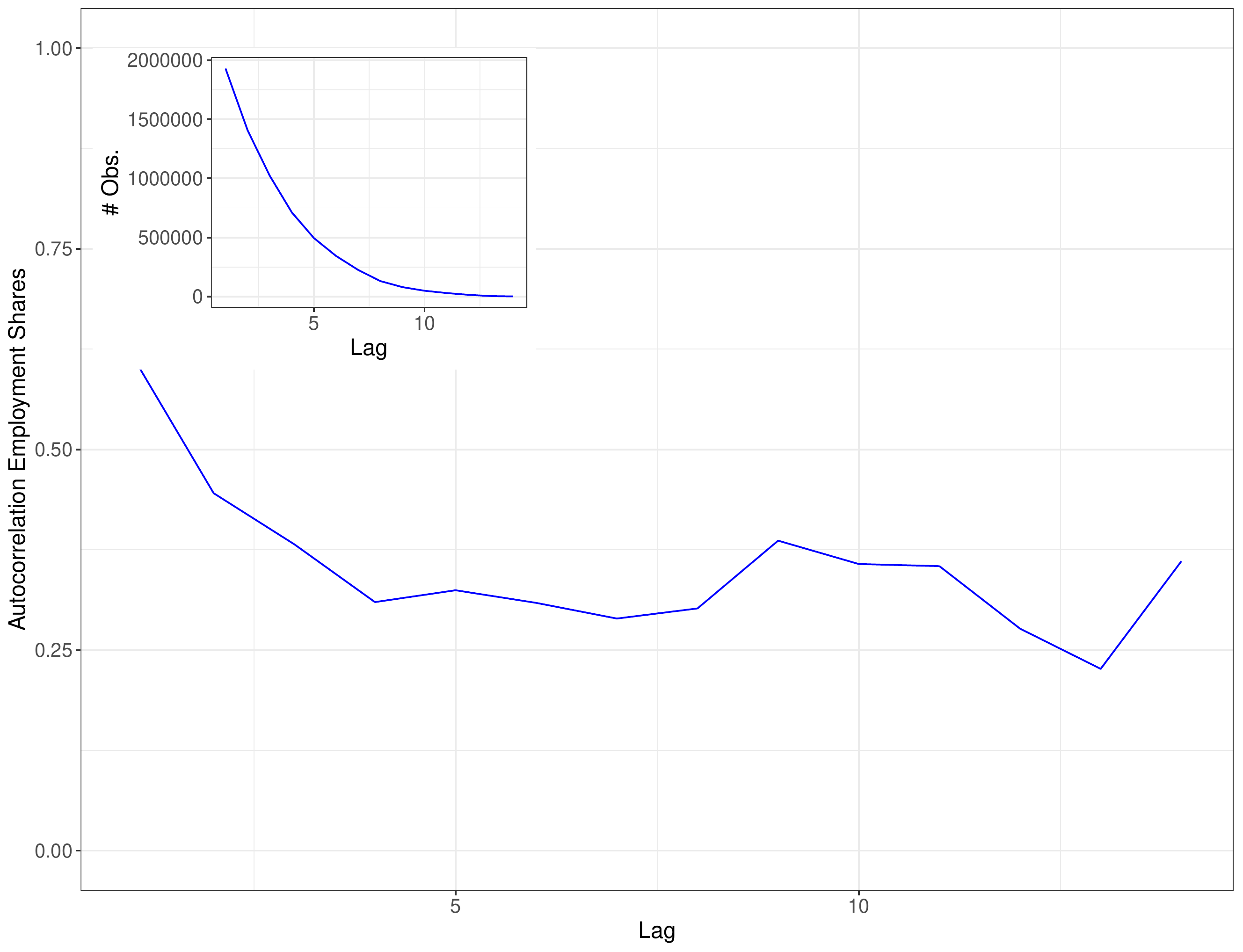}}
\subfloat[Value added]{\includegraphics[width=0.5\textwidth]{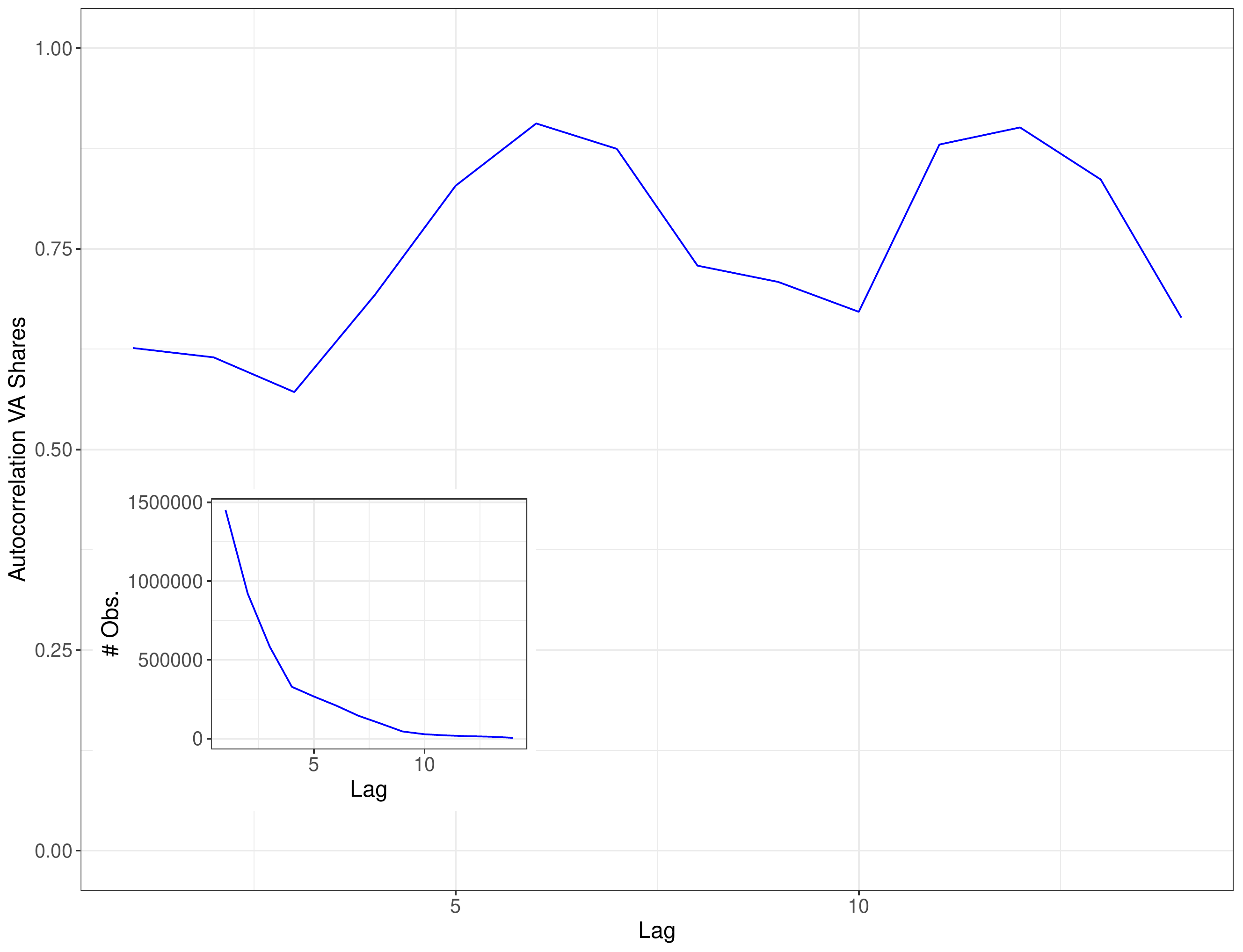}}
\caption{Autocorrelations of shares of employment and value added at the firm-level}
\label{fig:AC:firmlevel}
\end{figure}

\subsection{Autocorrelation of sectoral and micro-level quantities}
\label{sect:results:autocorrelations}

\citet{Metcalfeetal06} reported that industry sectors in their sample (the post-war USA) show differential growth in employment and in output - very much in line with our findings above - and that these have a significant degree of rigidity. The sectoral shares are highly autocorrelated and do not change quickly. Our results in Section \ref{sect:results:sectoral} and specifically in Fig. \ref{fig:Dev:2d:EMPL:LP} already point in the same direction. In this section, we briefly discuss autocorrelation sprectra of the relevant variables, employment and value added at the sectoral level in shares (Fig. \ref{fig:MAC:shares}) and in first differences of shares (Fig. \ref{fig:MAC:shares:diff}) as well as at the firm level (levels only shown in Fig. \ref{fig:AC:firmlevel}). The autocorrelation is 

\begin{equation}
 \varrho_X(\tau) = \frac{\mathop{\mathbb{E}}[(X_t-\mathop{\mathbb{E}}(X_t))(X_{t+\tau}-\mathop{\mathbb{E}}(X_{t+\tau}))]}{\sigma_{X_t}\sigma_{X_{t+\tau}}}
\end{equation}

where $\mathop{\mathbb{E}}(X)$ is the expectation and $\sigma^2_X$ is the variance of $X$. Note that we use broad range autocorrelations, i.e., we re-sample observations at different time periods to get more statistical power. The autocorrelation functions are therefore not functions of specific times, but merely of the time lag. The number of observations (and the rate of resampling) decreases with time lag. Since resampling does not allow us to use standard confidences, we bootstrap the standard errors (for 95\% confidence intervals) for the sectoral data.\footnote{We do not have sufficient computation power to obtain the bootstrap for the firm level.}  

We find very high and slowly decreasing levels of autocorrelation for the levels of the sectoral quantities, with employment decreasing slightly faster and with wider confidence margins (\ref{fig:MAC:shares}). For employment, positive autocorrelations are still present in the first differences (and possibly with no decay with increasing lags), while for value added this is not the case (Fig. \ref{fig:MAC:shares:diff}). That is, increasing employment shares tend to continue to increase and vice versa, potentially bringing about strong and increasing differences between sectors. For value added the short term trend can reverse while the levels evidently remain stable over long periods. Positive and high autocorrelations in employment and especially in value added are also present at the firm level (Fig. \ref{fig:AC:firmlevel}).

%\clearpage

\begin{figure}[h!]
\centering
\subfloat[Value added]{\includegraphics[width=0.5\textwidth]{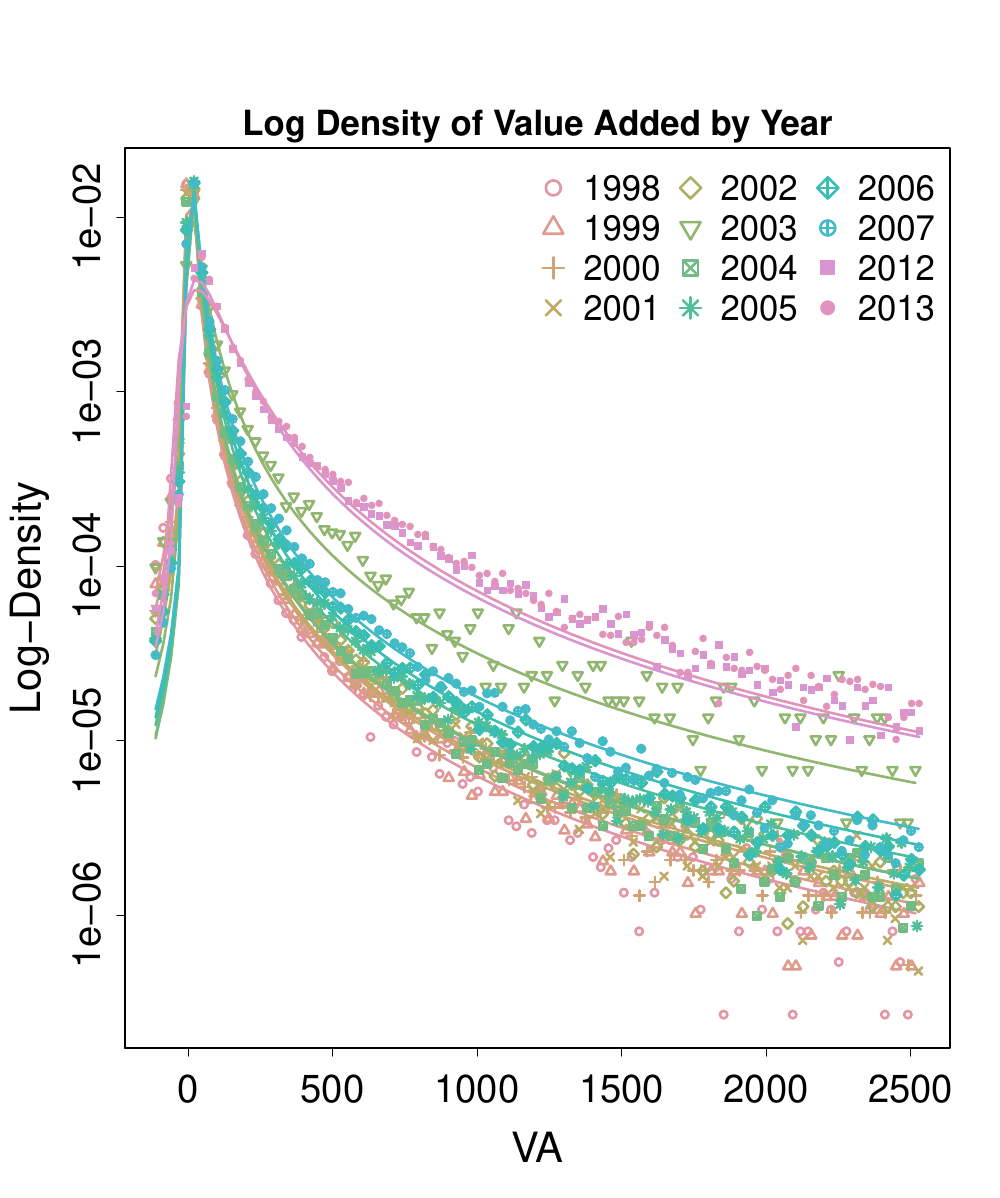}}
\subfloat[Value added change]{\includegraphics[width=0.5\textwidth]{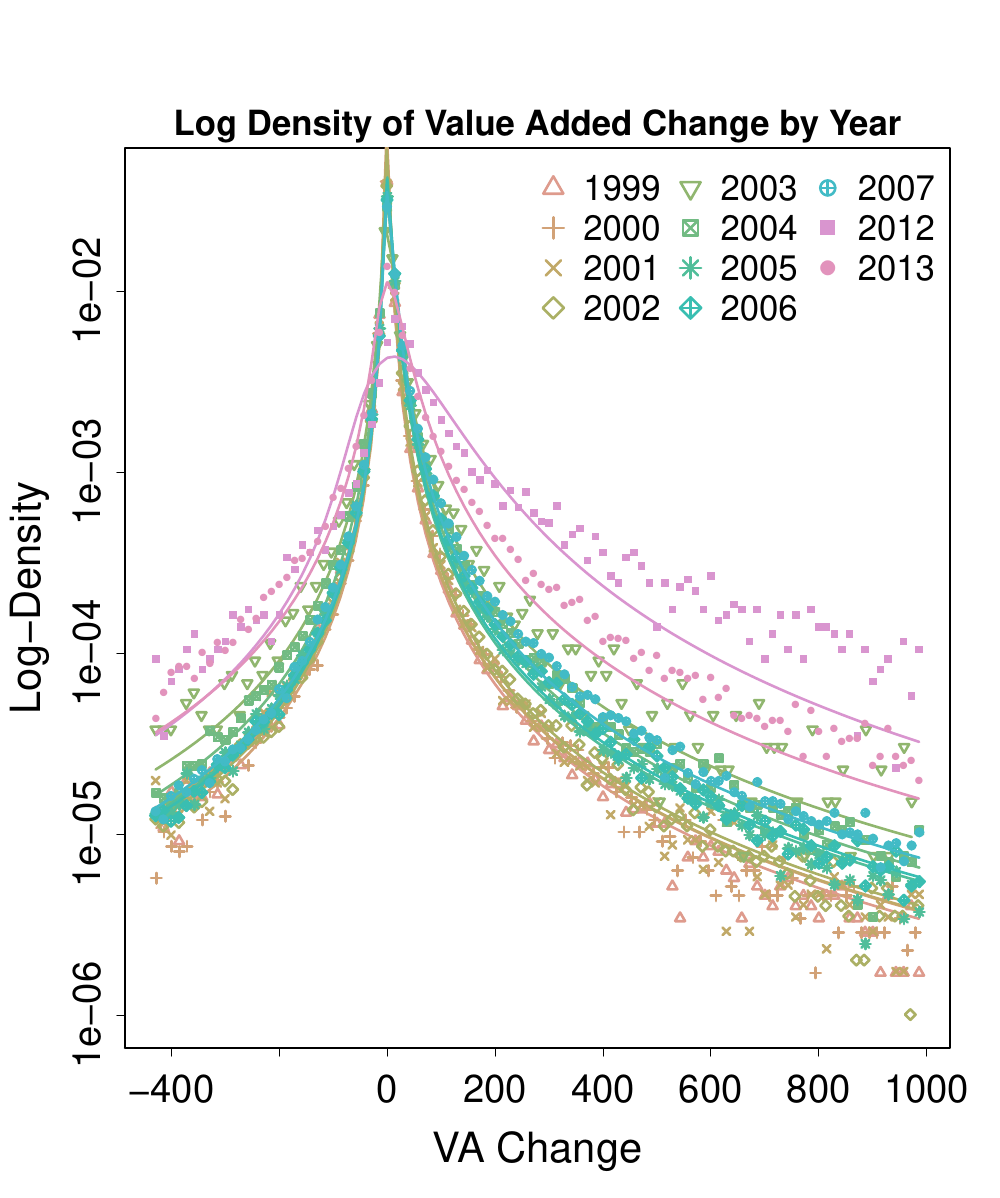}}\\
\subfloat[Value added growth]{\includegraphics[width=0.5\textwidth]{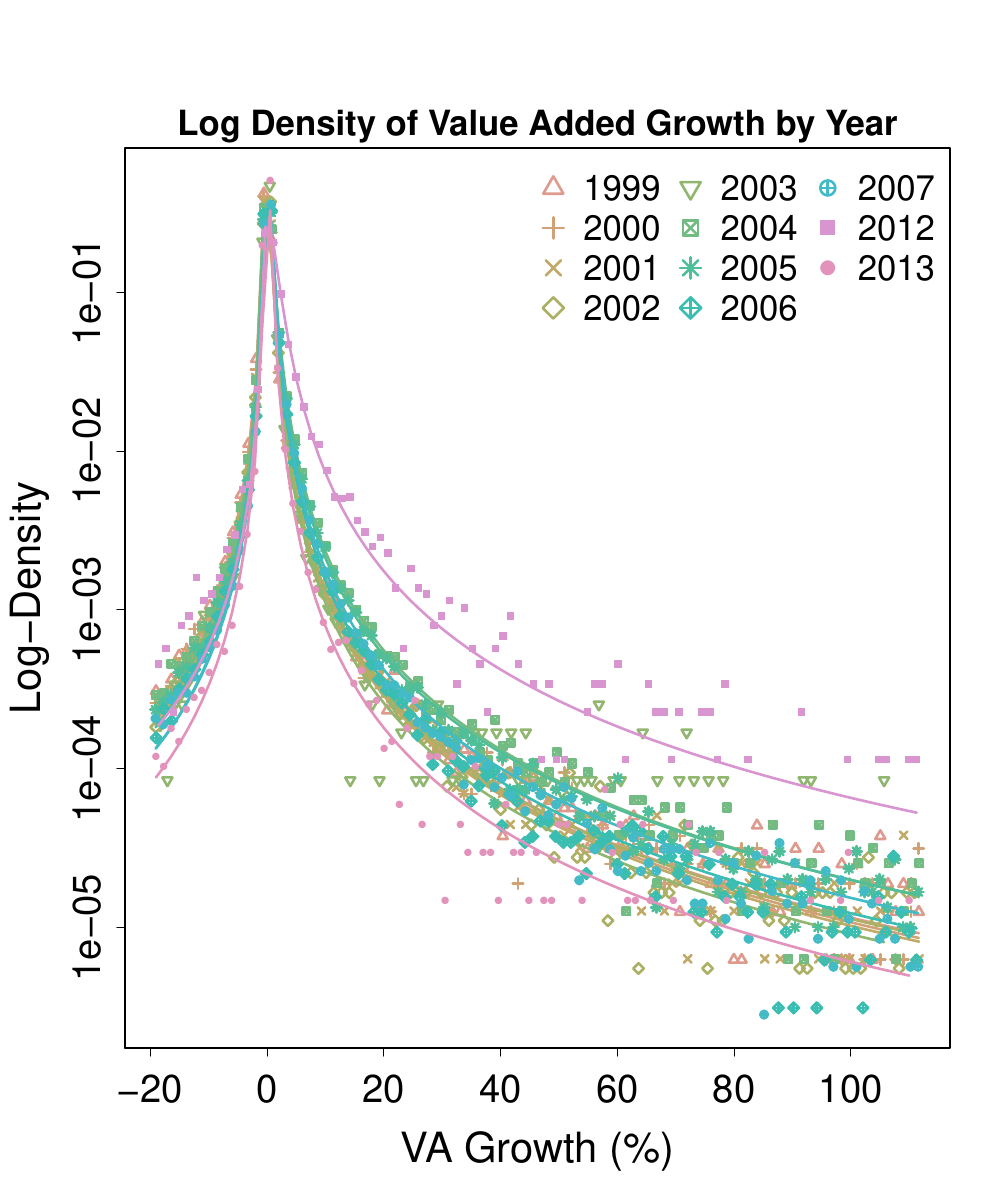}}
\subfloat[Labor productivity]{\includegraphics[width=0.5\textwidth]{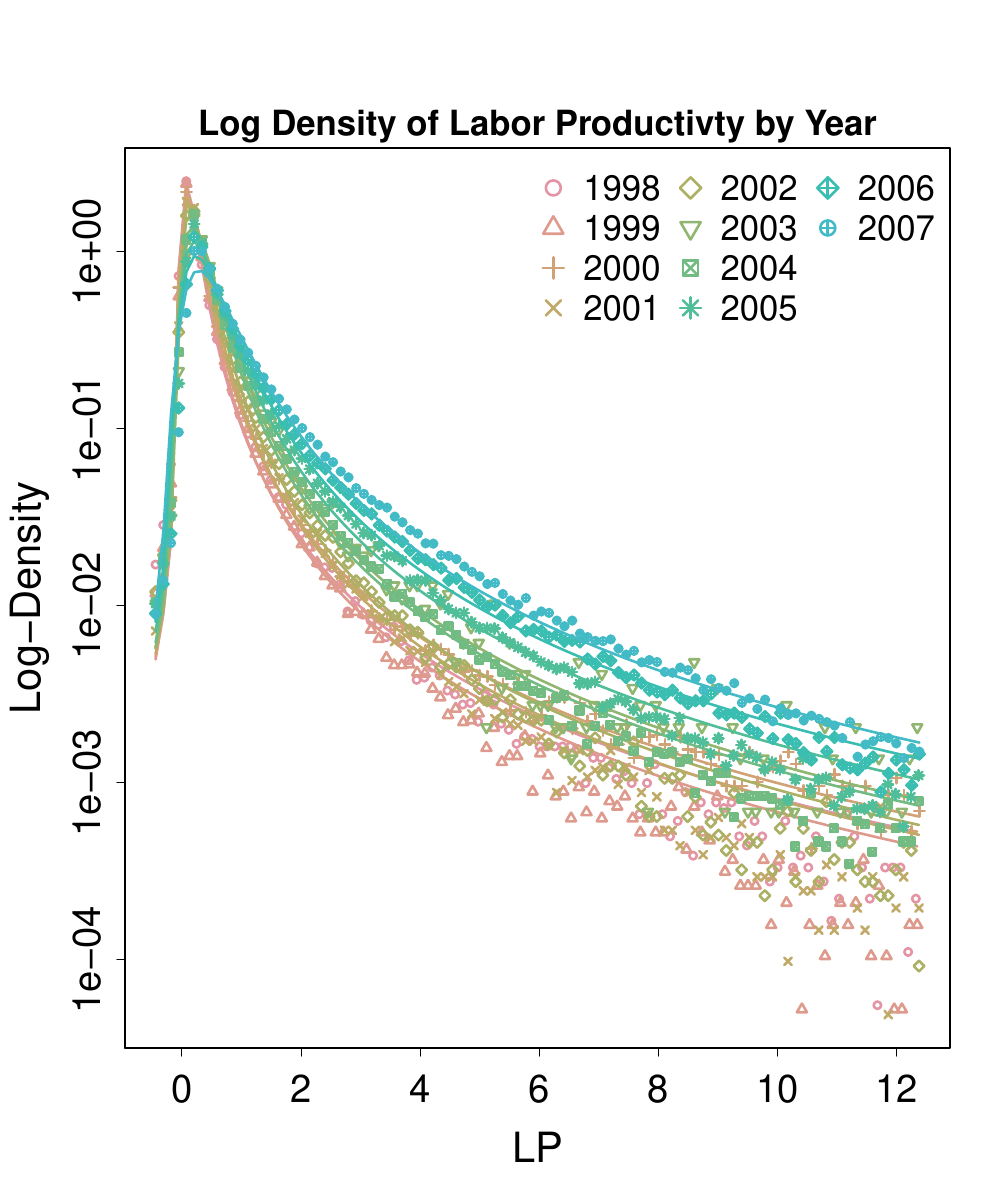}}
\caption{Density of value added and derived variables by year}
\label{fig:density:VA}
\end{figure}

\begin{figure}[tb!]
\centering
\includegraphics[width=0.5\textwidth]{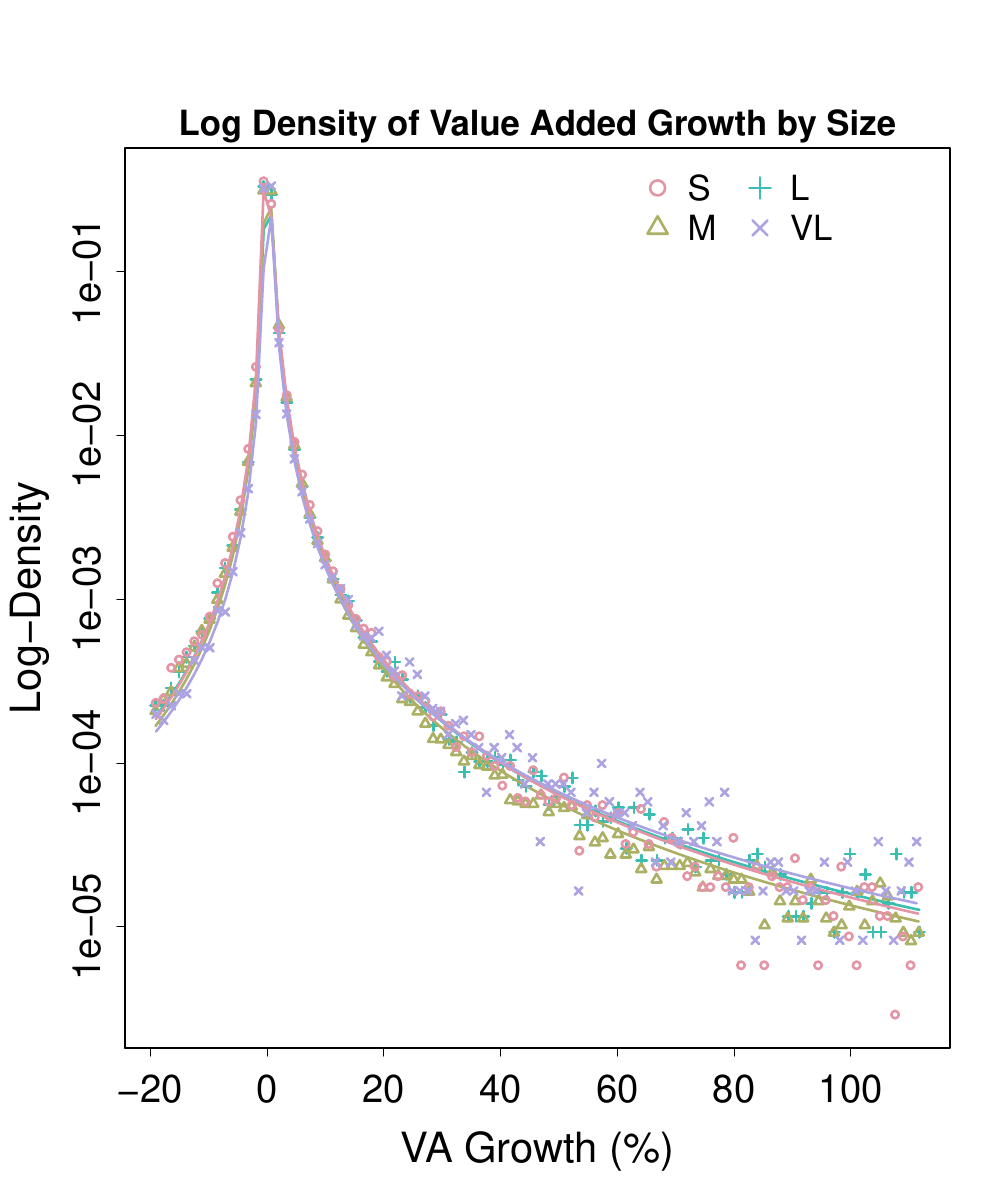}
\caption{Density of value added growth by firm size categories (S - small, M - medium sized, L - large, VL - very large).}
\label{fig:density:VAGrowth:Size}
\end{figure}

\begin{figure}[tb!]
\centering
\includegraphics[width=0.85\textwidth]{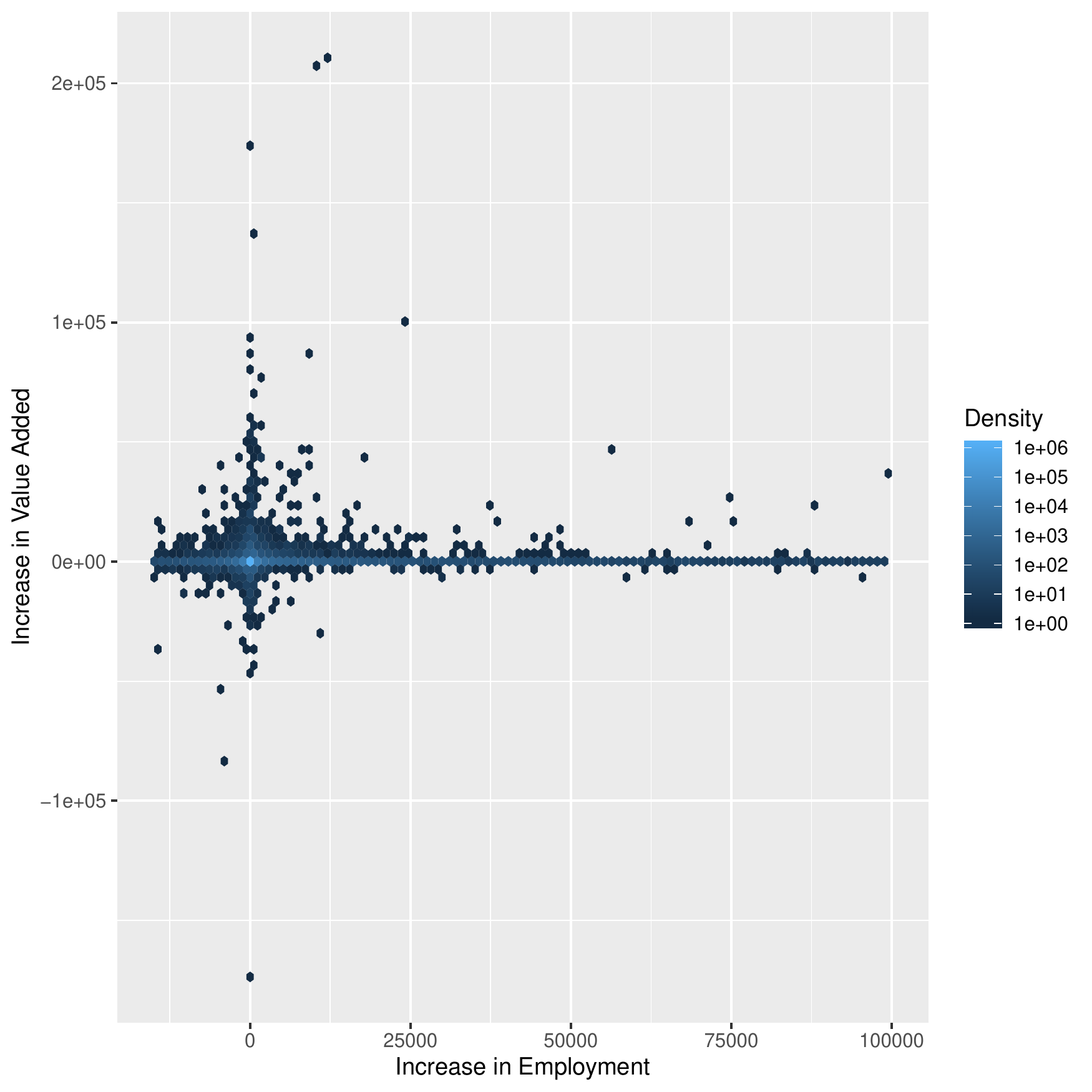}
\caption{2d histogram of $dl_i$ and $dy_i$ in CIE DB firm-level data}
\label{fig:2dhistogram}
\end{figure}

\subsection{Micro-level distributions and their functional forms}
\label{sect:results:distributions}

Elsewhere \citep{Yangetal19,Heinrichetal19}, we have shown that both labor productivities and labor productivity changes follow L\'{e}vy alpha-stable distributions. Specifically, the L\'{e}vy alpha-stable distribution is a much better distributional model than asymmetric exponential power distributions suggested for firm level growth (for which labor productivity or labor productivity changes may be proxies) elsewhere in the literature \citep{Bottazzi/Secchi06,Bottazzi/Secchi11}. We chose to focus on linear first differences instead of growth rates, as growth rates may be misleading and counter-intuitive for variables that can be negative like value added $y_k$ and labor productivity $q_k=y_k/l_k$ at the firm level.\footnote{For a detailed discussion, see \citet{Yangetal19}.} 

For the present study, employment and value added are equally relevant. While we may remain agnostic with respect to the specific distributional model of each of the variables, they are all heavy tailed. In the case of the value added as well as value added change, value added growth, and labor productivity (which is value added divided by employment, a quotient of two heavy tailed variables) the L\'{e}vy alpha stable seems to fit the distribution rather well, the densities and fits to L\'{e}vy alpha-stable are shown in Fig. \ref{fig:density:VA}. %It cannot possibly be a good fit for employment, as this is a one-tailed variable (there is no negative employment).

We perform finite moment tests using Trapani's approach \citep{Trapani16} and R's \verb|finity| package \citep{R::finity}, yielding infinite moments for the second moments in all relevant variables, see Table \ref{tab:finity-tests}. A consequence is that sample moments corresponding to the non-existing moments should not be used since they are contaminated by information about the sample sizes and do not offer an unbiased reading of the moment. There does not appear to be an average employment or a variance of labor productivity, value added, etc. This must be taken into account for any regressions run with these variables: There is a strong possibility that the errors inherit the heavy-tailedness of the variables, which would render OLS assumptions violated and OLS results invalid.

\begin{table}[hbt]
    \centering
    \caption{Finiteness tests for the first two moments (i.e., mean, variance) for value added ($Y$), value added change ($dY$), value added growth ($\dot{Y}$), value added ($L$), value added change ($dL$), value added growth ($\dot{L}$), labor productivity ($Q$), and labor productivity change ($dQ$) following Trapani's approach \citep{Trapani16} and R's finity package \citep{R::finity}. The second rows give the corresponding p-value: Finiteness is rejected if the p-value exceeds a threshold ($0.1$).}
    \label{tab:finity-tests}
\begin{tabular}{c c c c c c c c c}
\hline\hline
\textbf{Moment order} & \textbf{$Y$} & \textbf{$dY$} & \textbf{$\dot{Y}$} & \textbf{$L$} & \textbf{$dL$} & \textbf{$\dot{L}$} & \textbf{$Q$} & \textbf{$dQ$} \\
\hline\hline
  & 1937.39 & 1323.7 & 652.39 & 0.52   & 0   & 0   & 4750.84 & 1942.23 \\
1 & (0)     & (0)    & (0)    & (0.47) & (1) & (1) & (0)     & (0) \\
  & finite  & finite & finite & infinite & infinite & infinite & finite & finite \\\hline
  & 0       & 0      & NaN    & 0      & 0   & NaN & 0       & 0 \\
2 & (1)     & (1)    & (1)    & (1)    & (1) & (1) & (1)     & (1) \\   
  & infinite  & infinite & infinite & infinite & infinite & infinite & infinite & infinite \\
\hline\hline
 \end{tabular}
\end{table}

Finally, we investigate the bivariate dispersion and discover of the first differences in employment and value added. Fig. \ref{fig:2dhistogram} reveals that while there is some correlation (see Section \ref{sect:results:fabricant}), there is no string correspondence and the highest densities remain concentrated along the axes (i.e., change in one variable while the other remains approximately constant is the most frequent case).

\section{Regression analysis: the micro-macro connection in sectoral growth}   
\label{sect:regressions}

In section \ref{sect:results}, we have shown the characteristics of structural change for the PR China in the period of study (1998-2014) at both the sectoral and the micro-level. We have discussed the likely distributions of data, connections between variables, the stark difference between employment growth and output (value added) growth, and have discussed some implications and interpretations. We have also demonstrated that there is, unsurprisingly, a connection between sectoral and micro-level variables. We will now investigate this connection in more detail and will for this return to the evolutionary approach discussed in section \ref{sect:model:evol}.

\subsection{Model specification}

 Following the model in Eq. \ref{eq:regression-replicator}, we will in this section run regressions on firm level growth $\dot{x_i}$ (where $x_i$ can be employment $l_i$ and value added $y_i$) with sectoral level growth $\dot{x_k}$ as predictors in a linear (additive) combination with other possible predictors. We will continue to highlight differences between growth processes in employment and in value added. We avoid using multiple highly correlated predictors (as shown in  see Fig \ref{fig:heatmap:micro}) to avoid multicolinearities. This leads us to avoid capital and wage bill (correlated with value added), revenue (correlated with value added and employment), as well as wages, profitability, and capital intensity (correlated with labor productivities). 
 
 We do, however, include three predictors besides the sectoral growth: labor productivity, $q_i$, labor productivity change $dq_i$, and firm age $a_i$. 
 
 There is a broad literature connecting productivities and profitabilities to either growth potential or to realized growth \citep{Goddardetal04,Coad07,Bottazzietal08}; the coefficient is expected to be positive.\footnote{There are different conjectures regarding the causality. \citet{Goddardetal04} found profitability to impact growth while \citet{Coad07} hypothesized that growth may give managers slack for efficiency- and profitability-enhancing reorganization.}

Furthermore, especially the evolutionary literature expects growth to depend on relative performance or relative capabilities of the firm. In many contributions to evolutionary economics, fitness terms, as they appear in equation \ref{eq:standard-replicator} and similar models, are identified as productivity \citep{Nelson/Winter82,Silverberg/Lehnert93,Mulderetal01}.\footnote{In other models, the output price takes the role of the fitness \citep{Nelson/Winter82}; the two options may be connected via an appropriate demand function. Models frequently include more complex dynamics with labor and capital productivity or with technological change through innovation and imitation.} For a discussion of the empirical literature, see \citet{Coad07}. In the present case, working with relative productivity is not appropriate for two reasons: First, computing relative productivity would require computing the first moment. However, as the variable is heavy tailed, that moment may in samples be unreliable, since even if it exists, it will converge only slowly with sample size. Second, we only have sample data, not the full population. In light of this, it appears appropriate to use the first difference of labor productivity as a proxy. Under a wide range of assumptions changes in productivity should lead to dynamic adaptations of the growth process in evolutionary models. Labor productivity change is available for a sizeable subset of our observations.\footnote{The drop in the number of observations between models (1) and (2) in table \ref{tab:regression:heavy} is partly due to this.} 

A substantial literature also connects firm ages to growth (cf. \citet{Coad07}), often with a focus on potentially more dynamic startups \citep{Coad07,Pugsleyetal19}. However, a weakening of this has been observed in the recent past \citep{Pugsleyetal19}. Crucially for our study, the effect of firm age has also been found to be reversed for developing countries, specifically India, with older firms having better growth prospects \citep{Das95}.

Finally, effects of size, geography, sector, and other context may also be expected \citep{Coad07}, some studies going as far as ascribing a crucial impact on the distribution of growth rates to firm size \citep{Bottazzietal19}. While a visual inspection of our data (see figure \ref{fig:density:VAGrowth:Size}) does not confirm a significant impact of size, we will include fixed effects for size categories (small, medium sized, large, very large), firm types\footnote{We use the standard categorization for Chinese firms also employed by, e.g., \citet{yu2015institutional}, in SOE, collective owned, private, shareholding, other domestic, foreign, and Hong Kong, Macao or Taiwan owned firms.}, years, provinces, and also sectors, although sector fixed effects may interfere with the sectoral growth estimator ($\dot{x_k}$).

\begin{figure}[tb!]
\centering
\includegraphics[width=0.85\textwidth]{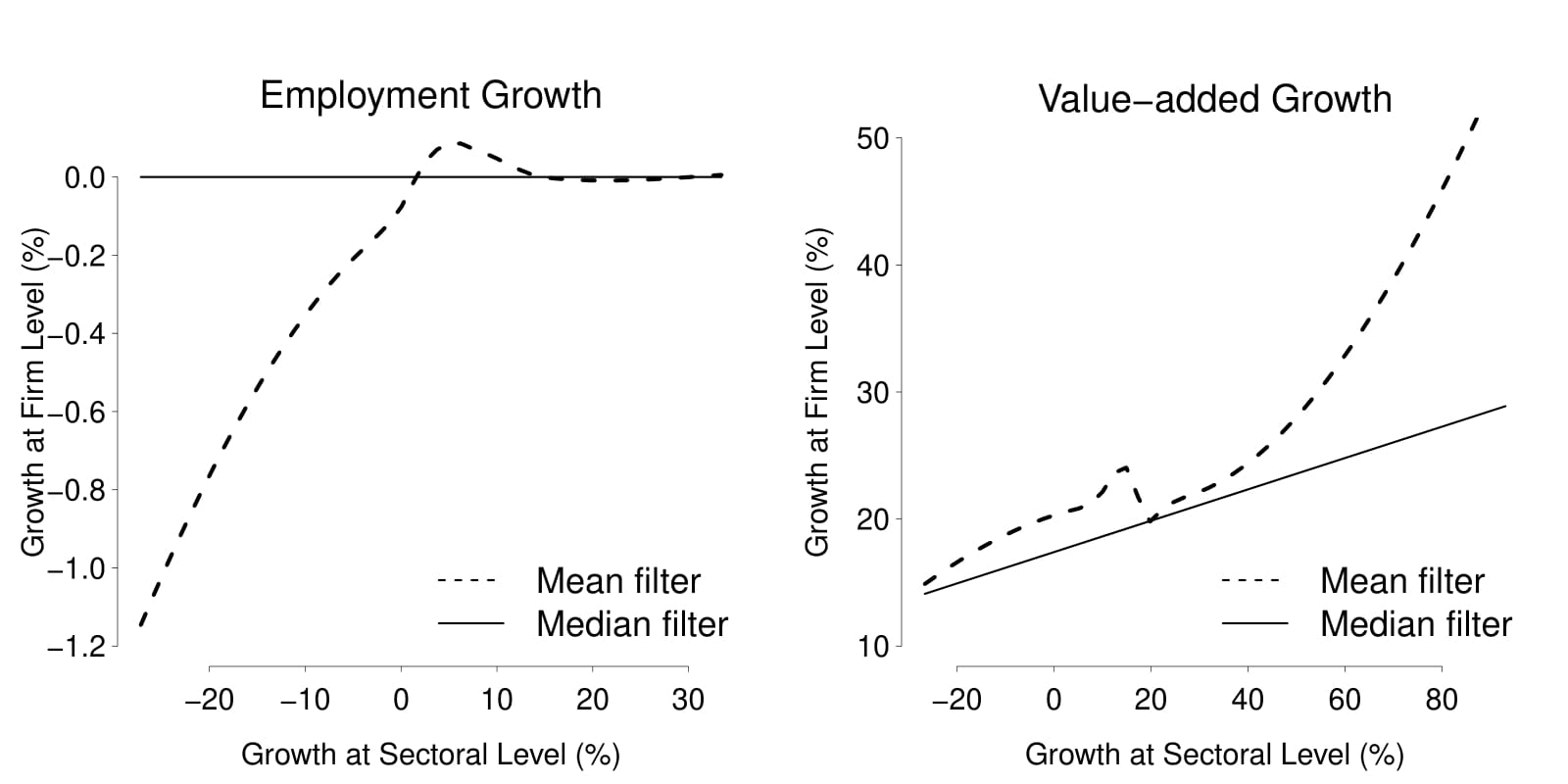}
\caption{Filtered mean and median of growth rate of employment (left panel) and value added (right panel) at the firm-level conditional on the corresponding employment and value added growth at the sectoral level}
\label{fig:filtered_mean_median}
\end{figure}

Figure \ref{fig:filtered_mean_median} shows the filtered mean and median of growth rate of employment (left panel) and value added (right panel) at the firm-level conditional on the corresponding employment and value added growth at the sectoral level, respectively.  The mean filter picks up a positive relationship between the firm-level and sector level employment growths. However, this result is somewhat misleading considering that the firm-level distribution of employment growth is heavy-tailed so its mean is less informative than the median. When we use the median filter, the positive relationship disappears. For the value-added case, both mean and median filters pick up a positive relationship between the firm-level and sectoral level growths. The following regression attempts to look into the relationship between the firm-level and sectoral growth of employment and value-added in a more systemic manner. %Considering the heavy-tailedness of the variables, we need to use a regression model that is robust to extreme values. We choose to use Cauchy error model. We compare the results to the OLS in Appendix \ref{app:regressions}, Tables \ref{tab:regression:lm:E} and \ref{tab:regression:lm:VA}. 

Starting from equation \ref{eq:regression-replicator} and taking into account hypotheses from the literature and characteristics of the data as explained in the present section, we consider the following models:

\begin{align}
\label{eq:lm50}
  \dot{x_{i}}=& \beta_0 + \beta_1 \dot{s_k} + \varepsilon \\
\label{eq:lm51}
  \dot{x_{i}}=& \beta_0 + \beta_1 \dot{s_k} + \beta_2 q_i + \beta_3 dq_i + \beta_4 a_i + \varepsilon \\
\label{eq:lm52}
  \dot{x_{i}}=& \beta_0 + \beta_1 \dot{s_k} + \beta_2 q_i + \beta_3  dq_i + \beta_4 a_i + \\
\notag        & YearFE + ProvinceFE + FirmTypeFE + FirmSizeFE + \varepsilon \\
\label{eq:lm53}
  \dot{x_{i}}=& \beta_0 + \beta_1 \dot{s_k} + \beta_2 q_i + \beta_3  dq_i + \beta_4 a_i + \\
\notag        & YearFE + ProvinceFE + FirmTypeFE + FirmSizeFE + SectorFE + \varepsilon 
\end{align}
where we run the regressions for both employment growth ($\dot{x_i}$ and $\dot{x_k}$ then being $\dot{l_i}$ and $\dot{l_k}$) and for value added growth as the appropriate indicator of output growth at the firm level ($\dot{x_i}$ and $\dot{x_k}$ then being $\dot{y_i}$ and $\dot{y_k}$). The $FE$ variables are fixed effects.

\subsection{Estimation results}

Performing OLS fits of models \ref{eq:lm50}, \ref{eq:lm51}, \ref{eq:lm52}, and \ref{eq:lm53} for both employment growth and value added growth reveals that the residuals of each of the regressions belong to a heavy-tailed distribution with infinite second moment. As a consequence, the underlying assumptions of OLS fitting are violated and the regression results have to be treated as unreliable. The results are reported for comparison in tables \ref{tab:regression:lm:E} and \ref{tab:regression:lm:VA} in Appendix \ref{app:regressions}. The last lines of the tables also give the results of the finiteness test following \citet{Trapani16} for the second moments.\footnote{The R package finity \citep{R::finity} was used.}

Consequently, the models were fitted using robust regression with Cauchy distributed errors\footnote{The Cauchy distribution is a special case of the L\'{e}vy alpha-stable with parametrization $(1, 0, \gamma, \delta)$.} using the R package \verb|heavy| \citep{R::heavy}. Table \ref{tab:regression:heavy} gives the the results of robust regressions for employment and value added growth. % (1) is the baseline regression of firm-level growth ($\dot{l_i}$ and $\dot{y_i}$ respectively) on the sectoral growth ($\dot{l_k}$ and $\dot{y_k}$ respectively) with no other covariates. (2) adds three covariates, labor productivity $q_i$, labor productivity change $dq_i$, and firm age $a_i$, (3) adds a few contral variables as fixed effects: firm-type, firm-size, province, and year indices, and (4) adds sector fixed effect. 

%%%%%%% JH 
\begin{sidewaystable}[!p] 
\centering 
\footnotesize
\vspace*{12cm}
  \caption{Robust regression results for models (1), (2), (3), (4) with firm-level employment growth $\dot{l_i}$ (left hand side) and value added growth $\dot{y_i}$ (right hand side) as dependent variable.} 
  \label{tab:regression:heavy} 
\hspace*{-4.5cm}
\begin{tabular}{@{\extracolsep{-15pt}}lD{.}{.}{-3} D{.}{.}{-3} D{.}{.}{-3} D{.}{.}{-3} p{1cm} D{.}{.}{-3} D{.}{.}{-3} D{.}{.}{-3} D{.}{.}{-3} } 
\\[-1.8ex]\hline 
\hline \\[-1.8ex] 
 & \multicolumn{9}{c}{\textit{Dependent variable:}} \\ 
\cline{2-10} 
\\[-1.8ex] & \multicolumn{4}{c}{Employment Growth} && \multicolumn{4}{c}{Value-Added Growth}  \\ 
\cline{2-5} \cline{7-10} 
\\[-1.8ex] & \multicolumn{1}{c}{(1) (Eq. \ref{eq:lm50})} & \multicolumn{1}{c}{(2) (Eq. \ref{eq:lm51})} & \multicolumn{1}{c}{(3) (Eq. \ref{eq:lm52})} & \multicolumn{1}{c}{(4) (Eq. \ref{eq:lm53})} && \multicolumn{1}{c}{(1) (Eq. \ref{eq:lm50})} & \multicolumn{1}{c}{(2) (Eq. \ref{eq:lm51})} & \multicolumn{1}{c}{(3) (Eq. \ref{eq:lm52})} & \multicolumn{1}{c}{(4) (Eq. \ref{eq:lm53})}\\  \\ 
\hline \\[-1.8ex] 
 \hline
Constant  & 0.0000  & 0.0062   & 0.0094   & 0.0127    && 0.1772   & 0.1978   &  0.15    & 0.1767   \\ 
          &(0.0000) & (0.0001) & (0.0007) & (0.0009)  && (0.0008) & (0.0008) & (0.0027) & (0.0032) \\ 

        & & & & \\ 
Sectoral growth 
          & 0.0000 & 0.0101   & 0.0157   & 0.018     && 0.1944   & 0.1264   & 0.0374   & -0.0034  \\ 
          & (0.0001)& (0.0014) & (0.0018) & (0.0025)  && (0.0043) & (0.0034) & (0.0038) & (0.0043) \\ 
     & & & & \\ 
Labor productivity
          &         & 0.005    & 0.0051   &  0.0048   &&          & -0.2476  & -0.2425  & -0.2519  \\
          &         & (0.0000) & (0.0000) & (0.0000)  &&          & (0.0001) & (0.0000) & (0.0000) \\ 
     & & & & \\ 
Labor productivity change 
          &         & -0.0081  & -0.0077  & -0.0075   &&          & 1.1604   & 1.1274   & 1.1287   \\ 
          &         & (0.0000) & (0.0000) & (0.0000)  &&          & (0.0001) & (0.0001) & (0.0001) \\ 
     & & & & \\ 
Firm age  &         & -0.0004  & -0.0004  & -0.0004   &&          & -0.002   & -0.0022  & -0.0022  \\ 
          &         & (0.0000) & (0.0000) & (0.0000)  &&          & (0.0000) & (0.0000) & (0.0000) \\ 
    & & & & \\ 
   \hline \\[-1.8ex] 
Observations       & 1,928,610 & 943,916  & 943,916  & 943,916   && 965,698    & 943,439    & 943,439    & 943,439 \\ 
Degrees of freedom & 1,928,608 & 943,911  & 9438,63  & 943,845   && 965,696    & 943,434    & 943,386    & 943,368 \\ 
Log-likelihood     & 877,346   & -128,178 & -96,490  & -95,971   && -1,195,808 & -1,061,605 & -1,038,828 & -1,037,930 \\ 
   \hline
\hline \\[-1.8ex] 
%\textit{Note:}  & \multicolumn{4}{r}{$^{*}$p$<$0.1; $^{**}$p$<$0.05; $^{***}$p$<$0.01} \\ 
\end{tabular}
\end{sidewaystable}

All coefficients across all models are statistically significant at the $0.01$\% level except for the coefficient of $\dot{y_k}$ in model (4) (Eq. \ref{eq:lm53}) for value added growth. 

Sectoral growth coefficients are in every case positive. Firm-level growth proves to be associated with sectoral growth in the same direction as per our hypothesis from Section \ref{sect:model:evol}. In all employment growth models this coefficient is significant, albeit small. %Employment growth is evidently determined much stronger by other influences besides its sectoral growth aspect. 
For value added, the coefficient becomes non-significant when sector fixed effects are included, but is otherwise positive and significant. This indicates that there is indeed a sectoral effect - the effect may just not have a linear form (cf. also Fig. \ref{fig:filtered_mean_median}). It may be subject to other sector-specific influences besides the sector level growth, although these influences are sufficiently close to the sector level growth to capture its impact on the dependent variable and render its coefficient non-significant. Sector fixed effects tend to be lower for most industry sectors (C sectors), than for either mining (B) or electricity and water distribution (D35, E36).

The coefficient for labor productivity is equally present and significant. It is positive but small for employment growth (higher labor productivity being associated with employment growth), but negative for value added growth. The latter is surprising but may constitute a saturation effect (if high productivity firms have mostly reached their capacity), while the positive impact of productivity is captured by the productivity growth term. The coefficient for labor productivity change is positive for value added growth, but negative and of relatively small magnitude for employment growth. We suspect that this reflects productivity increases through changes in capital intensity, which may occasionally be accompanied by a reduction or the labor force.

Firm age has a negative coefficient in all eight models, indicating that start-ups have higher growth potential than older firms.

Type\footnote{Private and foreign firms are associated with higher growth.}, size\footnote{Bigger firms are associated with higher growth.}, and year\footnote{This effect is dominated by data idiosynchrasies for particular years in the data set.} fixed effects are significant. Province fixed effects are only significant for some provinces.\footnote{Some of the most developed provinces (Shanghai, Fujian, Zhejiang etc.) are associated with lower employment growth but higher value added growth.}

%The key coefficient is the coefficient of sector level growth, $\dot{x_k}$, in each model. %For the employment growth.
%Since both firm-level and sector level growth are in percentage term, the coefficient is interpreted an increase in the firm-level growth by coefficient with a percentage increase in the sectoral growth. 
%
%The impact of sectoral employment growth on the firm-level employment growth is very limited as shown by the coefficient of $\dot{l_k}$ being close to zero across all four models. 
%
%The impact of sectoral value-added growth, $\dot{y_k}$, on the firm-level valued added growth, $\dot{y_i}$,  is ambiguous. It has relatively high impact in the models with no fixed effects. When various fixed effects are added, the coefficient becomes either small or insignificant. 

These results are dramatically different from those of OLS regressions, which we discuss in the Appendix \ref{app:regressions}. In the OLS regression, sectoral employment growth has a statistically significant impact on the firm-level growth across all 4 models as shown in Table \ref{tab:regression:lm:E}. In contrast, the impact of sectoral value-added growth on the firm-level value-added growth is positive but is statistically insignificant (Table \ref{tab:regression:lm:VA}). Various other effects are reversed compared to the robust regression.

\section{Discussion and conclusion}  
\label{sect:conclusion}

While structural change has been extensively studied in developed economies, not all aspects of it are well-understood for developing countries. Yes, it is obvious that modernization and the development of advanced industries\footnote{In other words, industries with a high Economic Complexity Index (ECI) should be developed. This is true although the ECI may not be a direct indicator of complexity but rather of similarity to other industries \citep{Mealyetal19}.} \citep{Hidalgoetal07,Hidalgo/Hausmann09} is a crucial factor and yes, institutional rigidities play a role. But why did the rapid development succeed in China and some other countries while it continues to be elusive in many cases? Can lessons be learned from China? Are policies applicable elsewhere? Did China's situation as a transformation economy moving from central planning to a market-based system play a part?

In this paper, we investigated the nature of China's structural change in the period 1998-2014 and any patterns and empirical regularities therein. We fitted a multi-level model inspired by considerations from evolutionary economics, and confirmed that growth at the micro-level and the sectoral level are connected. Sectoral level growth terms are a predictor of firm level growth. This may seem obvious, but it underlines the coherence of the model and the coefficients and the robustness to fixed effects shed light on the nature of this connection. For employment growth, the connection is robust and significant. For value added growth, the effect appears to be larger and remains robust and significant unless sector fixed effects are added, which then capture the sectoral effect better. This is likely due to either a nonlinear coupling of firm-level and sectoral terms or to the presence of other sector level specificities.

Second, we found that well-known correlation laws from sectoral decompositions (Fabricant's laws) fall apart at the micro-level. Crucially, there is no positive correlation between labor productivity and output variables (value added). In regressions, a significant association is found, but contrary to the sectoral level, it has a negative coefficient. %, although the term for the first difference is positive. 
The fact that the correlation law does not extend to the micro level inticates that they instead emerge from phenomena at some intermediate level. We conclude that it arises due to a property of the sectors (sector-specific rates of technological change, infrastructure, etc.), and is not a direct micro-level correspondence, a finding that corresponds to evolutionary economic theory \citep{Nelson/Winter75,Montobbio02}.

Third, we find that many variables at the firm level - including value added, employment, and labor productivity - follow heavy-tailed distributions, often two-sided L\'{e}vy alpha-stable distributions (cf. \citep{Yangetal19}), but occasionally other functional forms. This must be taken into account in econometric studies and generative models alike, if these variables are utilized. We concluded that robust regressions with Cauchy-distributed errors had to be run in order for our econometric models to yield valid results.

Fourth, we find that in growth processes in industry sectors, output growth tends to lead employment growth - at the sectoral level anyway.

Beyond these findings, we were able to confirm a range of known regularities at the sectoral level including Fabrciant's laws \citep{Fabricant42,Scott91} and Metcalfe et al.'s \citep{Metcalfeetal06} findings regarding autocorrelations of sectoral shares and dispersion measures (normalized Hirschmann-Herfindahl indices and entropy).

Our findings underline that modern economies are not as modular as it may seem at first glance. It is not possible to develop some particularly productive firms in isolation or even one sector without allowing a supporting infrastructure and firm ecosystem to develop. Nor is it advisable to think of firms as homogeneous entities such that a growing spread of productivities would be symptomatic for misallocation.

That said, it was in the Chinese case beneficial to focus modernization efforts on industry sectors in particular and to enable the reallocation of economic resources both by reforming SOEs (with unavoidable layoffs at a significant scale) and by permitting private entrepreneurship and foreign investment and strengthening the legal basis for this \citep{brandt2012creative}. The fact that sectoral output growth (and sectoral characteristics in the model in Eq. \ref{eq:lm53}) were found to be strongly associated to firm level growth might suggest that support for certain strategic sectors is crucial. This would be in line with Hidalgo et al.'s product space and economic complexity analysis \citep{Hidalgoetal07,Hidalgo/Hausmann09}. Other institutional factors as well as changes to patent law and incentives for innovation and R\&D were certainly also conductive of China's development \citep{Hu/Jefferson08}.

Whether these findings can easily be applied to other developing countries is doubtful because of the unique nature of China's economic system as both a developing economy and a transformation economy in a geographic region that is characterized by relative stability compared to other parts of our world. For many countries, micro-level data are furthermore not available, which makes it difficult to assess, to what extent the micro-levels of these economies are similar to either the Chinese one or those of developed countries. It should also be noted that our analysis excluded the input-output network within and across sectors, which may lead to additional important insights. With more detailed data, which may allow to confirm that processes for some countries are indeed similar to what is observed here for the PR China, it may become possible to recommend support for specific sectors or specific intensities of resource reallocation (that is, supporting labor and capital mobility without upending the economic and social structure of the country). 

%\newpage

\section*{Acknowledgements}
The authors would like to thank the participants of the 2019 annual meeting of the German Economic Association's Standing Field Committee for Evolutionary Economics as well as an anonymous reviewer for many helpful comments. All remaining mistakes are the authors'.

%% BibTeX users please use one of
%%\bibliographystyle{spmpsci}      % mathematics and physical sciences
%%\bibliographystyle{spphys}       % APS-like style for physics
%%\bibliographystyle{apalike}
%\bibliography{FabricantsLaws}
%\bibliography{FabricantsLaws}   % name your BibTeX data base

%\iffalse

%\fi

\clearpage
\appendix

\section{Variable definitions}
\label{app:variables}

The variables used in this paper and the corresponding symbols are summarized in table \ref{tab:variables}.

\begin{table}
\centering
\scriptsize  
\caption{Variable definitions as used in the paper.}
\label{tab:variables}
\begin{tabular}{p{4.5cm} p{1.5cm} p{7.7cm}}
\hline\hline
\textbf{Variable}  & \textbf{Symbol} & \textbf{Explanation} \\\hline\hline
Employment         & $L$, $l_i$ & Number of employees in the economy or in firm or sector $i$ \\
Value added        & $Y$, $y_i$ & Output- and income-type variable, especially for entities with non-zero intermediate inputs (firms, sectors) \\
Capital            & $K$, $k_i$ & Capital of the economy or of firm or sector $i$. If not stated otherwise, this is measured as total assets. In some tables, fixed assets are included for comparison. \\
Labor productivity & $Q$, $q_i$ & Value added per employee, $Q=Y/L$; $q_i=y_i/l_i$ \\
Capital intensity  &            & $K/L$; $k_i/l_i$ \\
Investment rate    & $\dot{K_t}$, $\dot{k_{i,t}}$  & $\dot{K_t}= \frac{K_t-K_{t-1}}{K_{t-1}}$, $\dot{k_{i,t}}= \frac{k_{i,t}-k_{i, t-1}}{k_{i,t-1}}$  \\
Wage bill          & $W$, $w_i$ & Total wages paid in the economy or in firm or sector $i$ \\
Average wage       &            & $W/L$; $w_i/l_i$  \\
Wage share         &            & Share of income paid as wages $W/Y$; $w_i/y_i$  \\
Share of variable $x_i$ & $s_{x,i}$   & $s_{x,i} = x_i / \sum_j x_j$    \\
Change of variable $x$ & $dx_t$  & $dx_t=x_t-x_{t-1}$ \\
Growth rate of variable $x$ & $\dot{x_t}$ & $\dot{x_t}= \frac{x_t-x_{t-1}}{x_{t-1}}$ \\
Normalized Hirschman-Herfindahl index of variable $x$ & & $\frac{\sum_{i=0}^Nx_i^2 - 1/N}{1-1/N}$ \\
Entropy of variable $x$ & & $-\sum_{i=0}^Ns_{x,i}\log s_{x,i}$  \\
Firm age          & $a_i$       & Age of firm $i$ in years \\
\hline\hline
\end{tabular}
\end{table}

%\section{JH:Temp Reg Tables}
\section{OLS Regressions}
\label{app:regressions}

This appendix shows OLS (Ordinary Lears Square) regressions corresponding to the robust regressions shown in Section \ref{sect:regressions} in Table \ref{tab:regression:heavy}. Regressions with employment growth as dependent variable are shown in Table \ref{tab:regression:lm:E}, those with value added growth as dependent variable are given in \ref{tab:regression:lm:VA}. The last lines in both tables report the finite moment test results for the crucial second moments (that is, for instance the variance) of the regression residuals. Trapani's \citep{Trapani16} finite moment test as implemented in the R package \verb|finity| \citep{R::finity} is used. For OLS regressions to deliver consistent results, the residuals should be close to Gaussian and certainly not heavy tailed corresponding to a distribution with infinite variance.

% Table created by stargazer v.5.2.2 by Marek Hlavac, Harvard University. E-mail: hlavac at fas.harvard.edu
% Date and time: Tue, Apr 28, 2020 - 14:52:44
% Requires LaTeX packages: dcolumn 
\begin{table}[!htbp] \centering \scriptsize
  \caption{OLS regression results for models (1), (2), (3), (4) with firm-level employment growth $\dot{l_i}$ as dependent variable} 
  \label{tab:regression:lm:E} 
\begin{tabular}{@{\extracolsep{5pt}}lD{.}{.}{-3} D{.}{.}{-3} D{.}{.}{-3} D{.}{.}{-3} } 
\\[-1.8ex]\hline 
\hline \\[-1.8ex] 
 & \multicolumn{4}{c}{\textit{Dependent variable:}} \\ 
\cline{2-5} 
\\[-1.8ex] & \multicolumn{4}{c}{Employment Growth} \\ 
\\[-1.8ex] & \multicolumn{1}{c}{(1) (Eq. \ref{eq:lm50})} & \multicolumn{1}{c}{(2) (Eq. \ref{eq:lm51})} & \multicolumn{1}{c}{(3) (Eq. \ref{eq:lm52})} & \multicolumn{1}{c}{(4) (Eq. \ref{eq:lm53})}\\ 
\hline \\[-1.8ex] 
 Sector employment growth & 1.267^{***} & 2.997^{***} & 1.149^{***} & 1.931^{***} \\ 
  & (0.114) & (0.315) & (0.397) & (0.553) \\ 
  & & & & \\ 
 Labor productivity   &  & -0.135^{***} & -0.130^{***} & -0.129^{***} \\ 
  &  & (0.003) & (0.003) & (0.003) \\ 
  & & & & \\ 
 Labor productivity change &  & -0.352^{***} & -0.351^{***} & -0.351^{***} \\ 
  &  & (0.004) & (0.004) & (0.004) \\ 
  & & & & \\ 
 Firm age &  & -0.024^{***} & -0.023^{***} & -0.025^{***} \\ 
  &  & (0.002) & (0.002) & (0.002) \\ 
  & & & & \\ 
  & & & & \\ 
 Constant & -0.338^{***} & -0.277^{***} & -0.092 & 0.069 \\ 
  & (0.012) & (0.032) & (0.167) & (0.195) \\ 
  & & & & \\ 
\hline \\[-1.8ex] 
Observations & \multicolumn{1}{c}{1,928,610} & \multicolumn{1}{c}{943,916} & \multicolumn{1}{c}{943,916} & \multicolumn{1}{c}{943,916} \\ 
R$^{2}$ & \multicolumn{1}{c}{0.0001} & \multicolumn{1}{c}{0.021} & \multicolumn{1}{c}{0.029} & \multicolumn{1}{c}{0.029} \\ 
Adjusted R$^{2}$ & \multicolumn{1}{c}{0.0001} & \multicolumn{1}{c}{0.021} & \multicolumn{1}{c}{0.029} & \multicolumn{1}{c}{0.029} \\ 
Residual Std. Error & \multicolumn{1}{c}{14.857 (df = 1928608)} & \multicolumn{1}{c}{20.428 (df = 943911)} & \multicolumn{1}{c}{20.351 (df = 943863)} & \multicolumn{1}{c}{20.346 (df = 943845)} \\ 
F Statistic & \multicolumn{1}{c}{123.204$^{***}$ (df = 1; 1928608)} & \multicolumn{1}{c}{5,180.543$^{***}$ (df = 4; 943911)} & \multicolumn{1}{c}{540.397$^{***}$ (df = 52; 943863)} & \multicolumn{1}{c}{407.435$^{***}$ (df = 70; 943845)} \\ 
\hline
Finite 2\textsuperscript{nd} moment test &&&&\\
$\chi^2$ Statistic& \multicolumn{1}{c}{0.0} & \multicolumn{1}{c}{0.0} & \multicolumn{1}{c}{0.0} & \multicolumn{1}{c}{0.0} \\ 
p-value & \multicolumn{1}{c}{1.0} & \multicolumn{1}{c}{1.0} & \multicolumn{1}{c}{1.0} & \multicolumn{1}{c}{1.0} \\ 
Finiteness & \multicolumn{1}{c}{infinite} & \multicolumn{1}{c}{infinite} & \multicolumn{1}{c}{infinite} & \multicolumn{1}{c}{infinite} \\ 
\hline 
\hline \\[-1.8ex] 
\textit{Note:}  & \multicolumn{4}{r}{$^{*}$p$<$0.1; $^{**}$p$<$0.05; $^{***}$p$<$0.01} \\ 
\end{tabular} 
\end{table}

% Table created by stargazer v.5.2.2 by Marek Hlavac, Harvard University. E-mail: hlavac at fas.harvard.edu
% Date and time: Tue, Apr 28, 2020 - 14:51:54
% Requires LaTeX packages: dcolumn 

% Table created by stargazer v.5.2.2 by Marek Hlavac, Harvard University. E-mail: hlavac at fas.harvard.edu
% Date and time: Tue, Apr 28, 2020 - 14:51:54
% Requires LaTeX packages: dcolumn 
\begin{table}[!htbp] 
\centering \scriptsize
  \caption{OLS regression results for models (1), (2), (3), (4) with firm-level value added growth $\dot{l_i}$ as dependent variable} 
  \label{tab:regression:lm:VA} 
\begin{tabular}{@{\extracolsep{5pt}}lD{.}{.}{-3} D{.}{.}{-3} D{.}{.}{-3} D{.}{.}{-3} } 
\\[-1.8ex]\hline 
\hline \\[-1.8ex] 
 & \multicolumn{4}{c}{\textit{Dependent variable:}} \\ 
\cline{2-5} 
\\[-1.8ex] & \multicolumn{4}{c}{Value Added Growth} \\ 
\\[-1.8ex] & \multicolumn{1}{c}{(1) (Eq. \ref{eq:lm50})} & \multicolumn{1}{c}{(2) (Eq. \ref{eq:lm51})} & \multicolumn{1}{c}{(3) (Eq. \ref{eq:lm52})} & \multicolumn{1}{c}{(4) (Eq. \ref{eq:lm53})}\\ 
\hline \\[-1.8ex] 
 Sector VA growth & 0.302 & 0.099 & 1.016 & 0.065 \\ 
  & (0.935) & (0.953) & (1.109) & (1.236) \\ 
  & & & & \\ 
 Labor productivity &  &  -0.844^{***} &  -0.845^{***} &  -0.844^{***} \\ 
  &  & (0.014) & (0.014) & (0.014) \\ 
  & & & & \\ 
 Labor productivity change &  & 1.988^{***} & 1.991^{***} & 1.990^{***} \\ 
  &  & (0.018) & (0.018) & (0.018) \\ 
  & & & & \\ 
 Firm age &  & -0.026^{***} & -0.035^{***} & -0.037^{***} \\ 
  &  & (0.008) & (0.010) & (0.010) \\ 
  & & & & \\ 
  Constant & -0.153 & 0.366^{*} & 0.812 & 1.250 \\ 
  & (0.182) & (0.215) & (0.785) & (0.911) \\ 
  & & & & \\ 
\hline \\[-1.8ex] 
Observations & \multicolumn{1}{c}{965,698} & \multicolumn{1}{c}{943,439} & \multicolumn{1}{c}{943,439} & \multicolumn{1}{c}{943,439} \\ 
R$^{2}$ & \multicolumn{1}{c}{0.00000} & \multicolumn{1}{c}{0.013} & \multicolumn{1}{c}{0.014} & \multicolumn{1}{c}{0.014} \\ 
Adjusted R$^{2}$ & \multicolumn{1}{c}{-0.00000} & \multicolumn{1}{c}{0.013} & \multicolumn{1}{c}{0.013} & \multicolumn{1}{c}{0.014} \\ 
Residual Std. Error & \multicolumn{1}{c}{94.874 (df = 965696)} & \multicolumn{1}{c}{95.097 (df = 943434)} & \multicolumn{1}{c}{95.089 (df = 943386)} & \multicolumn{1}{c}{95.088 (df = 943368)} \\ 
F Statistic & \multicolumn{1}{c}{0.104 (df = 1; 965696)} & \multicolumn{1}{c}{3,183.861$^{***}$ (df = 4; 943434)} & \multicolumn{1}{c}{248.861$^{***}$ (df = 52; 943386)} & \multicolumn{1}{c}{185.457$^{***}$ (df = 70; 943368)} \\ 
\hline
Finite 2\textsuperscript{nd} moment test &&&&\\
$\chi^2$ Statistic& \multicolumn{1}{c}{NaN} & \multicolumn{1}{c}{NaN} & \multicolumn{1}{c}{0.00002} & \multicolumn{1}{c}{0.00002} \\ 
p-value & \multicolumn{1}{c}{1.0} & \multicolumn{1}{c}{1.0} & \multicolumn{1}{c}{0.997} & \multicolumn{1}{c}{0.997} \\ 
Finiteness & \multicolumn{1}{c}{infinite} & \multicolumn{1}{c}{infinite} & \multicolumn{1}{c}{infinite} & \multicolumn{1}{c}{infinite} \\ 
\hline 
\hline \\[-1.8ex] 
\textit{Note:}  & \multicolumn{4}{r}{$^{*}$p$<$0.1; $^{**}$p$<$0.05; $^{***}$p$<$0.01} \\ 
\end{tabular} 
\end{table} 

\section{Sector classifications}
\label{app:sectorcodes}

Table \ref{tab:sectorcodes} gives the correspondence of sector codes in International Standard Industrial Classification Revision 4 (ISIC Rev.4) and the Gu\'{o}Bi\={a}o (\begin{CJK*}{UTF8}{bsmi}國標\end{CJK*}) 1994/2002 classification system. ISIC Rev.4 is used in the present paper and in the WIOD database, GB1994/2002 is used in the CIE DB.

\begin{table}[p]
\centering
\scriptsize  
\caption{Sector correspondence between ISIC (International Standard Industrial Classification) Rev. 4 and GB1994, GB2002 (guobiao versions 1994 and 2002) with explanations.}
\label{tab:sectorcodes}
\begin{tabular}{p{3.5cm} p{3.5cm} p{5.5cm}}
\textbf{ISIC Rev.4 (WIOD)} & \textbf{GB2002 (CIE DB)} & \textbf{Sector} \\\hline\hline
   A01 & not present  & Farming, hunting\\
   A02 & not present  & Forestry\\
   A03 & not present  & Fishing\\
   \hline
   \multirow{7}{*}{B}   & 06 & Mining (coal)\\
      & 07 & Mining (petrol, gas) \\
      & 08 & Mining (ferrous metals)\\
      & 09 & Mining (non-ferrous metals)\\
      & 10 & Mining (non-metal ores)\\
      & 11 & Mining (other)\\
      & 12 & Logging\\
   \hline
   \multirow{4}{*}{C10-C12} & 13 & Processing of agricultural food\\
    & 14 & Manufacturing of food\\
    & 15 & Manuf. of beverages\\
    & 16 & Manuf. of tobacco\\
   \hline
   \multirow{3}{*}{C13-C15} & 17 & Manuf. of textiles\\
    & 18 & Manuf. of apparel, footware\\
    & 19 & Manuf. of leather, fur products\\
   \hline
   C16 & 20 & Manuf. of wood, bamboo products\\
   C17 & 22 & Manuf. of paper\\
   C18 & 23 & Manuf. of recording media, printing\\
   ?   & 24 & Manuf. of culture, sports products\\
   C19 & 25 & Manuf. of fuel (petrol, coal, nuclear)\\
   C20 & 26 & Manuf. of chemicals\\
   C20 & 28 & Manuf. of chemical fibers\\
   C21 & 27 & Manuf. of medicine\\
   C22 & 29 & Manuf. of rubber\\
   C22 & 30 & Manuf. of plastics\\
   C23 & 31 & Processing of non-metal minerals\\
   \hline
   \multirow{2}{*}{C24} & 32 & Smelting, processing (ferrous metals)\\
       & 33 & Smelting, processing (non-ferrous metals)\\
   \hline
   C25 & 34 & Manuf. of metal products\\
   C26 & (40) inconsistent* & Manuf. of IT, optical, electronics\\
   C27 & (39) inconsistent* & Manuf. of electrical equipment\\
   C28 & 35 & Manuf. of general purpose machinery\\
   C29 & 36 & Manuf. of special purpose machinery\\
   C30 & 37 & Manuf. of transport equipment\\
   C31\_C32 & 21 & Manuf. of furniture\\
   C32 & ? & Other manufacturing \\
   C33 & ? & Repair and installation \\
   \hline
   \multirow{2}{*}{D35} & 44 & Production, distribution of electricity\\
       & 45 & Production, distribution of gas\\
   \hline
   E36 & 46 & Production, distribution of water\\
   E37-E39 & not present & Sewage, waste \\
   F &   not present & Construction\\
   G45 & not present & Trade of motor vehicles\\
   G46 & not present & Wholesale trade\\
   G47 & not present & Retail trade\\
   H49 & not present & Land transport\\
   H50 & not present & Water transport\\
   H51 & not present & Air transport\\
   H52 & not present & Warehousing\\
   H53 & not present & Postal and courier services\\
   I & not present & Accommodation, food service\\
   J58 & not present & Publishing\\
   J59\_J60 & not present & Film, Broadcasting\\
   J61 & not present & Telecommunications\\
   J62\_J63 & not present & IT, programming, information\\
   K64 & not present & Financial service\\
   K65 & not present & Insurance\\
   K66 & not present & Auxiliary financial service\\
   L68 & not present & Real estate\\
   M69\_M70 & not present & Legal, accounting, consulting\\
   M71 & not present & Architecture, engineering\\
   M72 & not present & Scientific R\&D\\
   M73 & not present & Advertising, market research\\
   M74\_M75 & not present & Other professional, scientific\\
   N & not present & Administrative\\
   O84 & not present & Public administration, defence\\
   P85 & not present & Education \\
   Q & not present & Health\\
   R\_S & not present & Arts, entertainment, other services\\
   T & not present & Households as employers\\
   U & not present & Extraterritorial organizations\\\hline
\end{tabular}\\
GB1994 and GB2002 are largely consistent for the sectors covered by the CIE data base and sector labels are also consistent over time with the exception of sectors 38-43. The ISIC and guobiao classification systems do not align 100\%, question marks indicate fields where the correspondence is unclear (note that these are minor sectors).
\end{table}

\section{Extended correlation laws in macro-level data}

\begin{figure}[tb!]
\centering
\includegraphics[width=1.\textwidth]{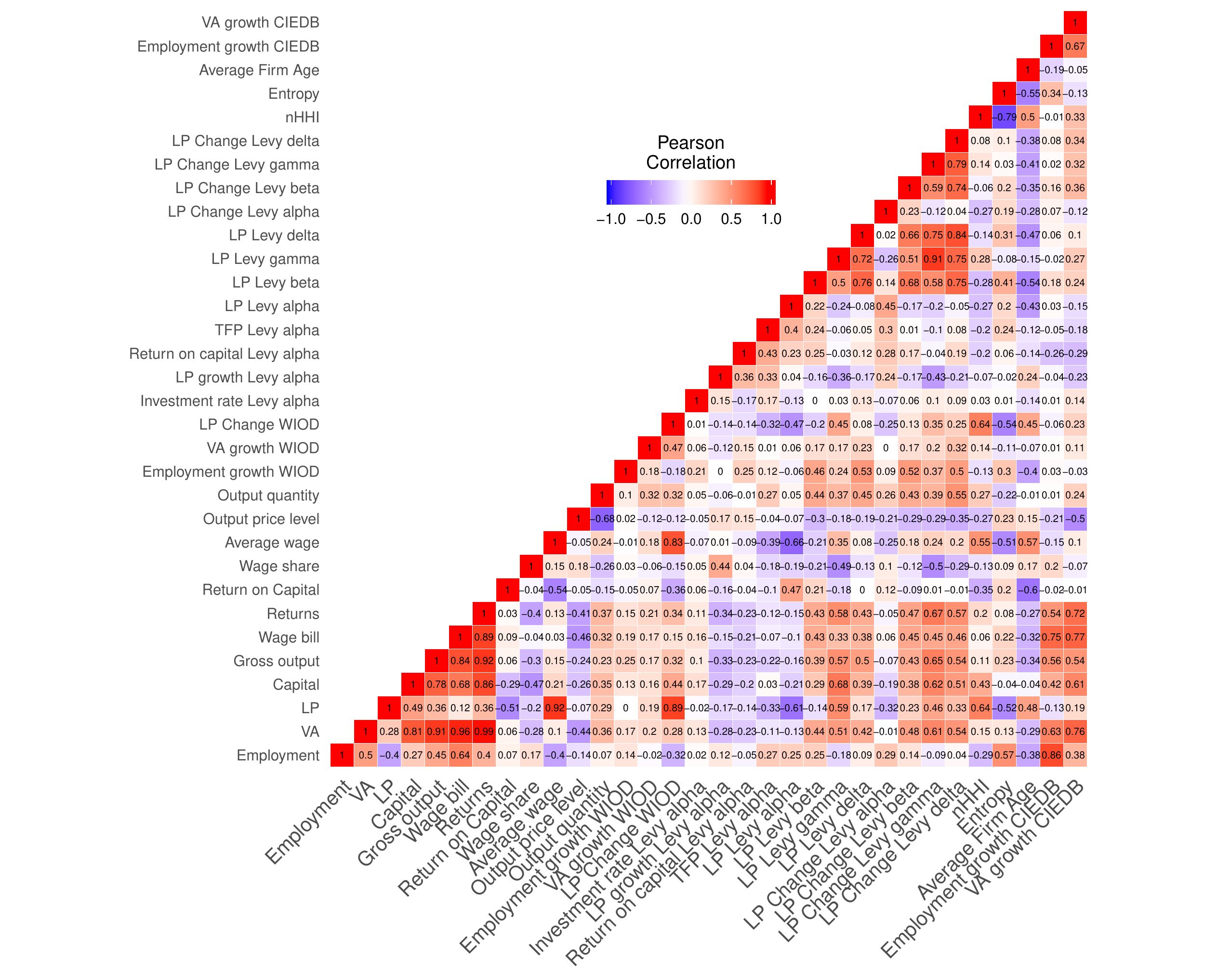}
\caption{Correlations in macro-level data (levels)}
\label{fig:heatmap:macro:level:extensive}
\end{figure}

\begin{figure}[tb!]
\centering
\includegraphics[width=1.\textwidth]{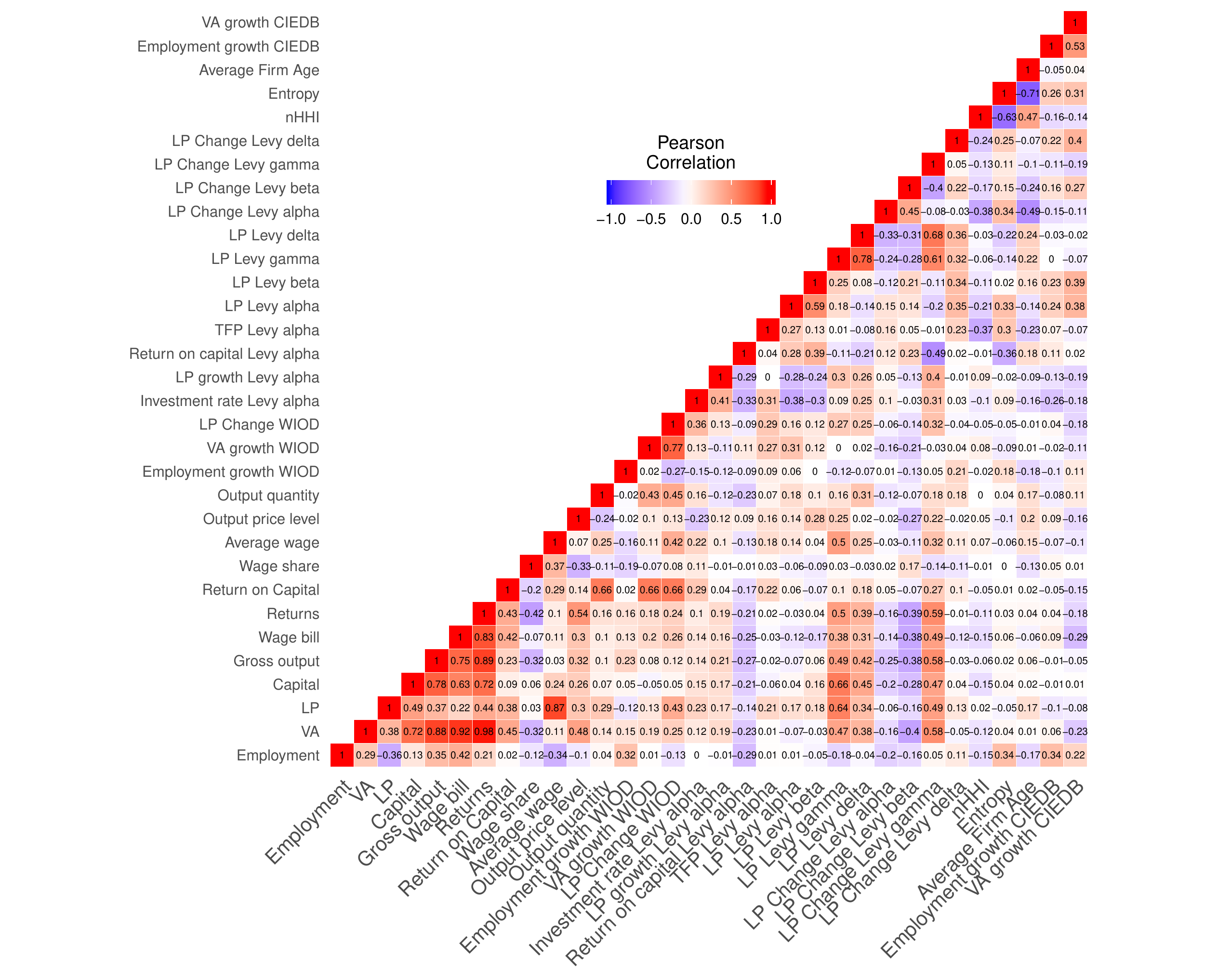}
\caption{Correlations in macro-level data (first differences)}
\label{fig:heatmap:macro:fd:extensive}
\end{figure}

Figures \ref{fig:heatmap:macro:level:extensive} and \ref{fig:heatmap:macro:fd:extensive} show a more more extensive set of pairwise correlations in sectoral data. Findings are discussed in Section \ref{sect:results:fabricant} in the main text.


\begin{figure}[tb!]
\centering
\includegraphics[width=0.85\textwidth]{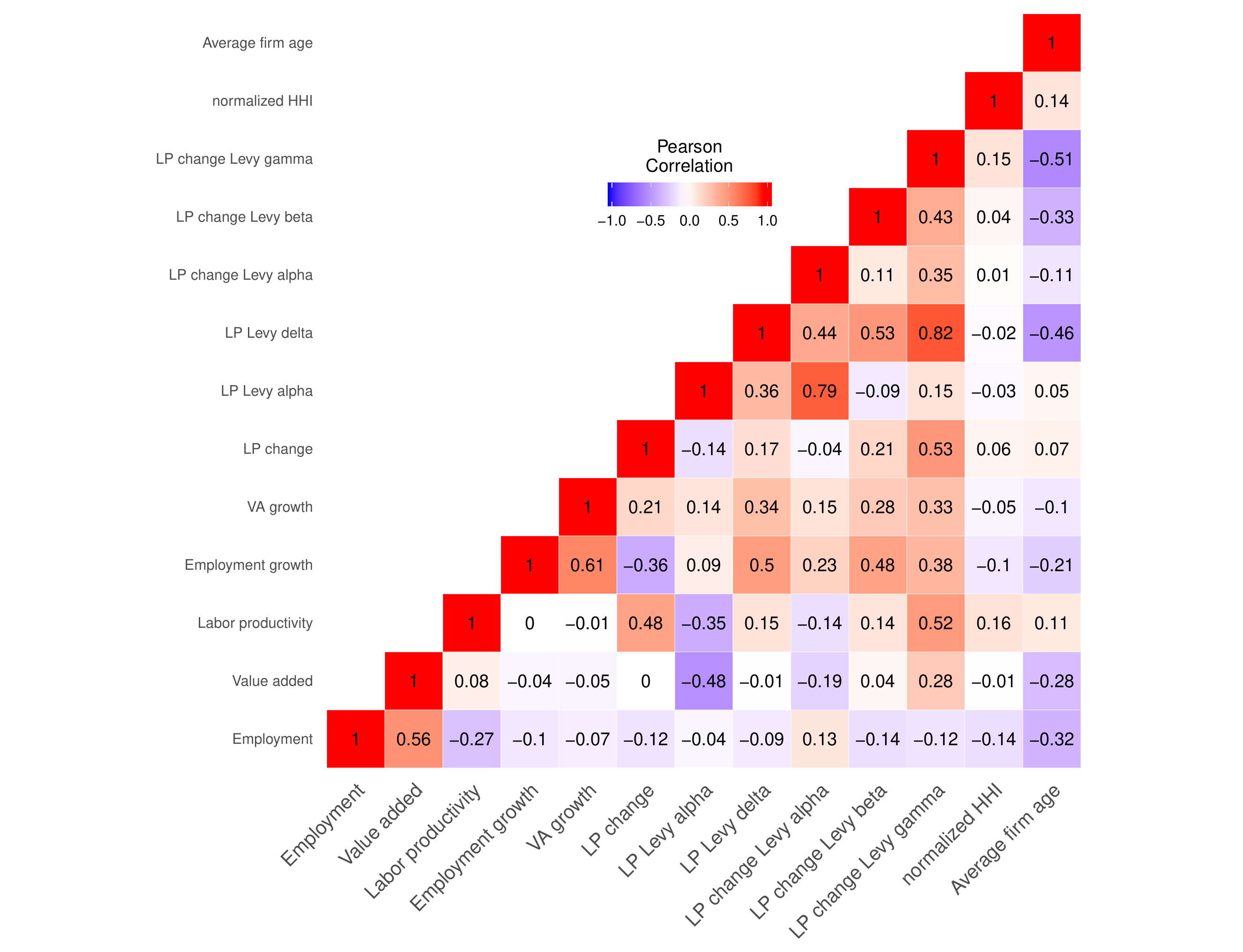}
\caption{Correlations in macro-level data and sectoral characteristics derived from firm-level data (L\'{e}vy alpha stable fit parameters for labor productivity and labor productivity change as well as the normalized Hirschman-Herfindahl index and the average firm age) (levels)}
\label{fig:heatmap:macromicro:level}
\end{figure}

\begin{figure}[tb!]
\centering
\includegraphics[width=0.85\textwidth]{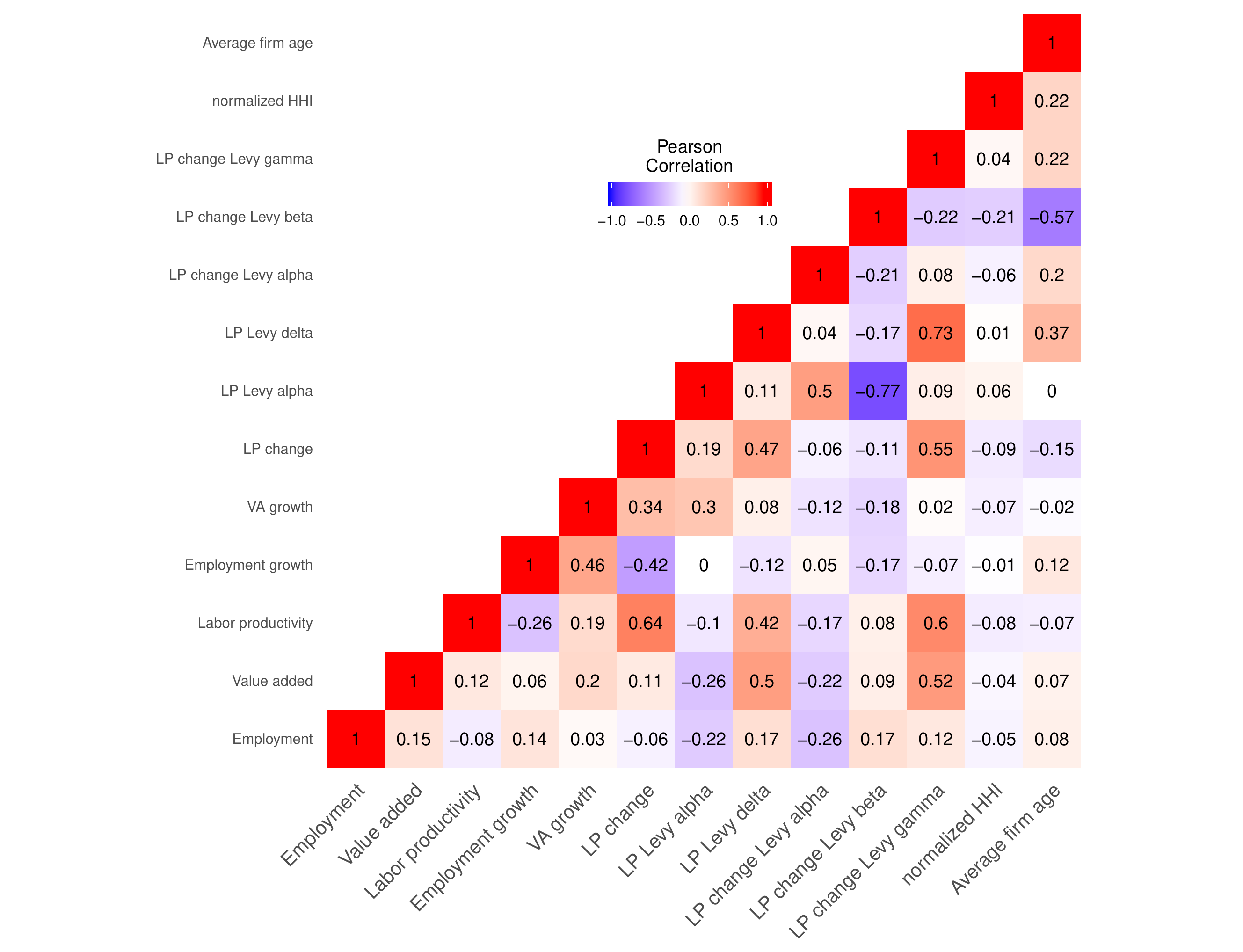}
\caption{Correlations in macro-level data and sectoral characteristics derived from firm-level data (L\'{e}vy alpha stable fit parameters for labor productivity and labor productivity change as well as the normalized Hirschman-Herfindahl index and the average firm age) (first differernces)}
\label{fig:heatmap:macromicro:fd}
\end{figure}

Further to the findings discussed in Section \ref{sect:results:fabricant} in the main text, we demonstrate in Figures \ref{fig:heatmap:macromicro:level} and \ref{fig:heatmap:macromicro:fd} that there is a strong correspondence between sector level characteristics and properties of the distributions of micro level data. To represent the distributions of the micro-level data, we choose the coefficients of L\'{e}vy alpha-stable distributional model fitted to distributions of labor productivity and labor productivity change. We select these two variables, because we have established that these quantities do very likely indeed follow L\'{e}vy alpha-stable distributions with respect to both the CIE DB data set \citep{Heinrichetal19} and a larger data set of European firms from Orbis Europe \citep{Yangetal19}. 

\begin{figure}[h!]
\centering
\subfloat[Dispersion of resources between sectors: Normalized Hirschmann-Herfindahl index of shares of capital, value added, and employment in sectoral data (WIOD)]{\label{fig:Dev:disp:macro}\includegraphics[width=0.5\textwidth]{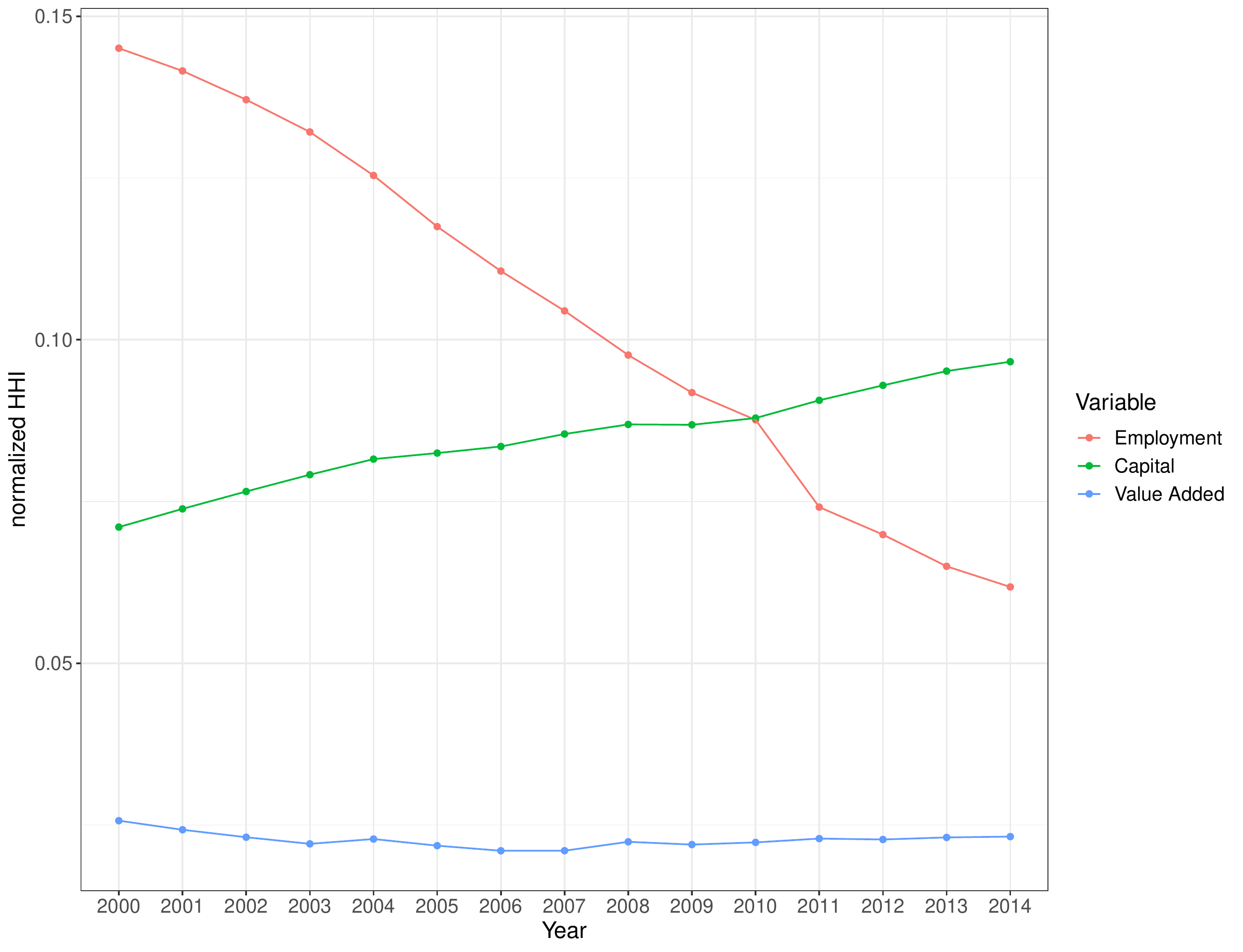}}
\hspace*{0.4cm}
\subfloat[Development of normalized Herfindahl-Hirschman-indices by sector in firm level data]{\label{fig:Dev:disp:sectoral}\includegraphics[width=0.5\textwidth]{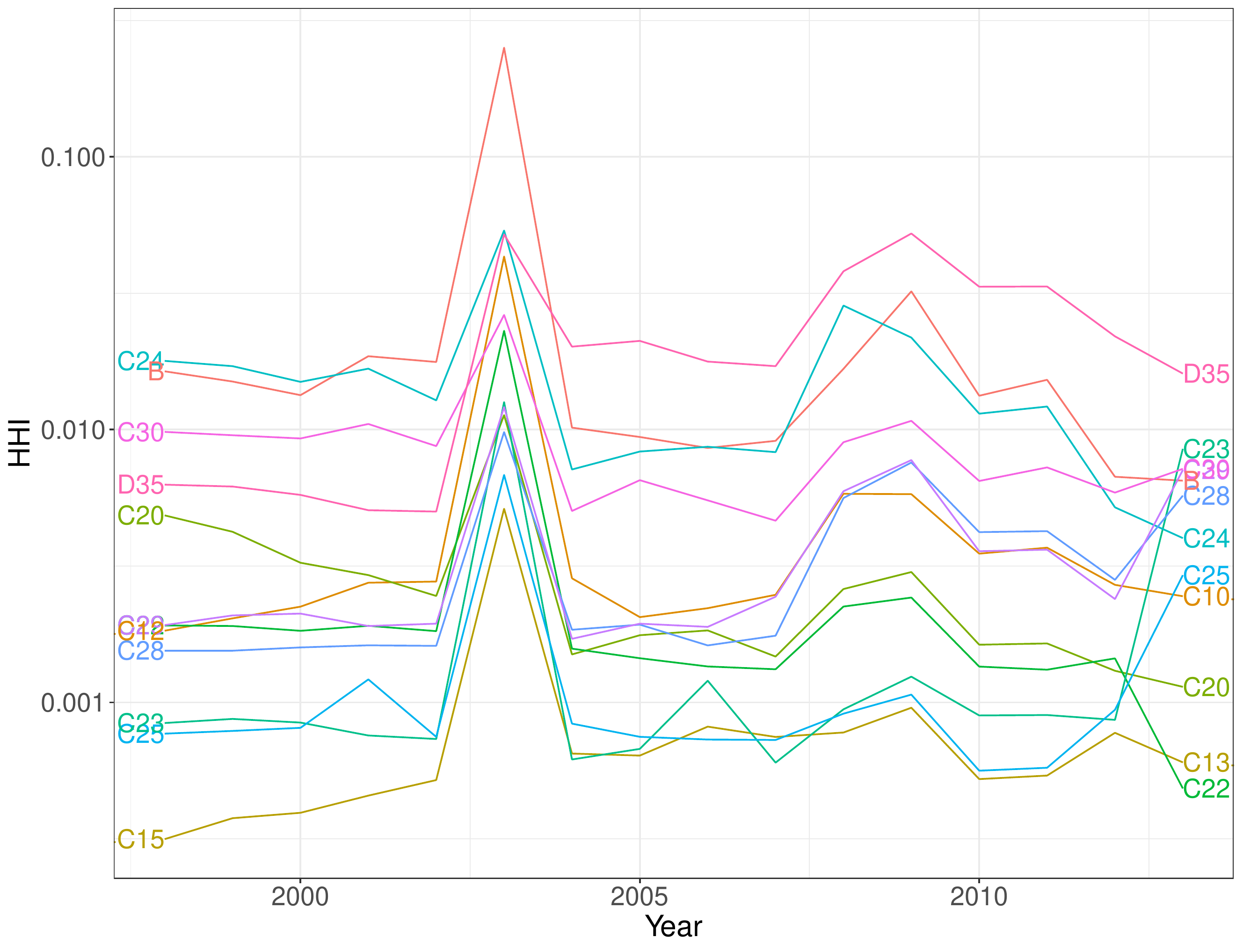}}
\caption{Development of the inter-sectoral and intra-sectoral dispersion of resources.}
\label{fig:Dev:disp}
\end{figure}

\iffalse
  \begin{figure}[h!]
  \centering
  \subfloat[Normalized Hirschmann-Herfindahl index]{\includegraphics[width=0.5\textwidth]{11_aggregated_HHI.pdf}}
  \subfloat[Entropy]{\includegraphics[width=0.5\textwidth]{11_aggregated_Entropy.pdf}}
  \caption{Development of inter-sectoral dispersion of shares of capital, value added, and employment}
  \label{fig:Dev:disp}
  \end{figure}
  
  \begin{figure}[tb!]
  \centering
  \includegraphics[width=0.85\textwidth]{09_ISICR4_Sectoral_Accounts__BGnHHI_TOAS.pdf}
  \caption{Development of normalized Herfindahl-Hirschman-indices by sector in firm level data}
  \label{fig:Dev:sectoral:nHHI}
  \end{figure}
\fi

\section{Dispersion of resources among and within sectors}
\label{sect:results:nHHI}

Finally, we consider the dispersion of resources across sectors. Again following \citet{Metcalfeetal06}, we use the normalized Hirschman-Herfindahl index (HHI) of capital, employment and value added,

We can use the HHI at the sectoral level in spite of it's property of tracking the second moment, because we are dealing with an aggregated level (with a biased aggregation) in which the moments may not be infinite any longer and because we are dealing with the full population of sectors in the economy. 

The figure (Fig. \ref{fig:Dev:disp:macro}) reveals that the dispersion in the three variables is not identical and develops in different directions. While the HHI of value added indicates a roughly constant dispersion, the distribution of the work force becomes more even (in parts because the primary sector is losing employment) and the distribution of capital becomes less even.

Regarding intra-sectoral dispersion of resources, Figure \ref{fig:Dev:disp:sectoral} seems to indicate that the sectors retain different but persistent levels of dispersion with energy, mining, and transport equipment manufacturing (D35, B, C30) at the least even side. However, considering the heavy-tailedness of variables, it is possible that these estimates are again biased by the sample sizes, and more so than the HHI is in itself. The entropy (not shown in the figure) shows a similar picture, however.

\clearpage
\section{Consistency checks: Sectoral accounts and firm-level data}

The following illustrations are to serve as consistency checks demonstrating that the firm-level data from CIE DB are while noisy, still coherent and sufficiently reliable to be used in analyses of structural change. We also compare firm-level data with aggregated sectoral data, show regional variability in firm level data, and add autocorrelation spectra of dispersion measures.

\subsection{Development of individual quantities (employment and value added) in sectoral accounts and chained firm level data}

\begin{figure}[h!]
\centering
\subfloat[Value added]{\includegraphics[width=0.5\textwidth]{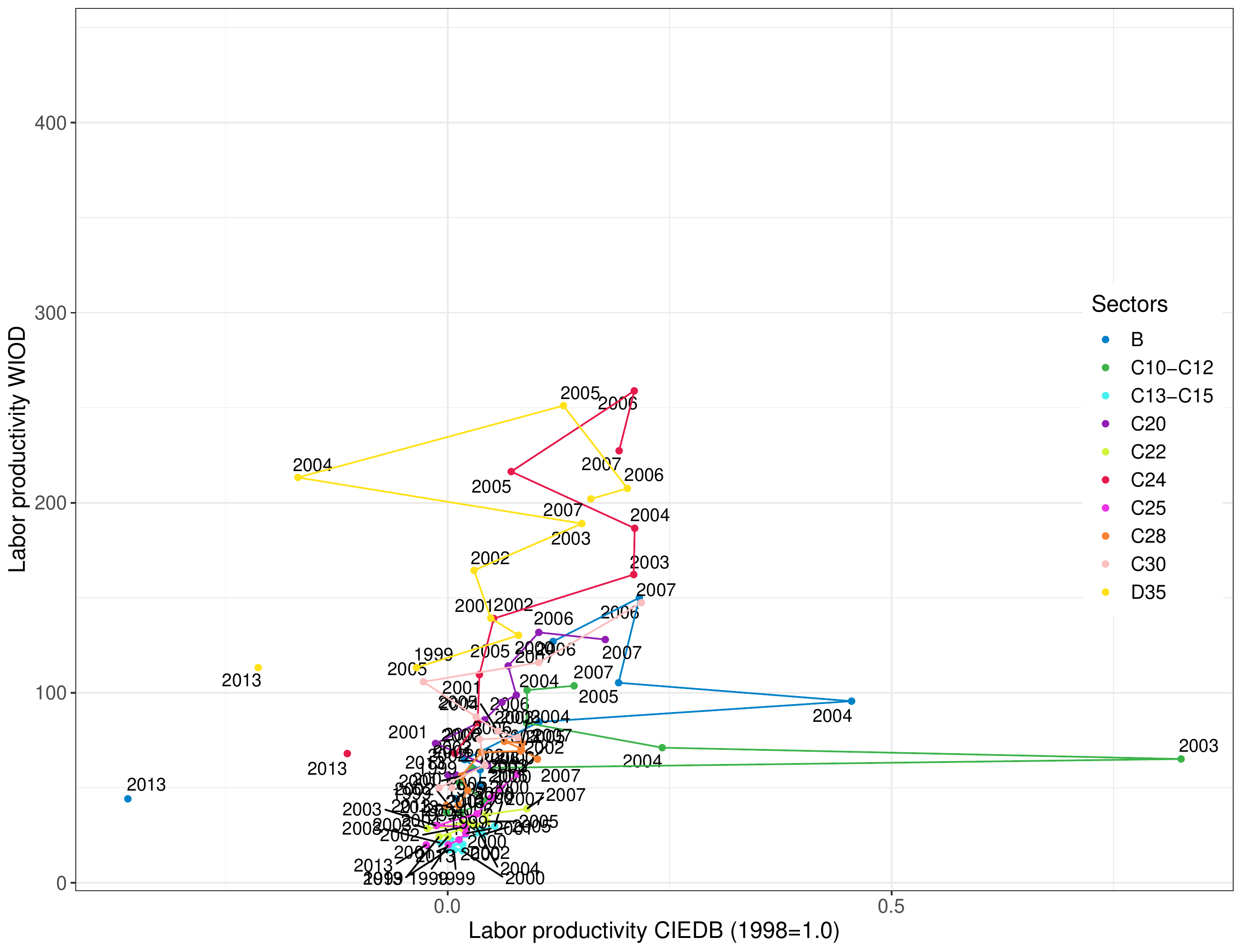}}
\subfloat[Employment]{\includegraphics[width=0.5\textwidth]{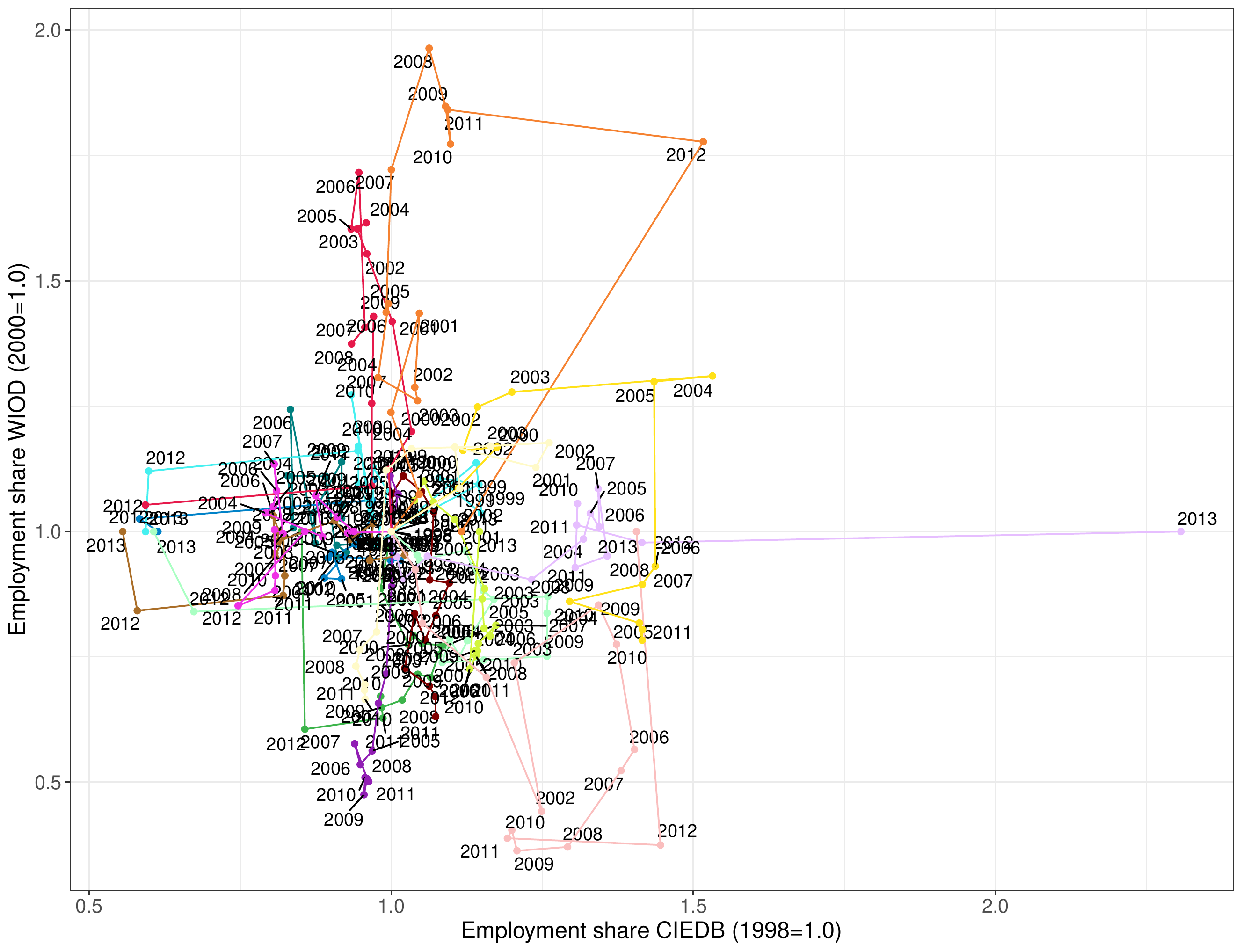}}
\caption{Development of value added and employment by sector in aggregated vs. firm-level chained panel data}
\label{fig:Dev:2d:consistencycheck}
\end{figure}

The left panel of Fig. \ref{fig:Dev:2d:consistencycheck} shows the development of value added shares by industry sector in aggregated sectoral data and firm-level data. The fact that these developments are by no means identical (they would then happen along the 45\textdegree line) shows how noisy the firm-level data is when compared to the aggregate. However, it can also be seen that for most sectors, the developments in firm-level and aggregate data move in the same direction. The right panel of the figure shows the same comparison for employment share growth.

\subsection{Sector shares in firm-level data}
\begin{figure}[tb!]
\centering
\includegraphics[width=0.85\textwidth]{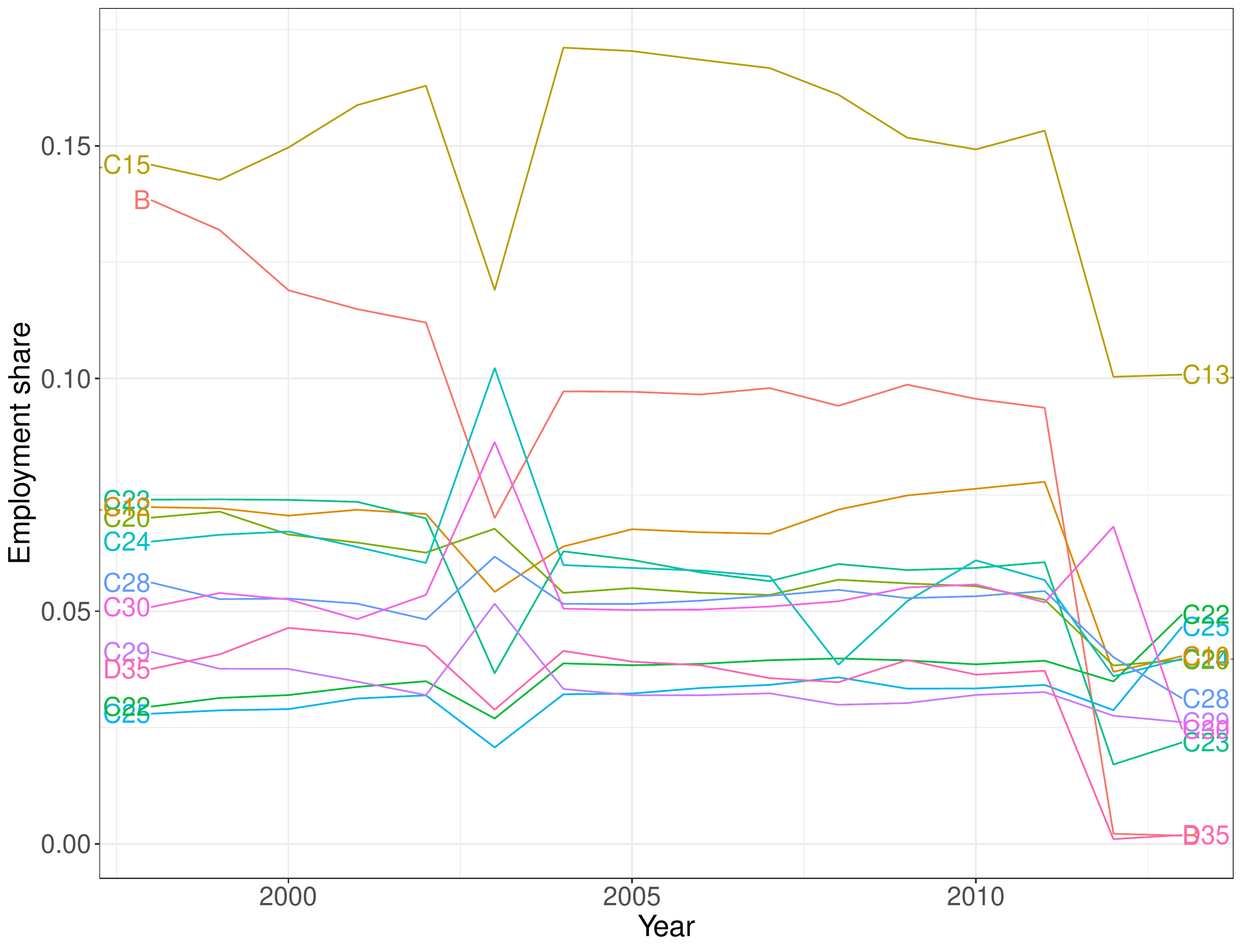}
\caption{Development of employment shares by sector}
\label{fig:Dev:sectoral:EMPL}
\end{figure}

\begin{figure}[tb!]
\centering
\includegraphics[width=0.85\textwidth]{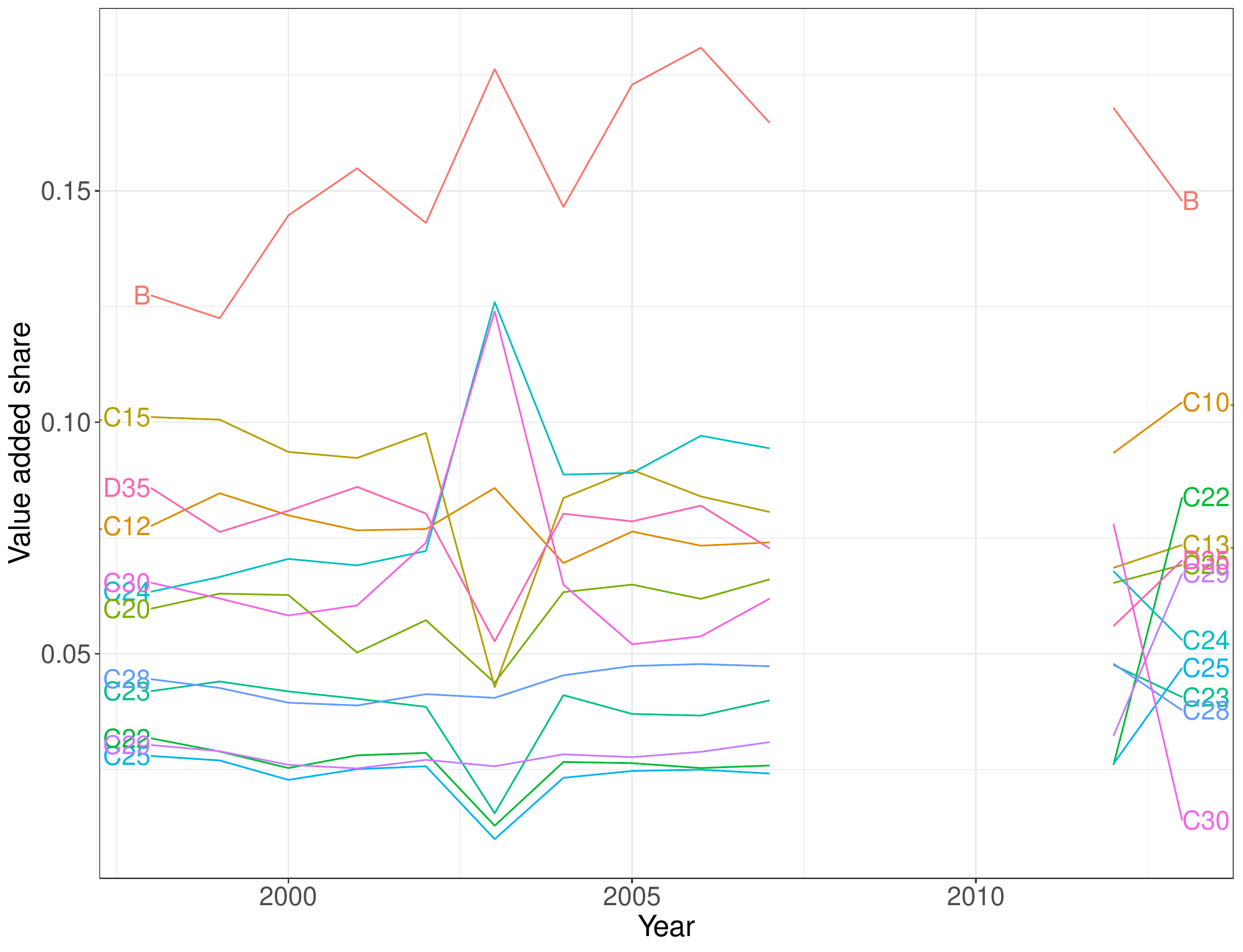}
\caption{Development of value added shares by sector}
\label{fig:Dev:sectoral:VA}
\end{figure}

Figure \ref{fig:Dev:sectoral:EMPL} shows the development of employment shares by industruy sector in firm-level data. It can be seen that albeit noisy, the panel data is consistent. Figure \ref{fig:Dev:sectoral:VA} shows the same for value added shares.

\iffalse
\subsection{Interdependent development}

Fig. \ref{fig:Dev:2d:consistencycheck:interdep}

\begin{figure}[h!]
\centering
\subfloat[Macro-level data (sectoral accounts)]{\includegraphics[width=0.5\textwidth]{Development_IND_MACRO_Employment_VA.pdf}}
\subfloat[Firm-level data (chained panel)]{\includegraphics[width=0.5\textwidth]{Development_ISICR4_BG_Employment_VA.pdf}}
\caption{Development of employment vs. value added by sector}
\label{fig:Dev:2d:consistencycheck:interdep}
\end{figure}
\fi

%\clearpage
\subsection{Regional dispersion}
\begin{figure}[tb!]
\centering
\includegraphics[width=0.85\textwidth]{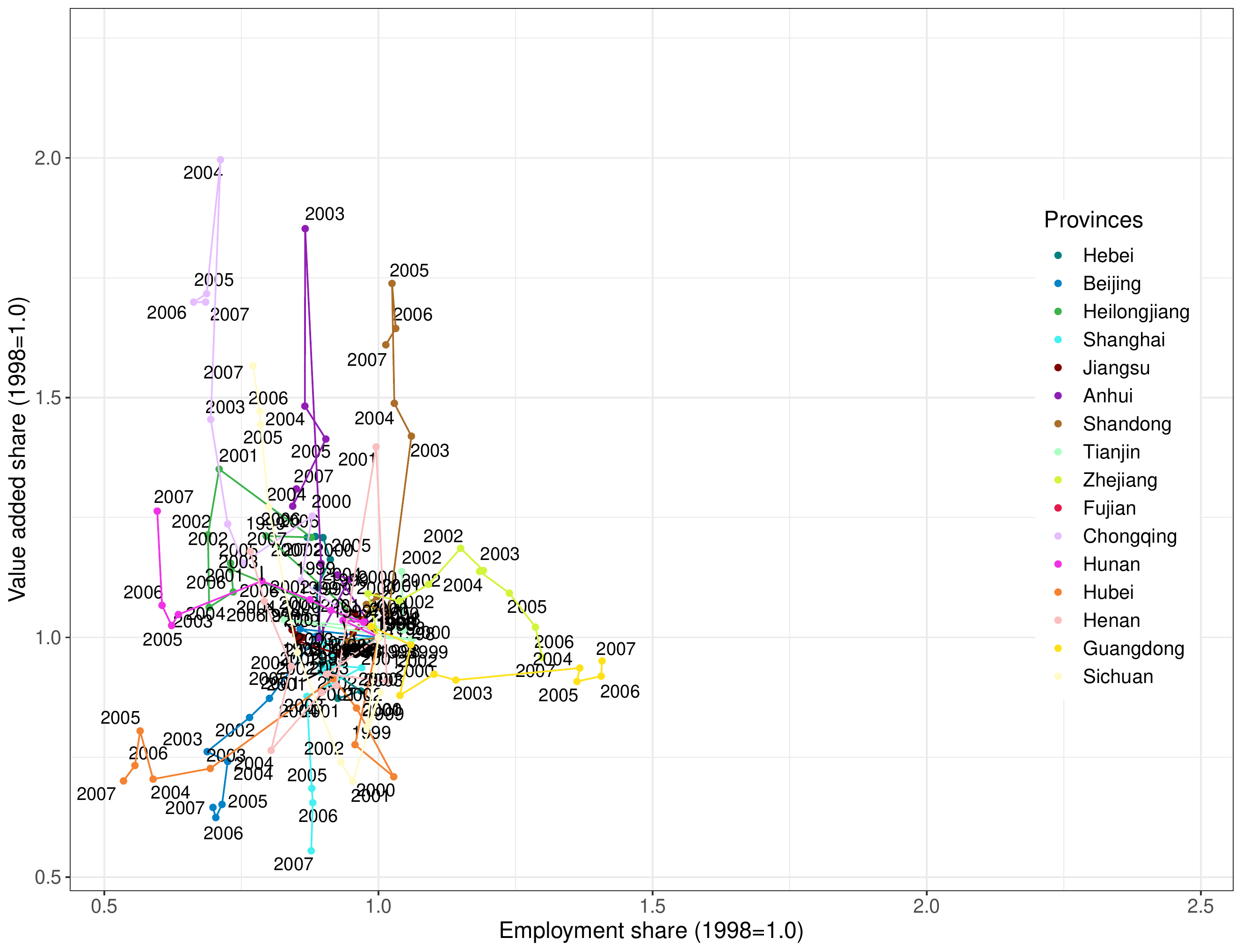}
\caption{Development of employment vs. value added by province (chained panels)}
\label{fig:Dev:2d:firmlevel:EMPL:VA}
\end{figure}

\begin{figure}[tb!]
\centering
\includegraphics[width=0.85\textwidth]{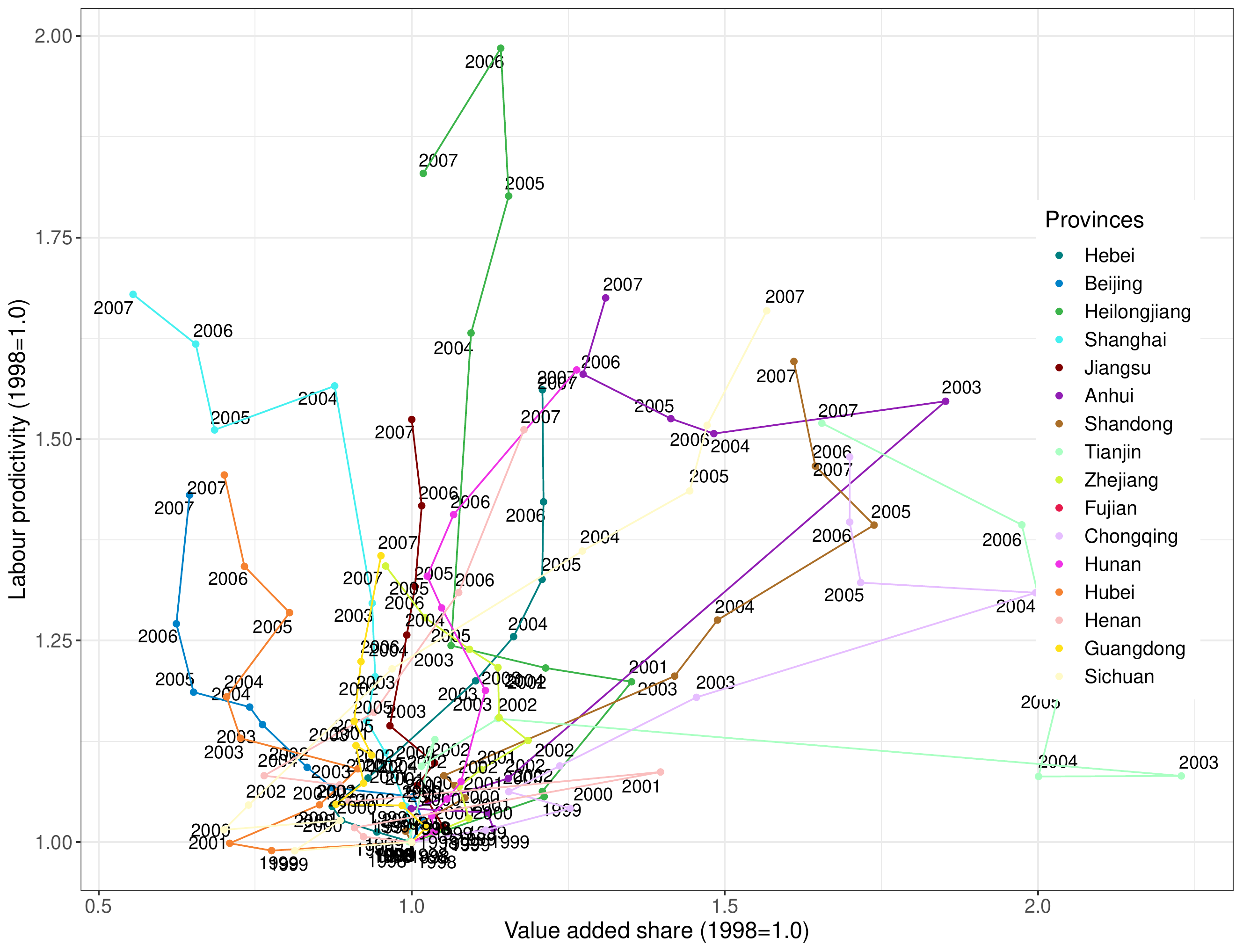}
\caption{Development of value added vs. labor productivity by province (chained panels)}
\label{fig:Dev:2d:firmlevel:VA:LP}
\end{figure}

\begin{figure}[tb!]
\centering
\includegraphics[width=0.85\textwidth]{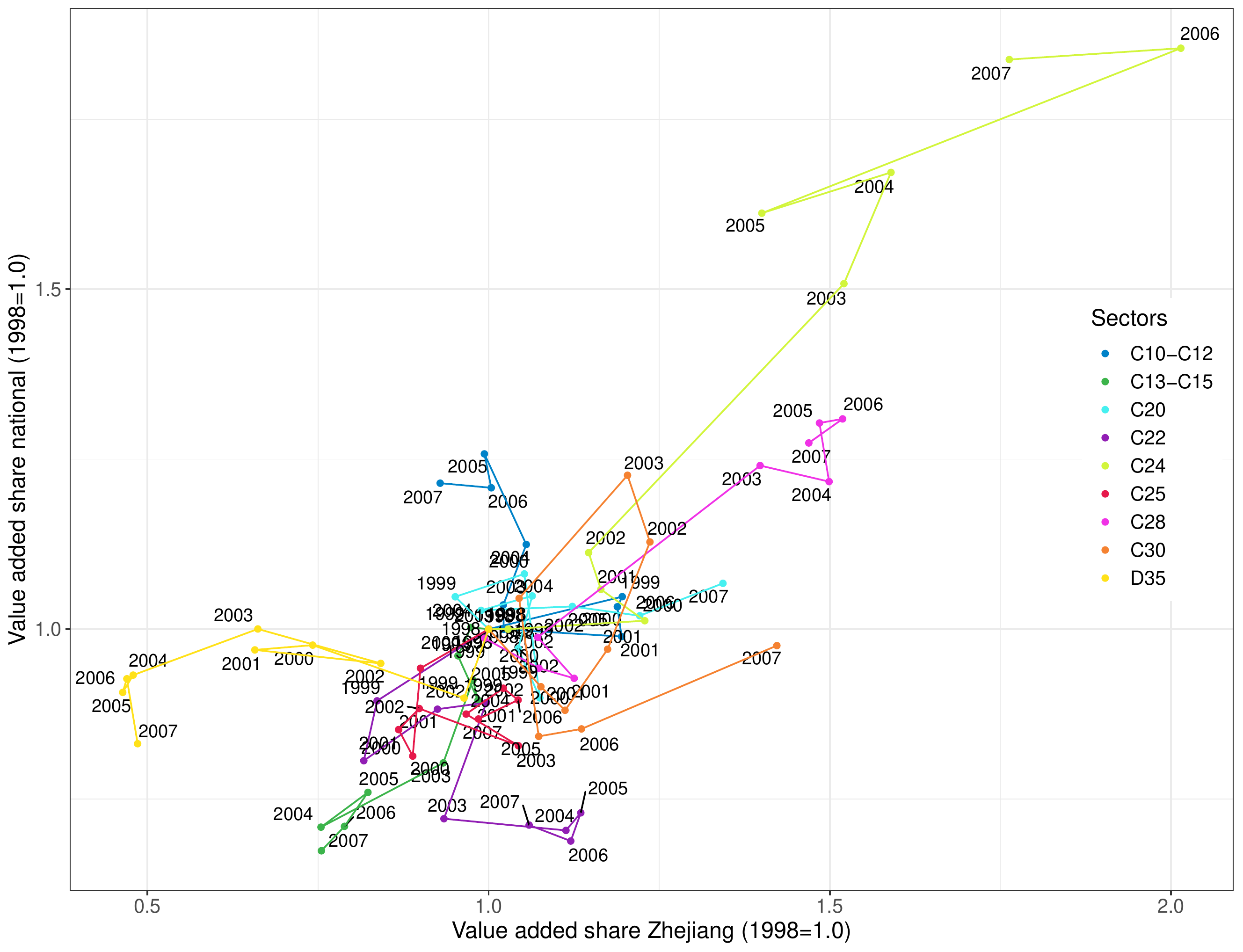}
\caption{Development of value added nationally vs. province Zhejiang (chained panels)}
\label{fig:Dev:2d:Zhejiang:VA}
\end{figure}

\begin{figure}[tb!]
\centering
\includegraphics[width=0.85\textwidth]{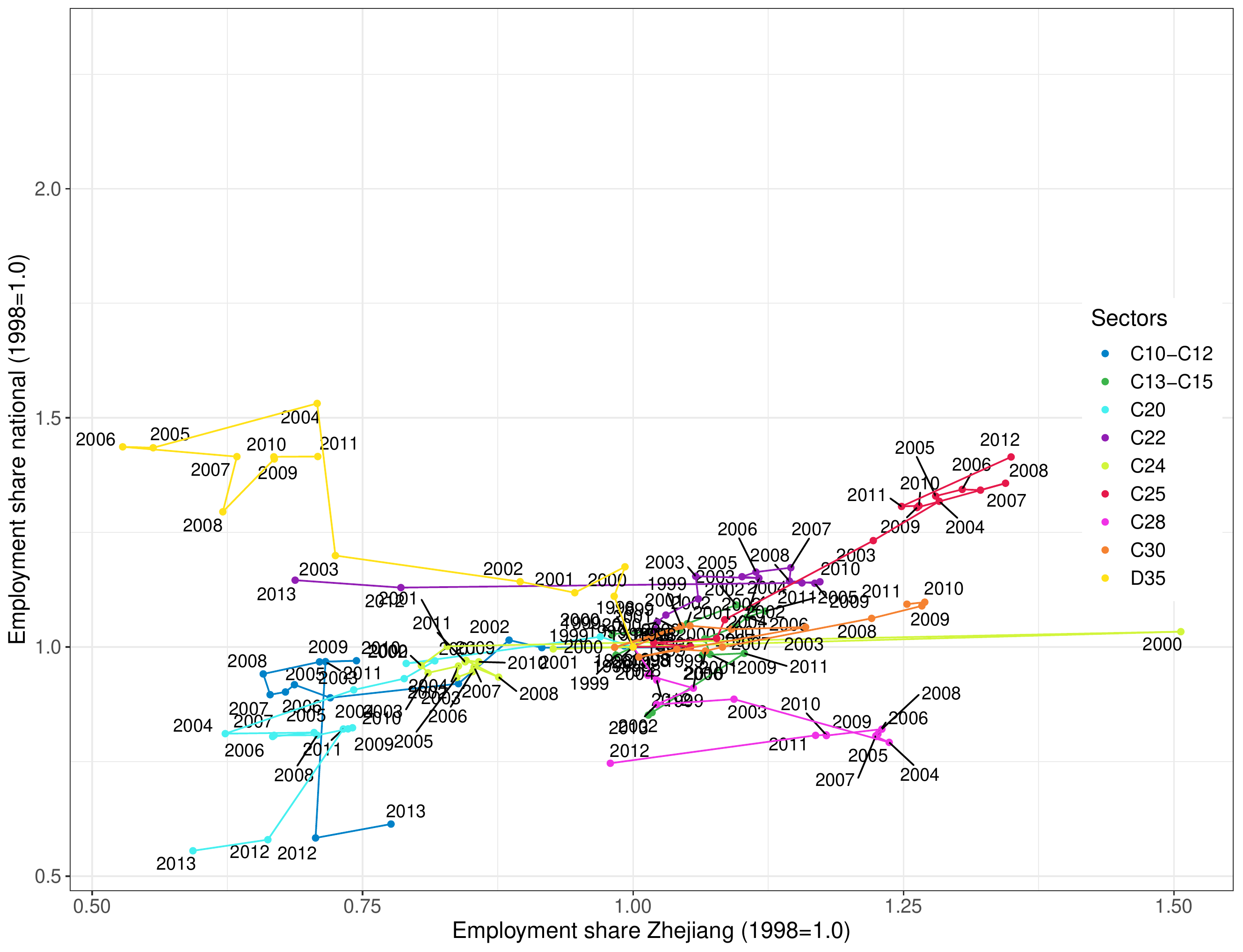}
\caption{Development of employment nationally vs. province Zhejiang (chained panels)}
\label{fig:Dev:2d:Zhejiang:EMPL}
\end{figure}

Fig. \ref{fig:Dev:2d:firmlevel:EMPL:VA}, \ref{fig:Dev:2d:firmlevel:VA:LP} show the bivariate development of employment, value added, and labor productivity by province in firm-level data. Fig. \ref{fig:Dev:2d:Zhejiang:VA},  \ref{fig:Dev:2d:Zhejiang:EMPL} compare the development of employment and of value added in only one province, Zhejiang, with that at the national level (firm-level data).

\subsection{Autocorrelations of competitiveness}

\begin{figure}[h!]
\centering
\subfloat[Normalized Hirschmann-Herfindahl index]{\includegraphics[width=0.5\textwidth]{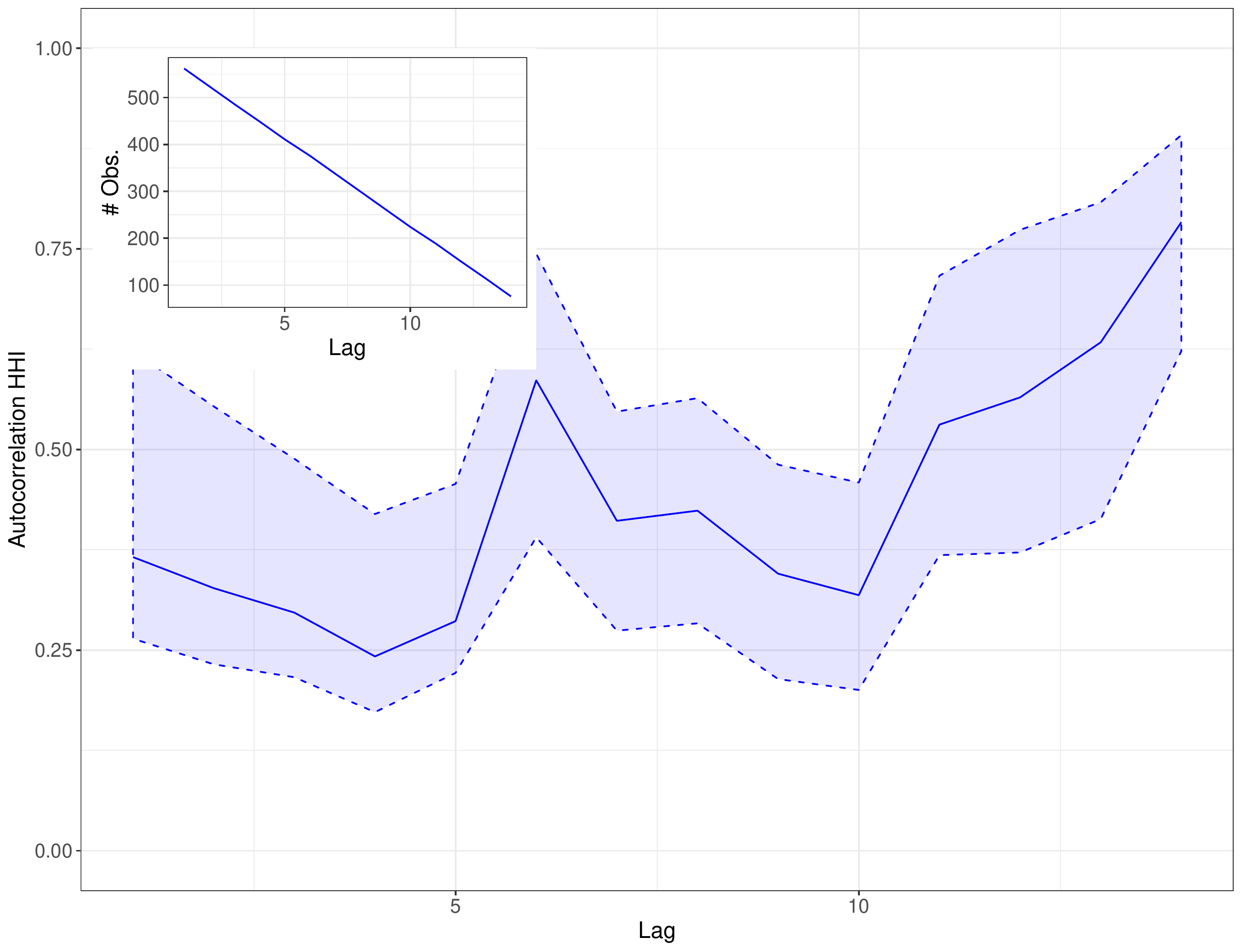}}
\subfloat[Entropy]{\includegraphics[width=0.5\textwidth]{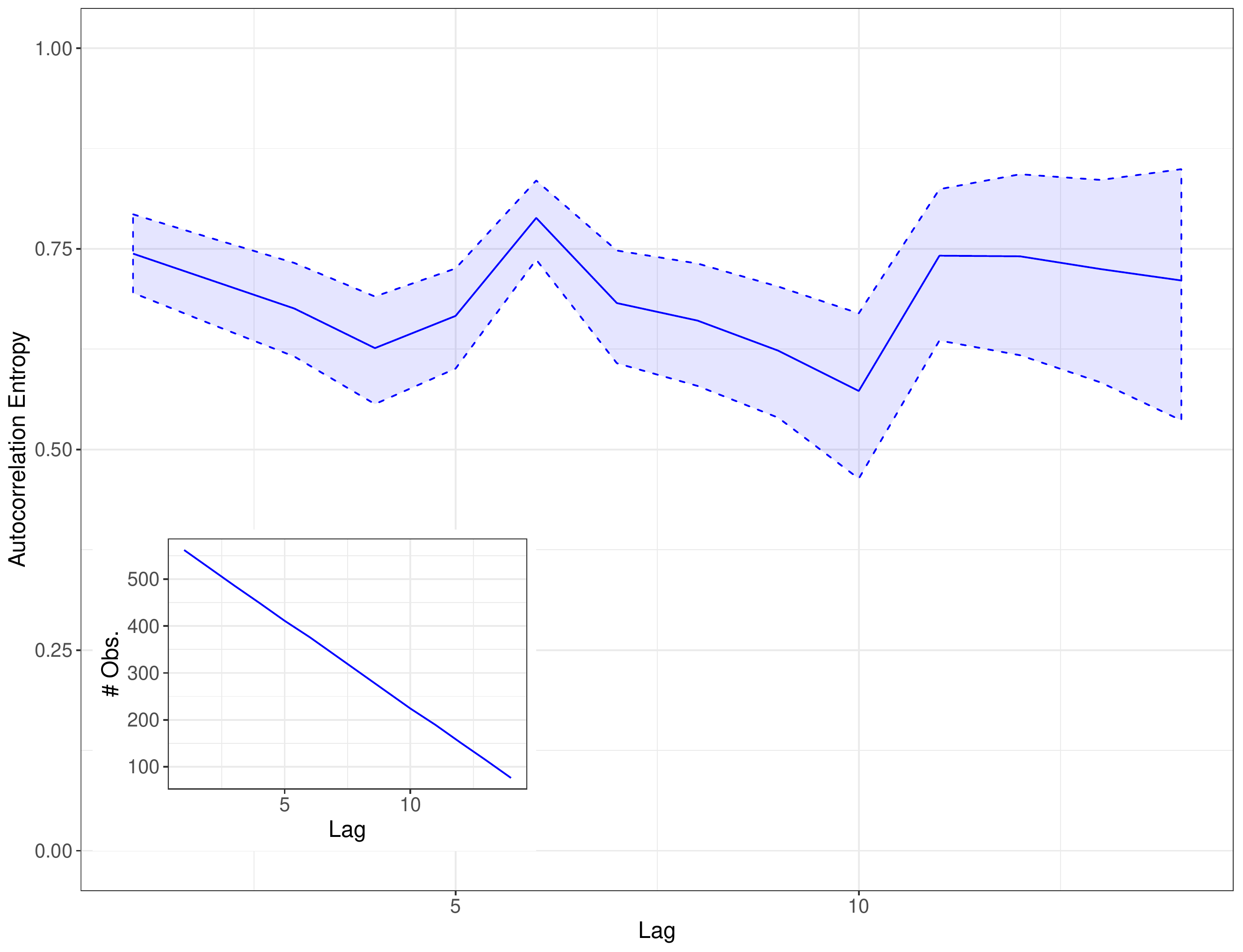}}
\caption{Autocorrelations of intra-sectoral dispersion measures of shares of total assets}
\label{fig:AC:disp}
\end{figure}

\begin{figure}[h!]
\centering
\subfloat[$\Delta$ Normalized Hirschmann-Herfindahl index]{\includegraphics[width=0.5\textwidth]{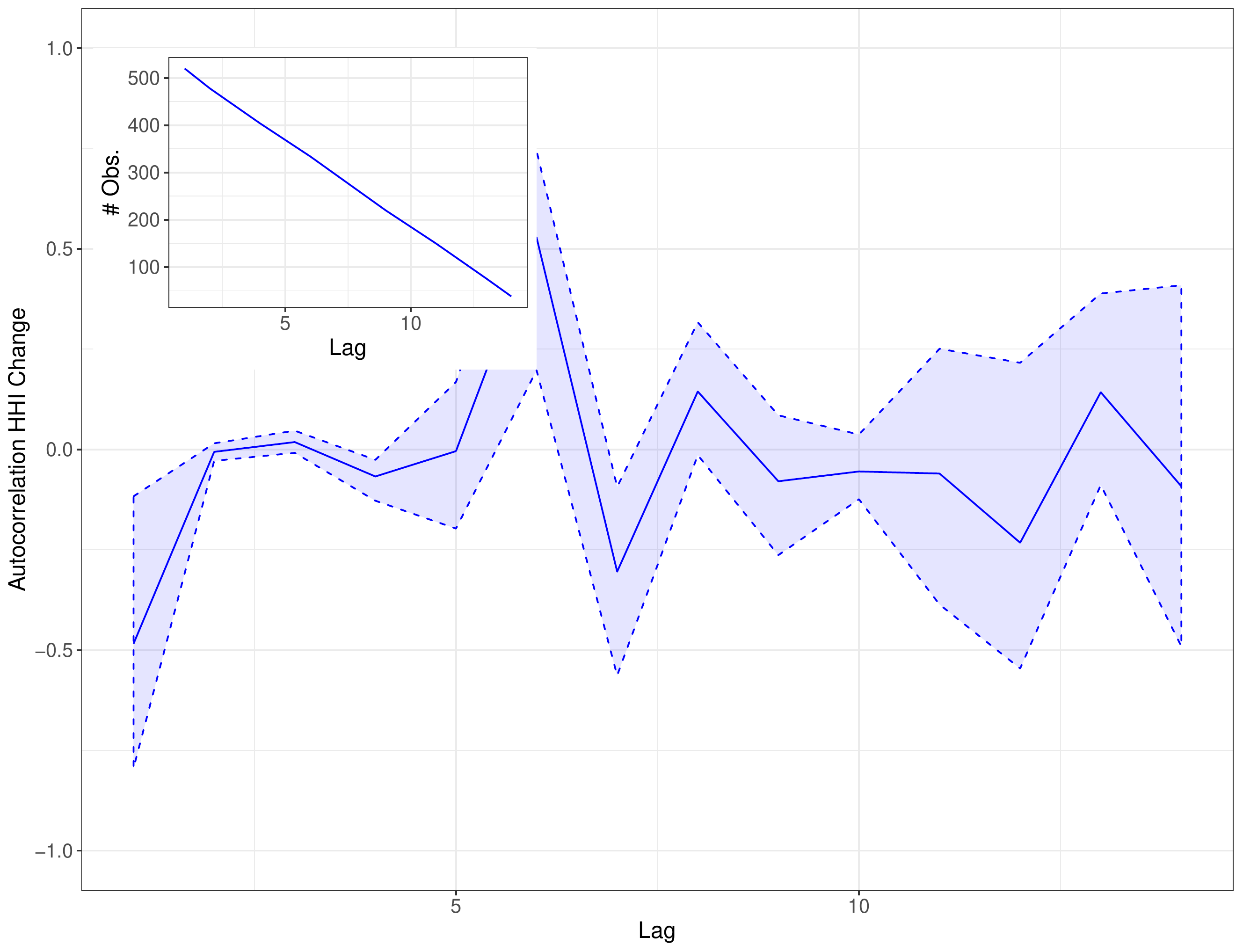}}
\subfloat[$\Delta$ Entropy]{\includegraphics[width=0.5\textwidth]{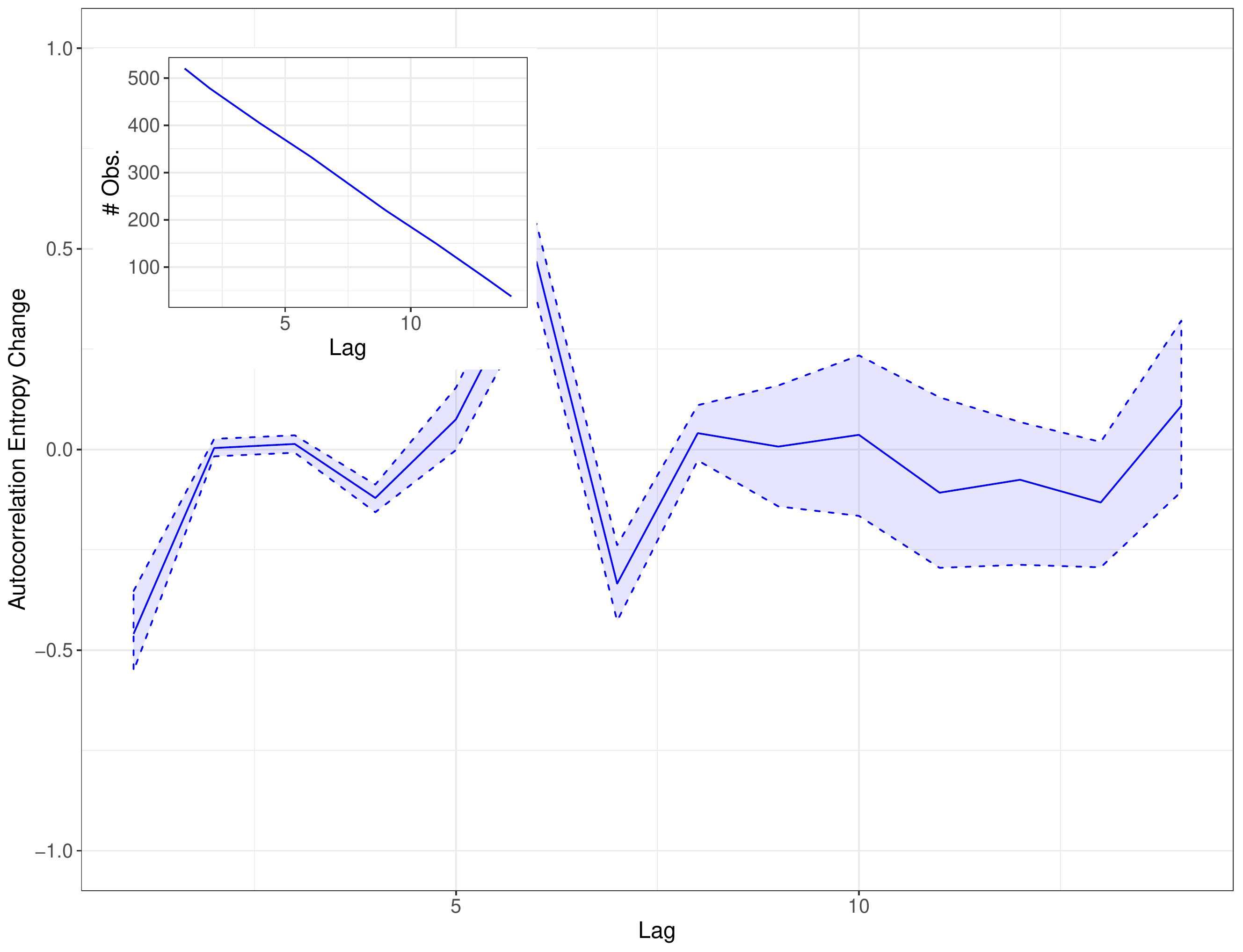}}
\caption{Autocorrelations of change of intra-sectoral dispersion (first difference of dispersion measures of shares of total assets)}
\label{fig:AC:disp:diff}
\end{figure}

Fig. \ref{fig:AC:disp} and \ref{fig:AC:disp:diff} show the autocorrelation spectra of dispersion measures (normalized Hirschmann-Herfindahl index and entropy) in levels (Fig. \ref{fig:AC:disp}) and first differences (Fig. \ref{fig:AC:disp:diff}) among sectors in terms of capital share (share of the economy's total capital).

\end{document}